\documentclass[aps,10pt,showpacs,preprintnumbers,nofootinbib]{revtex4-1}
\usepackage{graphicx}
\usepackage{dcolumn}
\usepackage{bm,amsfonts,amsthm,amsmath,amssymb}
\usepackage[]{epsfig,graphics}
\usepackage{comment}
\usepackage[T1]{fontenc}
\usepackage{lipsum}
\usepackage[utf8]{inputenc}
\usepackage{multirow}
\usepackage{color}
\usepackage{lastpage}
\usepackage{enumerate}
\usepackage{subfigure}
\usepackage{wasysym}
\usepackage{hyperref}
\usepackage{graphicx}
\usepackage{subfiles}
\usepackage{natbib}
\usepackage[normalem]{ulem}
\usepackage{float}

\hypersetup{
    bookmarks=true,         
    unicode=false,          
    pdftoolbar=true,        
    pdfmenubar=true,        
    pdffitwindow=false,     
    pdfstartview={FitH},    
    pdftitle={My title},    
    pdfauthor={Author},     
    pdfsubject={Subject},   
    pdfcreator={Creator},   
    pdfproducer={Producer}, 
    pdfkeywords={keyword1} {key2} {key3}, 
    pdfnewwindow=true,      
    colorlinks=false,       
    linkcolor=red,          
    citecolor=green,        
    filecolor=magenta,      
    urlcolor=cyan           
}

\newenvironment{itemize*}
  {\begin{itemize}
    \setlength{\itemsep}{0pt}
    \setlength{\parskip}{0pt}}
  {\end{itemize}}

\newenvironment{enumerate*}
  {\begin{enumerate}
    \setlength{\itemsep}{0pt}
    \setlength{\parskip}{0pt}}
  {\end{enumerate}}

\newenvironment{description*}
  {\begin{description}
    \setlength{\itemsep}{0pt}
    \setlength{\parskip}{0pt}}
  {\end{description}}

\def\ben{\begin{enumerate*}}
\def\een{\end{enumerate*}}
\def\bi{\begin{itemize*}}
\def\ei{\end{itemize*}}
\def\bd{\begin{description*}}
\def\ed{\end{description*}}
\def\be{\begin{equation}}
\def\ee{\end{equation}}
\def\bea{\begin{eqnarray}}
\def\eea{\end{eqnarray}}
\def\bfl{\begin{flushleft}}
\def\efl{\end{flushleft}}

\newcommand{\gsim}{\lower.7ex\hbox{$\;\stackrel{\textstyle>}{\sim}\;$}}
\newcommand{\lsim}{\lower.7ex\hbox{$\;\stackrel{\textstyle<}{\sim}\;$}}

\newcommand{\beq}{\begin{equation}}
\newcommand{\eeq}{\end{equation}}

\newcommand {\xmm} {\textsl{XMM-Newton}}

\newcommand {\swift} {\textsl{Swift}}
\newcommand {\rxte} {\textsl{RXTE}}

\newcommand {\nustar} {\textsl{NuSTAR}}
\newcommand {\suzaku} {\textsl{Suzaku}}
\newcommand {\integral} {\textsl{INTEGRAL}}

\begin{document}

\begin{flushright}
MI-TH-214\\
INT-PUB-21-004
\end{flushright}
\title{Axions: \\ From Magnetars and Neutron Star Mergers to Beam Dumps and BECs}

\author{Jean-Fran\c{c}ois Fortin}
\email{jean-francois.fortin@phy.ulaval.ca}
\affiliation{D\'epartement de Physique, de G\'enie Physique et d'Optique,\\Universit\'e Laval, Qu\'ebec, QC G1V 0A6, Canada}
\author{Huai-Ke Guo}
\email{ghk@ou.edu}
\affiliation{Department of Physics and Astronomy, University of Oklahoma, Norman, OK 73019, USA}
\author{Steven P. Harris}
\email{harrissp@uw.edu}
\affiliation{Institute for Nuclear Theory, University of Washington, Seattle, WA 98195, USA}
\author{Doojin~Kim}
\email{doojin.kim@tamu.edu}
\affiliation{Mitchell Institute for Fundamental Physics and Astronomy,
Department of Physics and Astronomy, Texas A\&M University, College Station, TX 77843, USA}
\author{Kuver Sinha}
\email{kuver.sinha@ou.edu}
\affiliation{Department of Physics and Astronomy, University of Oklahoma, Norman, OK 73019, USA}
\author{Chen Sun}
\email{chensun@mail.tau.ac.il}
\affiliation{School of Physics and Astronomy, Tel-Aviv University, Tel-Aviv 69978, Israel}

\date{\today}

\begin{abstract}
We  review topics in searches for axion-like-particles (ALPs), covering material that is complementary to other recent reviews. The first half of our review covers ALPs in the extreme environments of neutron star cores, the magnetospheres of highly magnetized neutron stars (magnetars), and in neutron star mergers. The focus is on possible signals of ALPs in the photon spectrum of neutron stars and gravitational wave/electromagnetic signals from neutron star mergers.  We then review recent developments in laboratory-produced  ALP searches, focusing mainly on accelerator-based facilities including beam-dump type experiments and collider experiments. We provide a general-purpose discussion of the ALP search pipeline from production to detection, in steps, and our discussion is  straightforwardly applicable to most  beam-dump type and reactor experiments. We end  with a selective look at  the rapidly developing field of ultralight dark matter, specifically the formation of Bose-Einstein Condensates (BECs).  We review the   properties of BECs of ultralight dark matter and bridge these properties with developments in  numerical simulations, and ultimately with their impact on  experimental searches.

\end{abstract}
\maketitle
\thispagestyle{empty}
\tableofcontents
\newpage
\thispagestyle{empty}

\section{Introduction}

Since its introduction  more than four decades ago \cite{Weinberg:1977ma, Peccei:1977hh, Wilczek:1977pj, Preskill:1982cy, Dine:1982ah, Kim:1979if}, the axion has come to occupy a central role in several important directions of research. Indeed, the term ``axion'', originally introduced by Wilczek to denote the particle associated with the Peccei-Quinn solution to the strong-$CP$ problem, has now transcended the original context in which it was introduced. In quantum field theory, the term can mean generic pseudoscalar Goldstone bosons described by a two-parameter model $(m_a, f)$, where $m_a$ is the mass of the particle and $f$ is the scale of a spontaneously broken chiral symmetry. In string theory, the axion may refer to general pseudoscalar matter fields (open string axions) or to fields  that are components of the complex degrees of freedom associated with compact extra dimensions (closed string axions). We will use the terms ``Axion-like-particle'' (ALP) and  ``axion'' interchangeably in this review, generally remaining agnostic about the specific context in which the particle arises.

Aspects of the physics of axions have been reviewed extensively by different authors over the years. Its connection to the strong-$CP$ problem was reviewed by Peccei in \cite{Peccei:2006as} and Kim et al.\ in \cite{Kim:2008hd}. Axions in the context of string theory have been explored in \cite{Svrcek:2006yi, Arvanitaki:2009fg, Cicoli:2012sz, Ringwald:2012hr}. The topic of axion inflation was reviewed in \cite{Pajer:2013fsa}. Many reviews on axion detection methods have also been published, based on the pioneering ideas of Sikivie (direct detection) \cite{Sikivie:1983ip} and Raffelt (indirect detection) \cite{Raffelt:1987im, Raffelt:1990yz}. The search for axions has certainly entered a golden age. Building on these  original ideas, numerous proposals to detect ALPs have been explored recently. These proposals span an astonishing range of experimental settings, from laboratory-based facilities to astroparticle physics.

The last few years have seen a proliferation of reviews on axions and ALPs. For the benefit of the reader, we briefly summarize the areas covered by them and comment on where the current review falls. On the more theoretical end of the spectrum are the reviews by Choi et al.\ \cite{Choi:2020rgn}, Hook's TASI lectures \cite{Hook:2018dlk}, Di Luzio et al.\ \cite{DiLuzio:2020wdo} and Marsh \cite{Marsh:2015xka}. Hook's lectures provide a thorough and intuitive introduction to the strong-$CP$ problem and the role of axions in solving it. Choi et al.'s review covers relatively new developments at the interface of string phenomenology and axions, such as the connection between the Weak Gravity Conjecture and axion field ranges \cite{Cheung:2014vva, Heidenreich:2015wga}. The review by  Di Luzio et al.\ covers the QCD axion in great detail, while the one by Marsh covers all aspects of axion cosmology, as well as many aspects of astrophysical searches for axions. 
On the more experimental end of the spectrum are several reviews by theorists, including Graham et al.\ \cite{Graham:2015ouw}, Irastorza et al.\ \cite{Irastorza:2018dyq}, and most recently Sikivie \cite{Sikivie:2020zpn}. Combined, these reviews provide an in-depth look at almost every aspect of current experimental searches for axions. 

The  current review is placed firmly at the experimental end of the spectrum. We will focus on a set of topics that are complementary to those covered by \cite{Graham:2015ouw, Irastorza:2018dyq, Sikivie:2020zpn}, covering the following topics:

\smallskip 

\noindent $(1)$ \textbf{Axions in extreme astrophysical environments:} The first half of the review, Sect.~\ref{sec:magnet} and Sect.~\ref{sec:merger},  will cover axions in the extreme environments of neutron star cores, the magnetospheres of highly magnetized neutron stars (magnetars), and in neutron star mergers. The focus will be on possible signals of axions in the photon spectrum of neutron stars and gravitational wave/electromagnetic signals from neutron star mergers. 

$(1a)$ For neutron stars, we will cover in detail the production of axions from the strongly degenerate nuclear matter in the core of the star.  This is a subject with a long history, going back to the 1980's \cite{Iwamoto:1984ir, PhysRevD.38.2338, Raffelt:1996wa} and becomes important in deriving axion constraints from supernovae, cooling neutron stars, and -- as we will emphasize -- in the emerging body of work studying the photon spectrum from neutron stars to constrain axions. In addition to reviewing the original literature, we will also discuss the effects of nucleon superfluidity on the production rate.  When the nuclear matter is not superfluid, axions are produced in the neutron star core via nucleon bremsstrahlung processes.  When one or both nucleon species are superfluid, the rate of the production due to bremsstrahlung is diminished, but a new production mechanism appears, where axions are created by the thermally-induced formation of nucleon Cooper pairs. These issues will be discussed in detail. We will also discuss several other issues, including changes in the emissivity coming from corrections to the one-pion-exchange approximation. 

One of the newer features that we will emphasize is the spectral features of the axions produced from the core. The question of spectral features  is not directly relevant for limits derived from cooling, which depend only on the integrated luminosity. However, they are relevant for constraints on axions stemming from studying the actual X-ray and gamma-ray spectra of neutron stars. The way that axions can be probed by studying such spectra is as follows. Once created, the axions escape from the interior of the neutron star  and have a probability of converting to X-ray and gamma-ray photons in the magnetic field of the  magnetosphere.  The resulting axion-converted photons constitute an exotic source of  emission from neutron stars different from any putative standard astrophysical processes. Features of the observed spectrum can thus be used to constrain the product of the coupling of axions to nucleons and photons. The focus of our attention will be on a class of highly magnetized neutron stars called magnetars, although the discussion would apply to other categories of neutron stars as well. 

Throughout, we will compare and contrast the conversion of axions to photons in large-scale magnetic fields, which is the standard method of probing them in astroparticle physics, versus the conversion process in the vicinity of neutron stars. We will provide a lightning review of some of the main trends in the former topic. It is hoped that the review will help the reader navigate, on the one hand, the field of particle physics in nuclear environments, and on the other hand, the fundamental physics of axions, especially in the context of X-ray and soft gamma-ray astronomy of neutron stars.

$(1b)$ We will review in detail neutron star mergers and the possibility of using them as a laboratory for the physics of axions, and perhaps hidden sectors in general. One of the distinguishing features of mergers is multi-messenger astronomy, and the hope is that gravitational wave, electromagnetic, and correlated data from mergers can be leveraged to investigate axions. This calls for collaboration between beyond-Standard-Model physicists, nuclear physicists, and experts on merger physics and simulations. Our hope is that this review will be a small step in furthering this collaboration.

When two neutron stars merge, the constituent nuclear matter - already quite dense - reaches temperatures of tens of MeV, comparable to those reached in a core-collapse supernova.  At these temperatures, the nuclear matter is no longer superfluid, and thus axions are primarily produced by nucleon bremsstrahlung. We review the calculation of the axion mean free path and  find that it is sufficiently long that axions would not be trapped anywhere in a neutron star merger.  Axions produced during the merger will free-stream through the nuclear matter and will take energy away from the remnant, cooling it.  We review the calculation of the cooling timescale due to axion emission and the results of a merger simulation which incorporated this cooling.  

While ultralight axions are not the focus of this part of our review, we will briefly indicate how a particular type of such species with a kilometer-scale Compton wavelength can impact the dynamics of the inspiral phase of a neutron star merger.

\smallskip

\noindent $(2)$ \textbf{Axions at accelerator-based experiments:} 
We  review recent developments in laboratory-produced axion or ALP searches in Sect.~\ref{sec:labsearch}, focusing mainly on accelerator-based facilities including beam-dump type experiments and collider experiments.
While astrophysical considerations afford excellent avenues to look into the axion parameter space, laboratory-based searches can constrain models of axions  in the most model-independent (hence conservative) fashion. These searches are not predicated upon  assumptions (e.g., stellar models) that astrophysical searches may take. 
Indeed, in the context of the (ultimately transient) PVLAS anomaly~\cite{Zavattini:2005tm} for which the preferred parameter region was already excluded by CAST~\cite{Zioutas:1998cc}, and in the wake of the more recent EDGES~\cite{Bowman:2018yin} and Xenon1T~\cite{Aprile:2020tmw} anomalies,  laboratory-based probes have received particular attention as they can provide the most stringent guidelines in constructing axion or ALP models.  

$(2a)$ The first half of Sect.~\ref{sec:labsearch} will be devoted to searches at beam-dump type and reactor experiments, including next-generation ones and related phenomenological studies. 
Many such experiments featuring highly intensified neutrino fluxes have begun their operation or are being seriously planned. Much of the literature has pointed out that they are capable of probing the ALP parameter space, and the associated experimental data will surge in the near future. Key specifications of these experiments are collected and tabulated in our review. We provide a general-purpose discussion of the ALP search pipeline from production to detection, in steps, and our discussion is  straightforwardly applicable to most of the beam-dump type and reactor experiments. 
For detection of ALPs, we cover not only the traditional search scheme that depends on ALP decays but newly proposed search channels (e.g., axion scattering and conversion channels).
We review the existing and future expected limits in all three channels, showing the complementarity among them. 

$(2b)$ The other half of Sect.~\ref{sec:labsearch} is reserved for a review of  recent developments in collider ALP searches. As colliders feature relatively large center-of-mass energy, they have played an important role in the search for MeV-to-TeV mass-range axions or ALPs. Moreover, models of ALPs interacting with SM heavy resonances, e.g., massive gauge bosons and Higgs, can be exclusively probed at colliders, especially through on-shell production of such resonances. 
As future proposed colliders are expected to produce them even more copiously, they will provide particular opportunities and richer phenomenology in the associated channels.
All these opportunities available at  existing and future colliders have been actively exploited over the past decade. 
We review this series of efforts and assort future energy-frontier colliders together with their key parameters. 

\smallskip

\noindent $(3)$ \textbf{Galactic and Stellar Bose-Einstein Condensates:} We will end this review with a selective look at  the rapidly developing field of ultralight dark matter (DM), specifically the formation of Bose-Einstein Condensates (BECs). Being scalars, ultralight dark matter exhibits unique collective properties that are different from heavy dark matter, such as behaving more like waves than point particles. This in turn leads to interesting dynamics and profiles, as well as constraints such as that from wave function stability. In Sect.~\ref{sec:BEC-sec} we look into these properties of BECs of ultralight dark matter and bridge them with numerical simulations, and ultimately with their impact on  experimental searches. We break this aspect of the review into two connected parts. 

$(3a)$ We start with the  requirement of consistency. We estimate the critical condition for BEC to form, which gives an upper limit on the ultralight dark matter mass. We briefly review the cosmological evolution of an ultralight scalar field and show the Jean's scale due to its quantum pressure. In addition, we also estimate the effect on the Jean's scale if the scalar field has a sizable quartic self-interaction.  After that, we discuss the stability requirement of the BEC system, which puts an upper limit on the size of the BEC structure. 

$(3b)$ We briefly go through the recent simulations of BEC dark matter. We show that the simulations indicate an interesting mass relation between the BEC core and the dark matter halo. 
In concluding this section, we quote a few constraints on ultralight dark matter on the galactic scale, including the results of cosmic microwave background (CMB) weak lensing, the Lyman-$\alpha$, stellar stream and strong lensing constraints on the halo mass function, Milky Way satellite counting, UV luminosity function, as well as the empirical BEC-halo mass relation. On the stellar scale, we point to a few references where novel signals could be used to probe boson stars.

\smallskip

Since much of the review is on search strategies, we will not cover any issues of the underlying  theory or model-building aspects of axions. Rather, we will simply start with the parts of the axion Lagrangian that are directly relevant for the search strategies we pursue: namely, its coupling to photons, electrons, and nucleons. 
\begin{equation}
\mathcal{L}_{\rm int} \supset \frac{1}{4}g_{a\gamma\gamma}aF_{\mu\nu}\tilde{F}^{\mu\nu}+ig_{aee}a\bar{\psi}_e\gamma_5\psi_e + G_{an} (\partial_\mu a)\bar{N}\gamma^\mu \gamma_5 N \label{eq:intlag}
\end{equation}
where $a$ denotes the ALP field, $F_{\mu\nu}$ the electromagnetic field strength tensor, $\tilde{F}^{\mu\nu}=(1/2)\epsilon^{\mu\nu\rho\sigma}F_{\rho\sigma}$ the dual electromagnetic field strength tensor, $\psi_e$ the electron field, and $N$ the nucleon field.  The reader is directed to the excellent theoretical reviews cited above for more details.

Before proceeding, we mention some of the numerous topics that will \textit{not} be covered in this review. Sect.~\ref{sec:magnet} will only scratch the surface of the vast topic of axion conversion in large-scale magnetic fields. Sect.~\ref{sec:labsearch} will skip over most of the classic laboratory-based techniques of  axion detection, which are covered in detail by the recent review of Sikivie \cite{Sikivie:2020zpn}. Likewise, Sect.~\ref{sec:BEC-sec} will omit many topics in the field of ultralight axions and dark matter, foremost among them the idea of superradiance.


\section{Astrophysical Searches: Axions and Neutron Stars \label{sec:magnet}}

This section will be structured as follows. In Sect.~\ref{largevsneut}, we will first contrast the conversion of axions in large-scale magnetic fields to their conversion in the localized magnetic fields near neutron stars. In Sect.~\ref{alpprodns}, we will  review the production of axions from the core of neutron stars. In Sect.~\ref{sec:magnetars}, we will then describe their fate in the magnetosphere, by describing the evolution equations of the axion-photon system and deriving expressions for the probability of conversion. We will use these results to show recently obtained constraints on axion couplings and discuss the possibility of using polarization to probe axions. Finally, in Sect.~\ref{egalp} we will  provide a lightning review of the conversion of axions in large-scale magnetic fields. 

\subsection{Axion conversions in large-scale magnetic fields versus near neutron stars}\label{largevsneut}

The conversion of ALPs to photons or vice versa in astrophysical magnetic fields constitutes a major search strategy for these particles. This conversion could happen in large-scale magnetic fields (a long-standing topic of study) or localized magnetic fields near compact objects like neutron stars and magnetars (a relatively newer topic of study). We discuss these topics in turn.

$(i)$ Galactic and extra-galactic case: In the traditional large-scale conversion scenario, 

the general  theme is as follows: Photons are emitted from a distant source and travel to the earth through intervening magnetic fields, where they undergo conversion to ALPs; this conversion results in spectral features of the source that may be discerned over background emission due to standard astrophysical processes, resulting in constraints on ALPs. The sources that have been studied in this framework are diverse: active galactic nuclei \cite{Conlon:2018iwn} including blazars \cite{Buehler:2020qsn} and  quasars, supernovae \cite{Calore:2020tjw} and in particular SN 1987A \cite{Payez:2014xsa}, red supergiant stars like Betelgeuse \cite{Xiao:2020pra}, etc. Much of this program has centered on ALP-photon  inter-conversion in the large-scale astrophysical magnetic fields that the ALP-photon system must traverse through in order to reach the earth. Depending on the location of the source where the ALP-photon system originates, these intervening  environments could include the magnetic fields of the host galaxy \cite{Conlon:2018iwn}, extragalactic space \cite{Kartavtsev:2016doq}, the Milky Way \cite{Day:2015xea} and  the specific environment at the origin, for example the field within the blazar jet  \cite{Dobrynina:2014qba}. The accurate modeling of these intervening large-scale magnetic fields -- in galactic and extragalactic environments and in active galactic nuclei, BL Lac jets, etc. -- presents a serious challenge in these endeavors. The literature has generally resulted in leading constraints on $g_{a\gamma\gamma}$ for ALP masses below $\sim \mathcal{O}(10^{-12})$ eV. 

$(ii)$ Localized case: ALP-photon production and inter-conversion near \textit{localized} sources, especially in the dipolar magnetic fields of neutron stars and magnetars, will be the main focus of this section. In these scenarios, ALPs present either in the  dark matter halo surrounding the neutron star (non-relativistic case) or emitted from the core of the neutron star (relativistic case) convert to photons in the dipolar magnetic field near the neutron star, yielding hard X-rays or gamma-rays in the relativistic case and radiowaves in the dark matter case, respectively. In contrast to the large-scale conversion case, the magnetic field is nearly critical in strength ($\sim 10^{14}$ G), is much more accurately known, and the distance traversed by the ALP during the conversion is small (typically a few thousand kilometers). This is a fundamentally different method of probing ALPs, in that the conversion occurs in the vicinity of the neutron star. The magnetic field is well-approximated by a dipole, in contrast to the galactic-scale conversion case, where the modeling of the field morphology is highly non-trivial. Due to the differences in the spatial dependence of the magnetic field between the two cases, the ALP-photon propagation equations have solutions whose parametric behaviors in the two cases are also  different. It therefore turns out that the two methods probe fundamentally different mass and coupling regimes of ALPs, with the localized case being sensitive to ALP masses below $\sim \mathcal{O}(10^{-5})$ eV. 

We now go on to a discussion of first the production of axions from the cores of neutron stars, and then their conversion in the magnetosphere.

\subsection{ALP production in neutron star cores} \label{alpprodns}
The dense matter in the core of neutron stars is strongly degenerate nuclear matter, consisting of Fermi seas of neutrons, protons, electrons, and muons \cite{glendenning2000compact,Shapiro:1983du}.  At the highest densities encountered in neutron stars, exotic phases of matter may appear (see the reviews and textbooks \cite{glendenning2000compact,Shapiro:1983du,Baym:2017whm,Pethick:2015jma,Alford:2007xm,Oertel:2016bki,Alford:2019oge}), but we do not discuss this possibility here.  As one moves from the core to the edge of the star, the density of the nuclear matter decreases and eventually the nuclear matter transitions to a solid crust, containing a Coulomb lattice of nuclei and a degenerate Fermi gas of electrons \cite{Douchin:2001sv,Oertel:2016bki,Chamel:2008ca}.  

Axions can be produced both in the core and the crust of the neutron star.  In the uniform nuclear matter of the neutron star core, the dominant production channels are the three nucleon bremsstrahlung processes $N+N'\rightarrow N+N'+a$, where $N$ and $N'$ are either neutrons or protons.\footnote{Nucleon bremsstrahlung is also the dominant way of producing axions in supernovae \cite{PhysRevD.38.2338,Carenza:2019pxu,Chang:2018rso,Lee:2018lcj,Raffelt:1996wa}.}  In the case where the nucleons are superfluid, these three processes can still proceed and produce axions, but the rate is Boltzmann suppressed, as we will discuss later in this section.  In addition, the presence of superfluid nucleons creates a new axion production process, from the formation of Cooper pairs.

In the solid crust of a neutron star, electron scattering off of nuclei $e^-+(Z,A)\rightarrow e^-+(Z,A)+a$ is likely to be the dominant axion production mechanism~\cite{Sedrakian:2018kdm,Iwamoto:1984ir}.\footnote{Electron-nucleus scattering is also the dominant axion production mechanism in white dwarfs \cite{Dessert:2019sgw,Giannotti:2017hny,Nakagawa:1987pga,Nakagawa:1988rhp} and red giant stars \cite{Giannotti:2017hny}.}  However, free neutrons in the neutron star crust exist and could be superfluid \cite{Sedrakian:2018ydt,Chamel:2008ca}, so the possibility of other axion production channels exists (for inspiration, see the neutrino pair production processes in the crust \cite{Yakovlev:2000jp}).  As it is expected to be subdominant to axion production in the core \cite{Sedrakian:2018kdm}, we will neglect axion production in the crust for the rest of this review.  Axion production processes are reviewed in \cite{Raffelt:1996wa,Raffelt:1990yz,Kolb:1990vq}.

\subsubsection{Superfluidity in neutron stars}
With the exception of right after their birth in a core collapse supernova \cite{Cerda-Duran:2018efz} or during a neutron star merger (see Sect.~\ref{sec:merger}), neutron stars are expected to be at sufficiently low temperature that the neutrons or the protons, or perhaps both, are in the superfluid phase.  Superfluid protons give rise to superconductivity, as protons are electrically charged.  The presence of superfluidity in the core of neutron stars has dramatic consequences for neutron stars, including their specific heat and rate of neutrino production.  Of consequence for this review, superfluidity can significantly alter the axion production rate in neutron star cores.

The nucleon-nucleon interaction is attractive under certain conditions, and thus the possibility exists for nucleons near the Fermi surface in degenerate nuclear matter to form Cooper pairs, as a consequence of the Cooper theorem \cite{Cooper:1956zz}.  Cooper pairing in a particular spin-angular momentum channel can occur between nucleons $N$ and $N'$ as long as the interaction is attractive at the energy scale of interest.  However, pairing between a neutron and a proton is unlikely in neutron stars due to the sizeable difference in the Fermi momentum between the two species~\cite{Sedrakian:1999cu}.\footnote{The Fermi momentum ratio $p_{Fn}/p_{Fp}=(n_n/n_p)^{1/3}=[(1-x_p)/x_p]^{1/3}\approx 2$ if the proton fraction $x_p\equiv n_p/n_B=0.1$.}  When the formation of Cooper pairs becomes possible, a gap at the Fermi energy appears in the single-particle energy spectrum of the paired species \cite{Bardeen:1957mv}.  The size of the energy gap varies with density, and at sufficiently high temperature the gap vanishes and the particle species is no longer in the superfluid phase (the non-superfluid phase is often called ``ungapped'').  A combination of nucleon-nucleon scattering data and nuclear theory calculations lead to our current understanding that the dominant pairing channels in neutron star matter are $^1S_0$ and $^3P_2$.  Superfluidity in nuclear matter and its consequences in neutron stars are reviewed in \cite{Sedrakian:2018ydt,Haskell:2017lkl,Page:2013hxa,Yakovlev:2000jp,Yakovlev:1999sk,landau1980course,10.1143/PTP.44.905}.  

Nucleon pairing can be constrained by studying the cooling of neutron stars, which, at least for the first hundred thousand years of the lifetime of the star, occurs predominantly through neutrino emission from the core \cite{Yakovlev:2004iq}.  Superfluidity strongly suppresses the rate of neutrino emission and thus slows the cooling of the neutron star.  In these studies, any possible contribution of axions to cooling is neglected.  The authors of \cite{Beznogov:2018fda} conducted simulations of neutron star cooling and compared the results to data from the supernova remnant HESS J1731-347.  Combining the results of their Markov Chain Monte Carlo analysis with the theoretical expectation that the proton $^1S_0$ critical temperature is higher than the neutron $^3P_2$ critical temperature \cite{Sedrakian:2018ydt,Haskell:2017lkl,Page:2013hxa}, they found that the critical temperature for proton singlet pairing must be larger than $4\times 10^9 \text{ K}$ in most of the neutron star core, while the critical temperature for neutron triplet pairing must be less than $3\times 10^8 \text{ K}$ in the entire core.  The authors of \cite{Beloin:2016zop} found similar results, comparing the results of their neutron star cooling simulations with cooling data from a wide variety of isolated neutron stars.  

Finally, although we will not discuss the neutron star crust in this review, it is expected that free neutrons in the inner crust of the neutron star pair in the $^1S_0$ channel \cite{Sedrakian:2018ydt,Chamel:2008ca}.  The phenomenological implications of this pairing are discussed in \cite{Haskell:2017lkl}.
\subsubsection{ALP emissivity}
\label{sec:magnetar_emissivity}
We will, in the coming paragraphs, develop the calculation of the axion emissivity from a neutron star core.  The emissivity measures the amount of energy lost via axion radiation per time, per volume.  Integrating the emissivity over the core gives the axion luminosity, which can be used to calculate the expected cooling rate of a neutron star due to axion emission \cite{Beznogov:2018fda,Sedrakian:2015krq,Sedrakian:2018kdm,Paul:2018msp,Hamaguchi:2018oqw}.  Later in this section, we will also be interested in the energy spectrum of the emitted axions, for the purpose of understanding the spectrum of the photons into which the axions have a probability of converting in the neutron star magnetosphere.  

In this section of the review, we focus on axion emission from the core of magnetars, although at the end of this section we briefly discuss more conventional neutron stars.  Magnetars are believed to be younger neutron stars (see Fig.~9 of \cite{Olausen:2013bpa}) with relatively high core temperatures and very strong magnetic fields.  While the surface temperature of a magnetar can be inferred from fitting the soft X-ray emission to a blackbody spectrum \cite{Turolla:2015mwa}, modeling is required to deduce the core temperature.  For standard neutron stars, this relationship is relatively well understood \cite{Potekhin:2015qsa,Yakovlev:2004iq}, but many magnetars have an anomalously high surface temperature, which would seemingly require their core temperature to be well above $10^9\text{ K}$.  Such a hot magnetar core is not believed to be sustainable over the lifetime of the magnetar, due to the large neutrino emissivity that would be generated \cite{Potekhin:2006iw,2012MNRAS.422.2632H}.  As a result, various mechanisms of surface heating due to the strong magnetic field have been proposed \cite{Beloborodov:2016mmx}.  Ordinarily, it is expected that the magnetar core is isothermal, due to its high thermal conductivity \cite{Pons:2019zyc}.  However, if it is the case that the high surface temperature is generated by magnetic heating in the interior of the star, then the core would no longer be isothermal \cite{Kaminker:2006ab}.  

Following \cite{Fortin:2021sst}, we will assume that the magnetar core temperature lies in the range $10^8 \text{ K}$ to a few times $10^9\text{ K}$.  In this case, the magnetar core temperature is expected to be larger than or comparable to the critical temperature of neutron triplet superfluidity, in which case it is common to consider superfluidity only in the proton singlet channel \cite{Fortin:2021sst,Leinson:2019cqv}, neglecting superfluidity of the neutrons.  The inclusion of neutron superfluidity is briefly discussed at the end of this section.

Following the lead of 
\cite{1979ApJ...232..541F}, the matrix elements for the three bremsstrahlung processes $N+N'\rightarrow N+N'+a$ are usually computed modeling the strong interaction of the two nucleons by assuming that the two nucleons exchange a pion.  This one-pion exchange (OPE) approximation \cite{OPE,Machleidt:2017vls} is used for many different types of nucleon bremsstrahlung processes including those that emit a neutrino-antineutrino pair \cite{1979ApJ...232..541F,Yakovlev:2000jp}, a CP-even scalar \cite{Dev:2020eam,Diener:2013xpa,Krnjaic:2015mbs,Lee:2018lcj,Ishizuka:1989ts}, a saxion \cite{Arndt:2002yg}, a Kaluza-Klein graviton \cite{Cullen:1999hc,Barger:1999jf}, or a dark photon \cite{Mahoney:2017jqk,Dent:2012mx}.

In the case of axion production, to which we now focus, the matrix element for $n+n\rightarrow n+n+a$ and $p+p\rightarrow p+p+a$ are \cite{PhysRevD.38.2338}
\begin{equation}
S\sum_{\text{spins}}\vert\mathcal{M}\vert^2 =C_\pi\frac{256}{3}(m_n/m_{\pi})^4 f^4 G_{an}^2\left[\frac{\mathbf{k}^4}{(\mathbf{k}^2+m_{\pi}^2)^2}+\frac{\mathbf{l}^4}{(\mathbf{l}^2+m_{\pi}^2)^2}+\frac{\mathbf{k}^2\mathbf{l}^2-3(\mathbf{k}\cdot \mathbf{l})^2}{(\mathbf{k}^2+m_{\pi}^2)(\mathbf{l}^2+m_{\pi}^2)}\right],\label{eq:nnmatrixelement}
\end{equation}
assuming that the axion-neutron and axion-proton couplings have the same strength.  In \eqref{eq:nnmatrixelement}, the symmetry factor is $S=1/4$ to account for the identical particles in both initial and final states.  The matrix element for $n+p\rightarrow n+p+a$ is
\begin{equation}
    S\sum_{\text{spins}}\vert\mathcal{M}\vert^2 =C_\pi\frac{1024}{3}(m_n/m_{\pi})^4 f^4 G_{an}^2\left[\frac{\mathbf{k}^4}{(\mathbf{k}^2+m_{\pi}^2)^2}+\frac{4\mathbf{l}^4}{(\mathbf{l}^2+m_{\pi}^2)^2}-2\frac{\mathbf{k}^2\mathbf{l}^2-3(\mathbf{k}\cdot \mathbf{l})^2}{(\mathbf{k}^2+m_{\pi}^2)(\mathbf{l}^2+m_{\pi}^2)}\right] ,\label{eq:npmatrixelement} 
\end{equation}
where the symmetry factor is simply $S=1$ due to the absence of identical particles in the initial or final states.  Although the matrix element \eqref{eq:npmatrixelement} was first computed in \cite{PhysRevD.38.2338}, a minus sign error was recently corrected in \cite{Carenza:2019pxu}.  In both \eqref{eq:nnmatrixelement} and \eqref{eq:npmatrixelement}, the factor $f\approx1$ is the pion-nucleon coupling constant \cite{1979ApJ...232..541F} and its definition introduces the extra neutron mass factor $m_n$.  As such, this factor is the vacuum neutron mass, not the effective neutron mass stemming from strong interactions in the magnetar core.  The three-vectors $\mathbf{k}=\mathbf{p_2}-\mathbf{p_4}$ and $\mathbf{l}=\mathbf{p_2}-\mathbf{p_3}$ represent the nucleon momentum transfer.  Finally, since the OPE approximation is known to overestimate the strength of the nuclear interaction, we introduce in \eqref{eq:nnmatrixelement} and \eqref{eq:npmatrixelement} a factor\footnote{The OPE potential is known to reproduce some, but not all, features of the nucleon-nucleon interaction \cite{greiner1996nuclear,bertulani2007nuclear}.  In \cite{Rrapaj:2015wgs}, the authors pointed out that the OPE approximation drastically underestimates the nucleon-nucleon scattering cross-section at low center-of-mass energy, and overestimates it by an increasingly large factor as the collision energy rises (see Fig.~3 in their paper).  To get an estimate of the discrepancy $C_{\pi}$ between the OPE approximation and QCD, the authors of \cite{Hanhart:2000ae} calculated the rate of $n+n\rightarrow n+n+a$ in the soft radiation approximation (SRA) where, as the axion energy goes to zero (technically, as the axion energy drops far below the center-of-mass energy of the nucleon scattering), the bremsstrahlung rate is directly related to the on-shell nucleon-nucleon scattering amplitude.  The calculation is model-independent, but neglects many-body effects.  This approach shows that the OPE approximation overestimates the emissivity by about a factor of 4 near nuclear saturation density, and so it has become common to choose $C_{\pi}=1/4$ \cite{Beznogov:2018fda,Fortin:2021sst}, although slightly different values are also used \cite{Sedrakian:2015krq,Lee:2018lcj}.  The SRA was also used to improve the OPE estimate of the emissivity of neutrinos \cite{Timmermans:2002hc,vanDalen:2003zw}, Kaluza-Klein gravitons and dilatons \cite{Hanhart:2000er}, and dark photons \cite{Rrapaj:2015wgs}.  

The above divergence is regulated by two effects.  For massive emitted particles, the energy of the emitted particle must be larger than its mass.  For both massive and massless emitted particles, the divergence is regulated by the finite decay width of the nucleon, the so-called Landau-Pomeranchuk-Migdal (LPM) effect \cite{Landau:1953gr,PhysRev.103.1811}, that kicks in only when the energy of the emitted particle is less than nucleon decay width.  Since the nucleon decay width is small for low temperatures, the LPM effect becomes important only when the temperature rises above 5-10 MeV \cite{vanDalen:2003zw}.} $C_\pi=1/4$.

The axion emissivity for any one of the nucleon bremsstrahlung processes is given by the phase space integral
\begin{align}
Q&= \int \frac{\mathop{d^3p}_1}{\left(2\pi\right)^3}\frac{\mathop{d^3p}_2}{\left(2\pi\right)^3}\frac{\mathop{d^3p}_3}{\left(2\pi\right)^3}\frac{\mathop{d^3p}_4}{\left(2\pi\right)^3}\frac{\mathop{d^3\omega}}{\left(2\pi\right)^3}\frac{S\sum \vert \mathcal{M}\vert^2}{2^5 E_1^*E_2^*E_3^*E_4^*\omega}\omega\label{eq:emissivity_integral}\\
&\times\left(2\pi\right)^4\delta^4(p_1+p_2-p_3-p_4-p_a)f_1f_2\left(1-f_3\right)\left(1-f_4\right).\nonumber
\end{align}
where $\omega$ is the energy of the axion and $f_i$ is the Fermi-Dirac factor of the appropriate nucleon.  The factors of energy in the denominator relate to the normalization of the nucleon wavefunctions.\footnote{When the nuclear interaction is described with a relativistic mean field theory, the nucleon energy denominators are given by $E^*=\sqrt{p^2+m_*^2}$, where $m_*$ is the Dirac effective mass of the nucleon \cite{glendenning2000compact}.  The nucleon wavefunctions and a discussion of calculating spin-summed matrix elements in a relativistic mean field theory are given in \cite{Roberts:2016mwj}, although \cite{Harris:2020qim,Reddy:1997yr,Fu_2008} also contain useful information.  In other treatments of the nucleons, the extreme nonrelativistic limit is taken, and the nucleon energy $E^*=m \approx 940 \text{ MeV}$ is chosen (see \cite{Dent:2012mx} and Appendix C of \cite{Harris:2020rus}).}  In magnetars, which contain strongly degenerate nuclear matter due to the fact that their temperatures lie well below 1 MeV, the emissivity can be calculated in the ``Fermi surface approximation'', described in \cite{Fortin:2021sst} (see also \cite{Harris:2020qim,Alford:2018lhf,Shapiro:1983du}), which assumes that nucleons near their Fermi surface provide the dominant contribution to the bremsstrahlung rate.  In degenerate nuclear matter where the protons, but not the neutrons, are superfluid, the axion emissivities from bremsstrahlung processes are
\begin{align}
    Q^0_{nn} &= \frac{31}{2835\pi}C_{\pi}(m_n/m_{\pi})^4f^4G_{an}^2p_{Fn}F(c)T^6 \label{eq:Qnn}\\
    Q^S_{pp} &= \frac{31}{2835\pi}C_{\pi}(m_n/m_{\pi})^4f^4G_{an}^2p_{Fp}F(d)T^6R_{pp}(n_B,T)\label{eq:Qpp} \\
    Q^S_{np} &= \frac{124}{2835\pi}C_{\pi}(m_n/m_{\pi})^4f^4G_{an}^2p_{Fp}G(c,d)T^6R_{np}(n_B,T).\label{eq:Qnp}
\end{align}
The derivations of these expressions are given in \cite{Fortin:2021sst}, as are the definitions of the functions $F$ and $G$, which depend on density through $c=m_{\pi}/(2p_{Fn})$ and $d=m_{\pi}/(2p_{Fp})$.  The functions $R_{np}$ and $R_{pp}$ arise from the gap in the proton energy spectrum, and Boltzmann suppress the corresponding bremsstrahlung rate by some function of $\exp{(-\Delta/T)}$, where $\Delta(n_B,T)$ is the superfluid gap in the proton energy spectrum.  Processes involving multiple ``gapped'' particles will experience additional suppression.  When the temperature rises above the critical temperature for proton pairing, the gap goes to zero and $R\rightarrow 1$.  The suppression factors $R$ are defined in \cite{Fortin:2021sst} and are further discussed in \cite{Yakovlev:1999sk}.

In nuclear matter below the critical temperature for proton pairing, Cooper-paired protons and unpaired protons coexist and at finite temperature, Cooper pairs are constantly breaking and reforming \cite{landau1980course,1976ApJ...205..541F}.  When a Cooper pair forms, at least $2\Delta$ of energy is liberated and can be carried by an emitted axion.  This process happens slowly far below the critical temperature, but the rate increases rapidly as the critical temperature is approached from below, since the fraction of protons in excited, unpaired states is higher.  Above the critical temperature, Cooper pairs cannot form.  The emissivity of axions due to $^1S_0$ proton Cooper pair formation is given by \cite{Keller:2012yr}
\begin{equation}
    Q_{CP} = \frac{G_{an}^2}{3\pi}\nu(0)v_{Fp}^2\Delta^2T^3 \int_0^{\infty}\mathop{dx} \frac{x^3}{(1+e^{x/2})^2}\frac{\theta(x-2\Delta/T)}{\sqrt{x^2-4\Delta^2/T^2}},\label{eq:Qcp}
\end{equation}
where $\nu(0)=m_p^Lp_{Fp}/(\pi^2)$ is the density of states at the proton Fermi surface \cite{coleman2015introduction}, the proton Fermi velocity is $v_{Fp}=\partial E_p/\partial p \vert_{p=p_{Fp}} = p_{Fp}/m_p^L$, and $m_p^L$ is the Landau effective mass of the proton \cite{Maslov:2015wba,Li:2018lpy}.

The authors of \cite{Fortin:2021sst} examine a selection of magnetars from the McGill magnetar catalog\footnote{The catalog is located at \href{http://www.physics.mcgill.ca/~pulsar/magnetar/main.html}{http://www.physics.mcgill.ca/~pulsar/magnetar/main.html} and more details on the catalog are provided in the associated paper \cite{Olausen:2013bpa}.} in an effort to constrain couplings of the axion to nucleons and photons.  As no mass measurements for the magnetars are available, the authors of \cite{Fortin:2021sst} assume that each magnetar is a $1.4M_{\odot}$ neutron star with nuclear matter described by the IUF equation of state \cite{Fattoyev:2010mx}.  Because the core temperatures are assumed to lie above $10^8\text{ K}$, neutron superfluidity is neglected, while $^1S_0$ proton pairing is included.  The zero-temperature proton gap $\Delta(T=0,n_B)$ is given by the parametrization (from \cite{Ho:2014pta}) of the calculation done by CCDK \cite{Chen:1993bam}, and is consistent with the constraints from neutron star cooling mentioned above.  Further details, including the relationship between the zero-temperature gap and the critical temperature as well as plots of the critical temperature and superfluid gaps throughout the $1.4M_{\odot}$ magnetar core are given in \cite{Fortin:2021sst}.  In the rest of this section, we will consider axion emission from this $1.4M_{\odot}$ magnetar.

\begin{figure}
\centering
\includegraphics[width=0.48\textwidth]{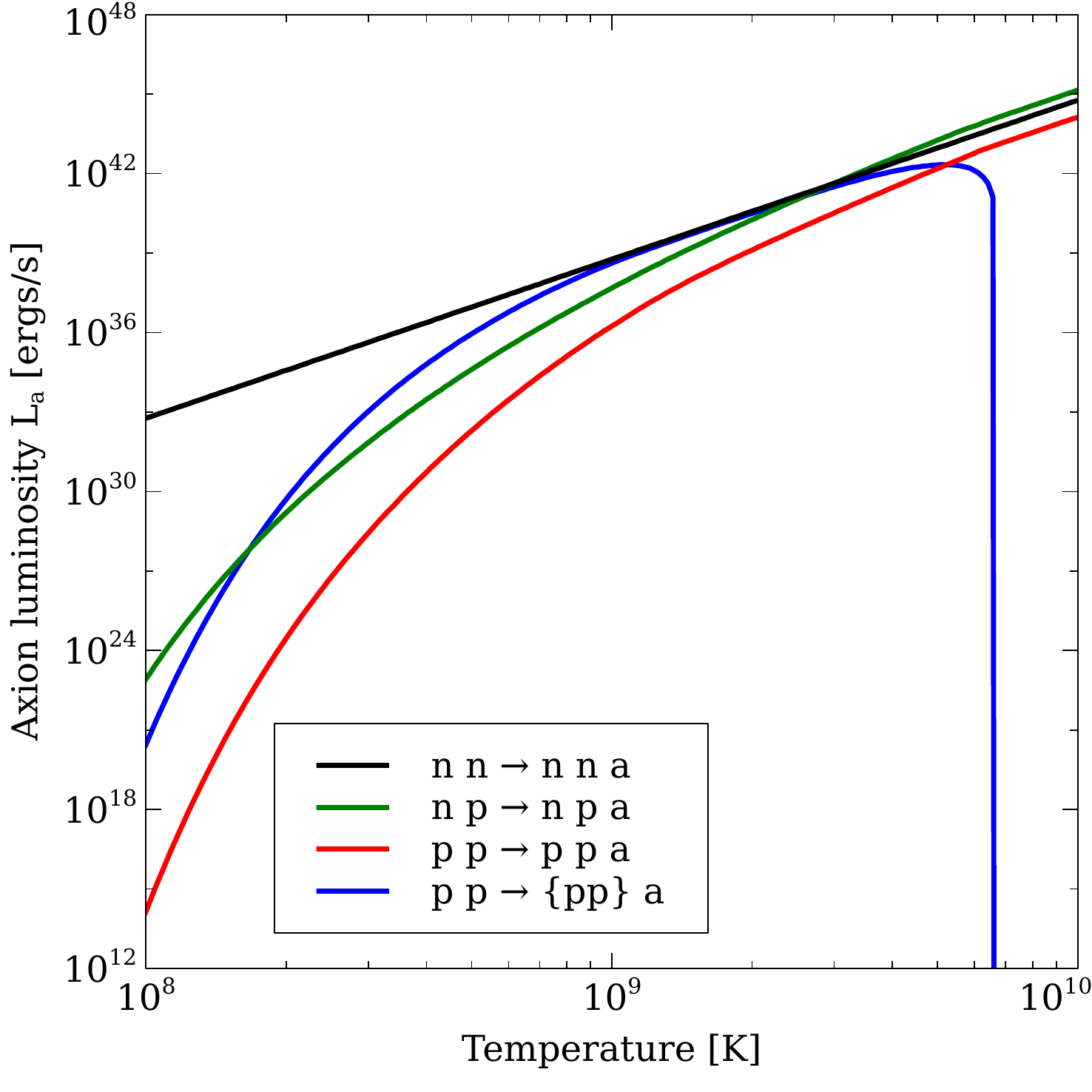}
\quad
\includegraphics[width=0.48\textwidth]{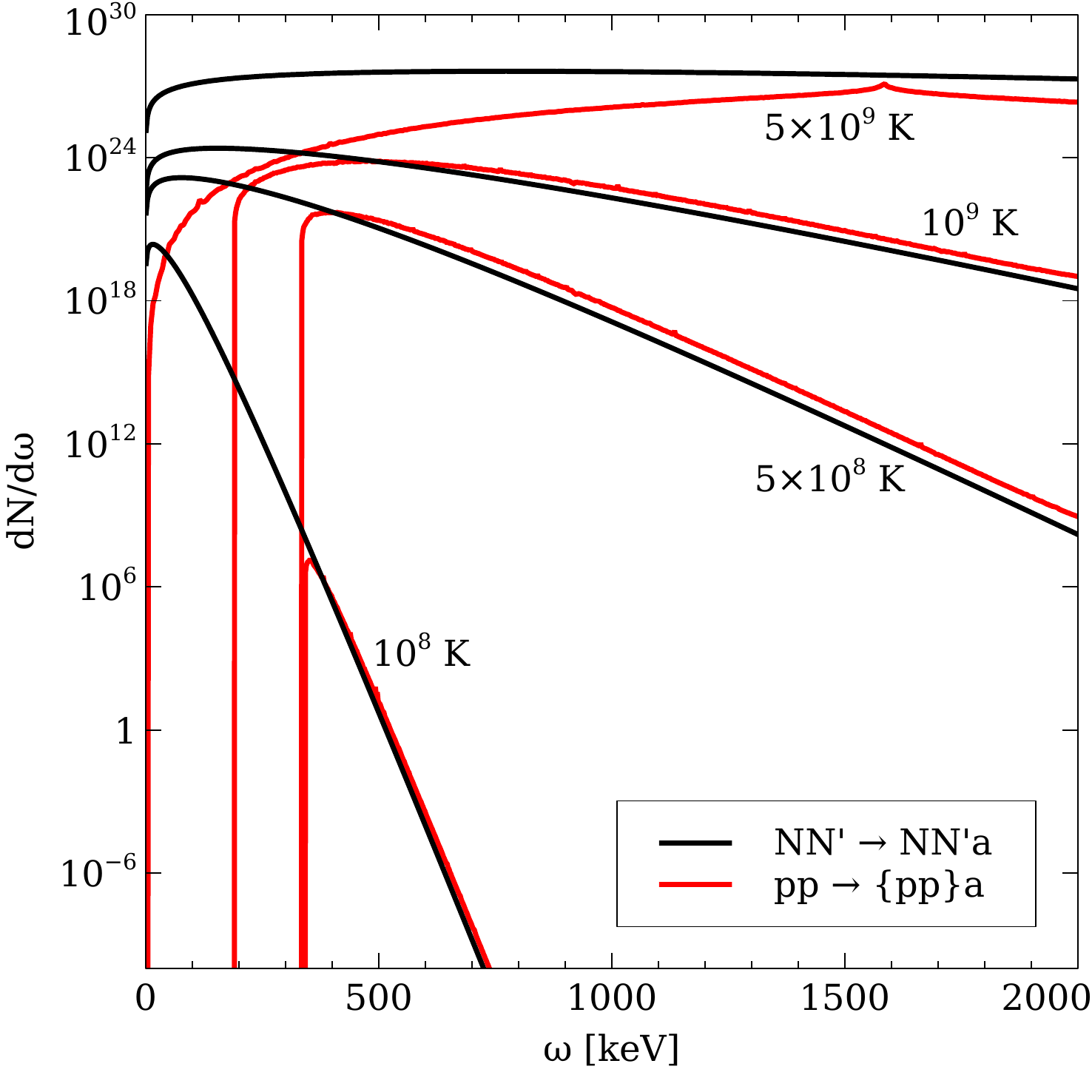}
\caption{Axion emission from the characteristic $1.4M_{\odot}$ magnetar with CCDK $^1S_0$ proton pairing discussed in the text.  In both panels, the axion-nucleon coupling constant has been chosen to be at the upper limit set by SN1987a $G_{an} = 7.9\times 10^{-10} \text{ GeV}^{-1}$ \cite{Graham:2015ouw}.  Left panel: Axion luminosity from the magnetar, split into contributions from three different nucleon bremsstrahlung processes $N+N'\rightarrow N+N'+a$ and from the formation of proton Cooper pairs $p+p\rightarrow \{pp\}+a$.  At low temperatures, the protons are superfluid and processes involving protons are strongly suppressed.  As the core temperature approaches the critical temperature in much of the magnetar, the Cooper pair formation process becomes more prominent.  When the core rises above $T\approx 7\times 10^9\text{ K}$, none of the star is superfluid (we ignore the crust, which might have a high critical temperature for $^1S_0$ neutron pairing) and the bremsstrahlung processes operate as they do in ungapped nuclear matter.  Right panel: 
Spectrum of axions emitted from the magnetar due to the three nucleon bremsstrahlung processes (combined) and Cooper pair formation.  The low-energy axion spectrum comes from the bremsstrahlung processes, but the Cooper pair formation process produces most of the high-energy axions, as long as the protons in most of the star are superfluid.  \label{fig:axion_emission_profiles}
}
\end{figure}

In the left panel of Fig.~\ref{fig:axion_emission_profiles}, we plot the luminosity of axions emitted from the $1.4M_{\odot}$ magnetar discussed above, split into the individual contributions of each of the three bremsstrahlung processes and the $^1S_0$ proton Cooper pair formation process.  When the temperature of the magnetar core is well less than the critical temperature (which is always above $10^9 \text{ K}$ in the CCDK model considered here), the Boltzmann suppression $\exp{(-\Delta/T)}$ for any process involving a proton is very strong, and so the neutron bremsstrahlung process dominates the axion production.  For core temperatures well less than $10^9 \text{ K}$, most protons near the Fermi surface are already Cooper paired, so the rate of axion production from the Cooper pair formation process is small.  As the core temperature approaches $10^9 \text{ K}$, the fraction of protons near the Fermi surface that are unpaired increases and therefore the Cooper pair formation process becomes an important source of axion production and competes with $n+n\rightarrow n+n+a$.  At the same time, while the protons are still superfluid, the Boltzmann suppression of $n+p\rightarrow n + p + a$ is not very large at these higher temperatures, and so it too becomes an important source of axions.

As the core temperature rises above $10^9 \text{ K}$, the protons in the center of the magnetar are no longer superfluid and once the core passes $T\approx 5\times 10^9 \text { K}$, protons in most of the neutron star are no longer Cooper-paired.  At this point, the rate of Cooper pair formation dramatically falls off and the three nucleon bremsstrahlung processes produce the vast majority of the axions.  For core temperatures above about $7\times 10^9 \text{ K}$, none of the core remains superfluid.  In ungapped nuclear matter, all three nucleon bremsstrahlung processes are relevant, though $n+p\rightarrow n+p+a$ produces the largest contribution to the axion emissivity.  Profiles of the axion emissivity throughout the $1.4M_{\odot}$ magnetar, for different values of the core temperature, can be seen in Fig.~2 of \cite{Fortin:2021sst}.
\subsubsection{ALP spectrum}
It is important to understand the energy spectrum of axions emitted from neutron star cores, because those axions, with some probability, convert into photons with the same energy (see Sect.~\ref{sec:magnetars}).  The axion spectrum and resultant photon spectrum will not be identical, as the conversion probability depends on the axion energy.  In the right panel of Fig.~\ref{fig:axion_emission_profiles} we plot $\mathop{dN_a}/\mathop{d\omega}$, the number of axions produced with energies between $\omega$ and $\omega+\mathop{d\omega}$ per time in the core of the $1.4M_{\odot}$ magnetar considered above, separating the emission into the contribution from the three bremsstrahlung processes and the contribution from Cooper pair formation.  The quantity $\mathop{dN_a}/\mathop{d\omega}$ is related to the differential luminosity of axions, which can be written as an integral of the axion emissivity over the core of the magnetar
\begin{equation}
    \frac{\mathop{dN_a}}{\mathop{d\omega}} = \frac{1}{\omega}\frac{\mathop{dL_a}}{\mathop{d\omega}} = \frac{1}{\omega}\int \mathop{d^3r}\frac{\mathop{dQ_a}}{\mathop{d\omega}} = \frac{4\pi}{\omega}\int_0^{R_\text{crust}}\mathop{dr}r^2\frac{\mathop{dQ_a}}{\mathop{d\omega}}. \label{eq:dNda}
\end{equation}
Expressions for the differential emissivity $\mathop{dQ_a}/\mathop{d\omega}$ coming from the integrated expressions Eqs.~\ref{eq:Qnn}, \ref{eq:Qpp}, \ref{eq:Qnp}, and \ref{eq:Qcp} are given in \cite{Fortin:2021sst}.

In degenerate nuclear matter, the nucleon bremsstrahlung process produces axions with an energy spectrum that peaks at around $\omega=2T$ \cite{Raffelt:1996wa}.  In contrast, the Cooper pair formation process produces axions only with energies greater than or equal to twice the superfluid gap, which is the binding energy of a Cooper pair.  In a section of the neutron star where the proton gap is $\Delta$, the spectrum of axions produced from Cooper pair formation is at its maximum very close to $\omega=2\Delta$.  When studying the spectrum of axions produced by Cooper pair formation in the star as a whole, it is clear that the spectrum will begin at $\omega=2\Delta_{\text{min}}$, where $\Delta_{\text{min}}$ is the minimum (but nonzero) value of the gap found in the magnetar.  

As can be seen in the right panel of Fig.~\ref{fig:axion_emission_profiles}, when the core has $T=10^8 \text{ K}$, the bremsstrahlung processes produce axions with a spectrum peaked at 17 keV.  At this temperature, the superfluid gap is about 160 keV at the center of the magnetar, and rises to 1 MeV near the crust (see Fig.~1 of \cite{Fortin:2021sst}).  Therefore, the axions produced by Cooper pair formation in the center of the star will have energies greater than 320 keV, and those produced from the outer regions of the star will have energies greater than 2 MeV.  Therefore, at this core temperature, all axions produced with energies below about 320 keV are from bremsstrahlung, while those produced with energies greater than 320 keV are produced about evenly from bremsstrahlung and from Cooper pair formation.  This trend holds for core temperatures of $5\times 10^8 \text{ K}$ and $10^9 \text{ K}$, however at $10^9\text{ K}$ the superfluid gap at the center of the magnetar is slightly smaller than at the lower temperatures displayed here, and thus the axion spectrum coming from Cooper pair formation starts at about 200 keV.  At these temperatures, the high-energy axion spectrum comes largely from Cooper pair formation.  When the core temperature is $5\times 10^9 \text{ K}$, only a tiny part of the star is superfluid, so bremsstrahlung dominates the axion production.  However, the Cooper pair formation can produce axions of arbitrarily low energy in this case, because the superfluid gap rises continuously from zero as, moving outward from the magnetar center, the protons transition from ungapped to gapped at about 8 km from the center of the magnetar.  

Since the authors of \cite{Fortin:2021sst} studied the hard X-ray spectrum coming from axion-photon conversion, only axions emitted with energies between 20-150 keV needed consideration.  Thus, the authors neglected axion production from the formation of $^1S_0$ proton Cooper pairs, because the axions emitted in this channel have energies above 200 keV (except in the case where the core temperature is $5\times10^9\text{ K}$, but in that case the channel is subdominant to axion production from bremsstrahlung since protons are no longer superfluid in much of the magnetar).

The above analysis of the total energy loss from axion emission and the spectra of the emitted axions followed the investigation of axion emission from magnetars in \cite{Fortin:2021sst}, which considered only proton singlet superfluidity due to the expected high core temperature of the magnetars.  If the magnetar core is colder than assumed in \cite{Fortin:2021sst}, then neutron superfluidity should be considered, which would strongly suppress the axion emission from bremsstrahlung processes.  In addition, the breaking of $^3P_2$ neutron Cooper pairs could also produce axions.  In an analysis of a group neutron stars called the Magnificent Seven, which (for the most part) have core temperatures less than the magnetars considered in \cite{Fortin:2021sst}, the authors of \cite{Buschmann:2019pfp} discussed the effects of including neutron superfluidity, although the main results of their study assume the absence of both superfluid neutrons and protons.  Neutron Cooper pairing and its effect on the production of light particles is discussed further in \cite{Yakovlev:1999sk,Yakovlev:2000jp,Sedrakian:2018ydt}.

\subsection{Conversion of ALPs to photons near neutron stars}
\label{sec:magnetars}

Having discussed the production of axions from neutron stars, we now go on to a discussion of the subsequent fate of these axions as they enter the magnetosphere. 

ALP production  in the neutron star core peaks at energies of a few keV to a few hundreds keV; the ALPs subsequently  escape and  convert to photons in the magnetosphere. The emissivity has a parametric dependence on the core temperature that goes as $\sim T^6$, and the conversion probability is enhanced by large magnetic fields. This makes  magnetars, with high core temperatures and strong magnetic fields, ideal targets for probing ALPs. Other systems -- such as magnetic white dwarfs \cite{Dessert:2019sgw} and  the Magnificent Seven \cite{Buschmann:2019pfp, Dessert:2019dos} -- have also been studied recently  in the same framework. Quiescent emission in the  hard X-ray and soft gamma-ray band from neutron stars will be our data of interest. There is a separate program of cold ambient axions from the halo converting to photons in the magnetosphere, but the putative signal in that case will lie in the radio band \cite{Hook:2018iia}. In our opinion, radio signals are complicated by plasma effects which must be carefully treated; such effects are negligible at the higher energies we will review.

The inter-conversion of ALPs and photons in the dipolar magnetic fields near neutron stars has only been fully understood recently, and semi-analytic solutions for the probability of conversion obtained \cite{Fortin:2018aom, Fortin:2018ehg, Fortin:2021sst, Lloyd:2020vzs}. We now summarize these results.

\subsubsection{ALP-photon conversion in dipolar magnetic fields}


The evolution equation for the ALP-photon system propagating in the magnetic field of a neutron star magnetosphere in terms of the dimensionless distance $x=r/r_0$ is given by \cite{Raffelt:1987im}
\begin{equation}
i\frac{d}{dx}\left(\begin{array}{c}a\\E_\parallel\\E_\perp\end{array}\right)=\left(\begin{array}{ccc}\omega r_0+\Delta_ar_0&\Delta_Mr_0&0\\\Delta_Mr_0&\omega r_0+\Delta_\parallel r_0&0\\0&0&\omega r_0+\Delta_\perp r_0\end{array}\right)\left(\begin{array}{c}a\\E_\parallel\\E_\perp\end{array}\right),
\label{EqnDiffMat}
\end{equation}
where
\begin{equation}
\Delta_a=-\frac{m_a^2}{2\omega},\qquad\Delta_\parallel=(n_\parallel - 1) \omega,\qquad\Delta_\perp=(n_\perp - 1) \omega,\qquad\Delta_M=\frac{1}{2} g_{a \gamma \gamma} B\sin\theta.
\label{propequa2}
\end{equation}
The ALP field has been represented by $a(x)$ while the parallel and perpendicular electric fields are denoted by $E_\parallel(x)$ and $E_\perp(x)$, respectively.  The energy is denoted by $\omega$, the neutron star radius by $r_0$, and  the angle between the direction of propagation and the magnetic field  by $\theta$. The magnetic field $B$ is given by a dipole with strength $B_0$ at the surface $r_0$.

In the strong magnetic fields near neutron stars, the photon refractive indices $n_\parallel$ and $n_\perp$ can depart significantly from their standard vacuum values of unity. The refractive indices can be derived from the photon polarization tensor 
and perturbative expansions are possible in various limits of the magnetic field $B$ and photon energy $\omega$. For ALPs with energies in the hard X-ray and soft gamma-ray range $\omega \sim \mathcal{O}(100-1000)$ keV that convert near the ``radius of conversion'' [that we define below in Eq.~\eqref{EqnConvRadius}], the appropriate limits are
\begin{equation}
 \frac{eB}{m^2_e} \equiv \frac{B}{B_c} \ll1 \quad\quad{\rm and}\quad\quad \left(\frac{B}{B_c}\right)^2\frac{\omega^2\sin^2\theta}{m^2_e}\ll1\,, \label{eq:pertreg-lc}
\end{equation}
and the corresponding indices of refraction are given by
\begin{align}
\left\{
 \begin{array}{c}
 \! n_\parallel \! \\
 \!  n_\perp \!
 \end{array}
\right\}
=1+\frac{\alpha}{4\pi}\left(\frac{B}{B_c}\right)^2\sin^2\theta\ \frac{2}{45}
\left\{
 \begin{array}{c}
7\\
4
 \end{array}
\right\} + {\cal O}\left((eB)^4\right)\,. \label{eq:refind}
\end{align}
Here, $B_c$ denotes the critical magnetic field: $B_c=m_e^2/e=4.413\times10^{13}\,\text{G}$,  where $e=\sqrt{4\pi\alpha}$ is the charge given in terms of the fine structure constant $\alpha\sim1/137$. 

The mixing angle $\varphi_\text{mix}$ between the ALP field and the perpendicular photon field can be extracted from Eq.~\eqref{EqnDiffMat} and is given by
\begin{equation} \label{mixingangle}
\tan(2\varphi_\text{mix})=\frac{2\Delta_M(x)}{\Delta_\parallel(x)-\Delta_a}.
\end{equation}
The mixing angle \eqref{mixingangle} at the magnetar surface is negligible, which led the classic paper by Raffelt and Stodolsky \cite{Raffelt:1987im} to dismiss this particular avenue of probing ALPs. However, it is clear that the $r$-dependence of $\Delta_M$ and $\Delta_\parallel$ lead to an eventual increase of the mixing angle away from the magnetar:  indeed, $\Delta_\parallel \sim 1/r^6$, while $\Delta_M \sim 1/r^3$.  Thus the denominator decreases faster than the numerator, and the conversion probability peaks near the radius of conversion where the off-diagonal $\Delta_M$ contribution  becomes the same order as the diagonal $\Delta_{\parallel}$ term. This is given by 
\be \label{EqnConvRadius}
r_{a\to\gamma}=r_0 \left(\frac{7\alpha}{45\pi}\right)^{1/6}\left(\frac{\omega}{m_a}\frac{B_0}{B_c}|\sin\theta|\right)^{1/3}.
\ee
For large conversion radius where $r_{a\to\gamma}\gg r_0$ which is typically the case, the probability of conversion takes a simple form in both the small and large $|\Delta_ar_{a\to\gamma}|$ regimes,
\bea \label{EqnPinfapproxanal}
P_{a\to\gamma}=\left(\frac{\Delta_{M}r_0^3}{r_{a\to\gamma}^2}\right)^2\times\begin{cases}\frac{\pi}{3|\Delta_ar_{a\to\gamma}|}e^{\frac{6\Delta_ar_{a\to\gamma}}{5}}&|\Delta_ar_{a\to\gamma}|\gtrsim0.45\\
\frac{\Gamma\left(\frac{2}{5}\right)^2}{5^\frac{6}{5}|\Delta_ar_{a\to\gamma}|^\frac{4}{5}}&|\Delta_ar_{a\to\gamma}|\lesssim0.45\end{cases}.
\eea
These expressions constitute semi-analytic solutions to the ALP-photon propagation system and can be used to obtain the final spectrum of photons coming from axion conversions.

\begin{figure}
\centering
\includegraphics[width=0.48\textwidth]{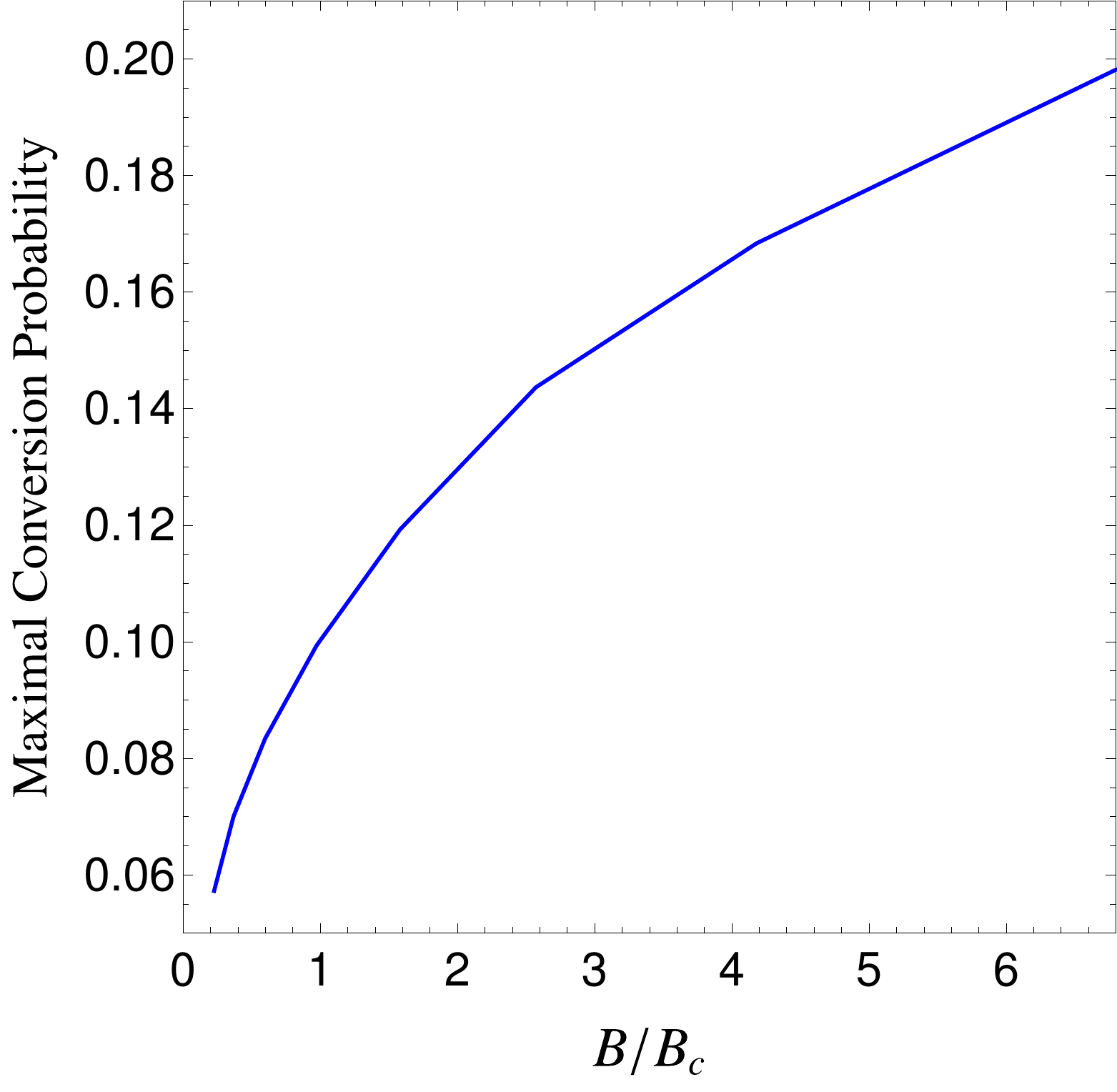}
\quad
\includegraphics[width=0.47\textwidth]{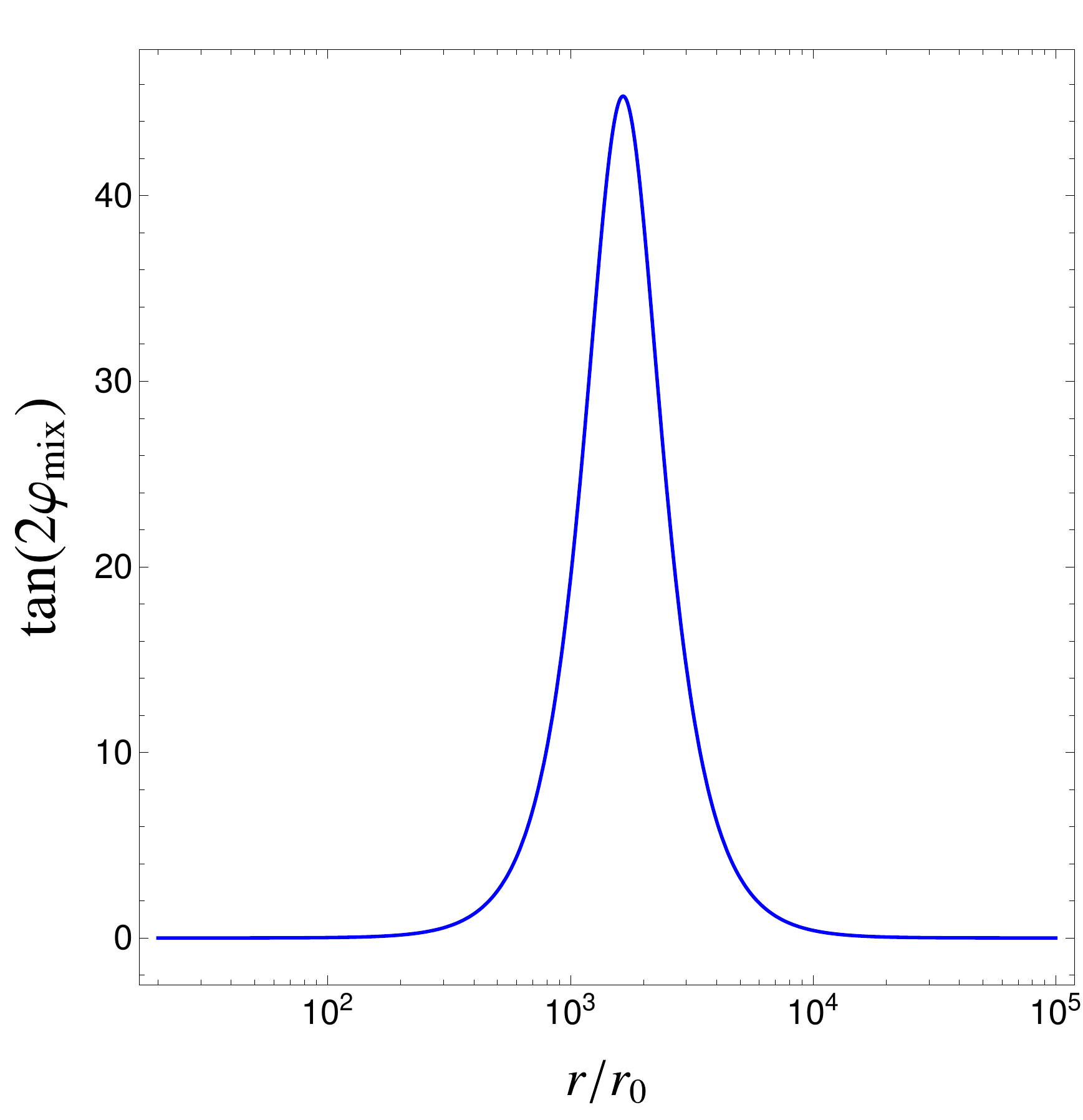}
\caption{
Left panel: The maximal conversion probability is shown as a function of $B_0/B_c$ where $B_0$ is the surface magnetic field and $B_c$ the critical QED field strength. Right panel: 
The mixing angle is shown as a function of the dimensionless distance $r/r_0$ from the magnetar surface. The benchmark values are chosen to be  $\omega=80\,\text{keV}$, $g_{a \gamma\gamma}=10^{-9}\,\text{GeV}^{-1}$ and 
$B_0=1.3\times 10^{14}$ G corresponding to the  magnetar 4U 0142+61.
\label{fig:FigEvol}
}
\end{figure}

\vspace{4mm}


\subsubsection{X-rays and gamma-ray signatures of ALPs: current constraints}


From Eq.~\eqref{eq:dNda}, the total energy of photon emission from ALP conversion in the magnetosphere is obtained by multiplying the ALP emission spectrum by the 
ALP to photon conversion probability~\cite{Fortin:2018ehg}:
\begin{eqnarray} \label{master1}
L_{a \rightarrow \gamma} = \int_0^{\infty} d \omega
\frac{1}{2 \pi} \int_0^{2 \pi} d \theta   \cdot \omega \cdot \frac{d N_a}{d \omega} \cdot P_{a \rightarrow \gamma} (\omega,\theta) ,
\end{eqnarray}
where $P_{a\to\gamma}$ can be obtained from Eq.~\eqref{EqnPinfapproxanal}.

The photon spectrum per area observed at the earth is then obtained by dividing $d L_{a \rightarrow \gamma}/d \omega$ by $4 \pi D^2$ with $D$ the
distance between the magnetar and the earth. The experimental data of the spectrum is usually expressed on the $\nu F_{\nu}$ plane, which is related to the above definition by
the following relation:
\begin{equation}\label{nuFnu}
\nu F_{\nu}(\omega) = \omega^2 \frac{1}{4 \pi D^2}
\frac{1}{\omega}\frac{d L_{a \rightarrow \gamma}}{d \omega}.
\end{equation}
This is the master equation that can be used to place constraints on $G_{an} \times g_{a\gamma\gamma}$. 

The most promising targets for this direction are magnetars, with their high core temperatures and strong magnetic fields.  Magnetars are a group of  neutron stars with dipole magnetic fields of  strengths up to $10^
{14}$--$10^{15}$~G (see \cite{Turolla:2015mwa, Kaspi:2017fwg} for recent reviews). Apart from   short X-ray bursts  and giant 
flares, they exhibit persistent emission in the $\sim 0.5$--$200$~keV band, with luminosity $L \sim 10^{31}$--$10^{36}$~erg~s$^{-1}$.  This  persistent emission is of interest for our purposes. It consists of a soft quasi-thermal component up to around 10~keV, and a very flat hard X-ray tail extending to beyond 200~keV. While the soft component is due to thermal emission from the surface, the hard X-rays are generally attributed to resonant inverse Compton scattering of thermal photons. 

 Data from    \integral~\cite{Papitto:2020tgi, Kuiper:2004ya, Kuiper:2006is,Hartog:2008tq,Hartog:2008tp, Mereghetti:2004sx, Molkov:2004sy, Gotz:2006cx}, \suzaku~\cite{Morii:2010vi,Enoto:2011fg, Enoto:2010ix, Enoto:2017dox}, \rxte ~\cite{Levine:1996du}, \swift ~\cite{Kuiper_2012}, \xmm~\cite{Rea:2009nh}, ASCA and \nustar~\cite{An:2013xui, Vogel:2014xfa, Younes:2017sme}  has revealed the hard X-ray component for around nine magnetars. Moreover, data from \integral\ and COMPTEL can be used to place upper limits on the soft gamma-ray emission from a subset of these magnetars. Experimental data in both bands can be compared to the theoretical spectrum coming from ALP-photon conversion, and constraints can be placed on the product of ALP couplings $G_{an} \times g_{a\gamma\gamma}$ by demanding that the theoretical spectrum obtained from Eq.~\eqref{master1} not exceed the observed spectrum in any energy bin. It is important to note that ALP-photon conversion is not being advanced as the underlying model of magnetar emission; in particular, no excess is being claimed over astrophysical ``background''. Rather, upper limits of $G_{an} \times g_{a\gamma\gamma}$ are being obtained by directly comparing with observational data.

\begin{figure}
\centering
\includegraphics[width=0.8\textwidth]{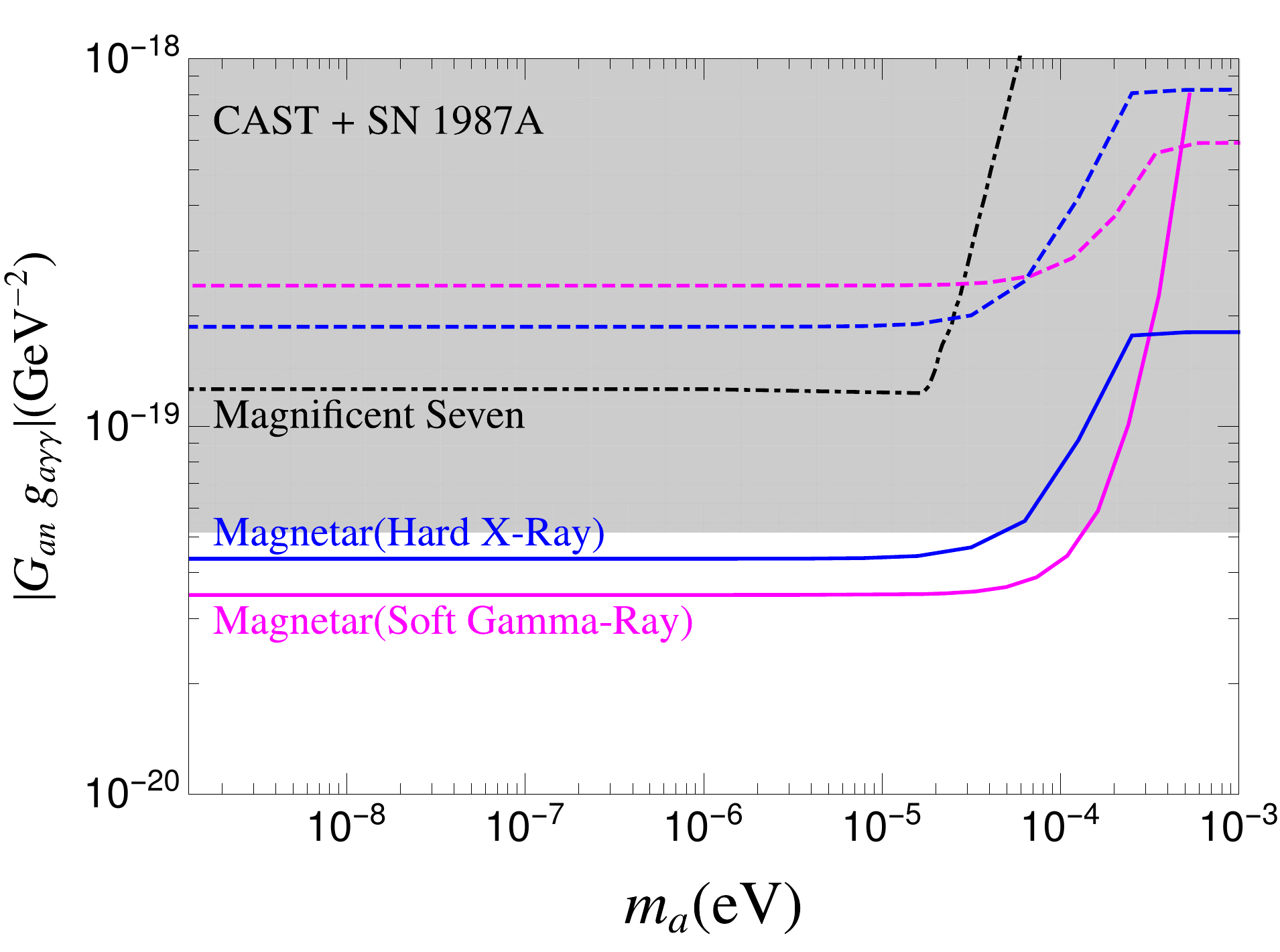}
\caption{\label{fig:magn1} 
The $95\%$ CL limits on $G_{an} \times g_{a\gamma\gamma}$ obtained by comparing Eq.~\eqref{nuFnu} against data. The blue lines are the limits obtained using the X-ray band (10$-$160 keV) from 8 magnetars with the $^1S_0$ proton superfluidity included~\cite{Fortin:2021sst}. The magenta lines use the soft gamma-ray band ($300$ keV$-$1 MeV) without including 
superfluidity~\cite{Lloyd:2020vzs}. For both analyses, two core temperatures of the magnetars are used:
$5\times 10^8 K$ (dashed) and $10^9K$ (solid). 
The limit corresponding to the dot-dashed line and denoted by Magnificent Seven is taken from \cite{Buschmann:2019pfp}, and the gray shaded region is the exclusion
derived from both CAST and SN 1987a from~\cite{Buschmann:2019pfp}.
}
\end{figure}

The constraints can be obtained by comparing the theoretical spectrum against data from magnetars in the hard X-ray (10-160 keV) and soft gamma-ray (300-1000 keV) bands. This analysis was performed in \cite{Fortin:2021sst, Lloyd:2020vzs}. The results  are displayed in Fig.~\ref{fig:magn1}, with the constraints coming from the analysis of the hard X-ray (soft gamma-ray) spectrum being depicted by the blue (magenta) curves. For both analyses, two core temperatures of the magnetars are used:
$5\times 10^8 K$ (dashed) and $10^9K$ (solid). It should be noted that \cite{Fortin:2021sst, Lloyd:2020vzs} performed the analysis for several magnetars. What is being depicted in  Fig.~\ref{fig:magn1} are the best case scenarios: magnetar 1E$\,$1547.0$-$5408 for the hard X-ray analysis, and magnetar 4U$\,$0142+61 for the soft gamma-ray analysis. Also depicted in Fig.~\ref{fig:magn1} is the analysis of the Magnificent Seven performed by  \cite{Buschmann:2019pfp}. The gray shaded region is the exclusion derived from combining current limits from the CAST experiment and SN 1987A. 

There are several lessons from Fig.~\ref{fig:magn1}. Firstly, the constraints coming from these studies are broadly competitive with those coming from the CAST experiment and SN 1987A and, in some  cases, better. Secondly, a better determination of the core temperature of magnetars is required to obtain definitive constraints. Thirdly, hard X-ray and soft gamma-ray data of magnetars and neutron stars from $Fermi$-GBM, AMEGO, and future telescopes may yield stronger constraints on the axion-photon coupling and should be a component of the fundamental physics case for such experiments. 

\begin{figure}\label{polfig}
\centering
\includegraphics[width=0.45\textwidth]{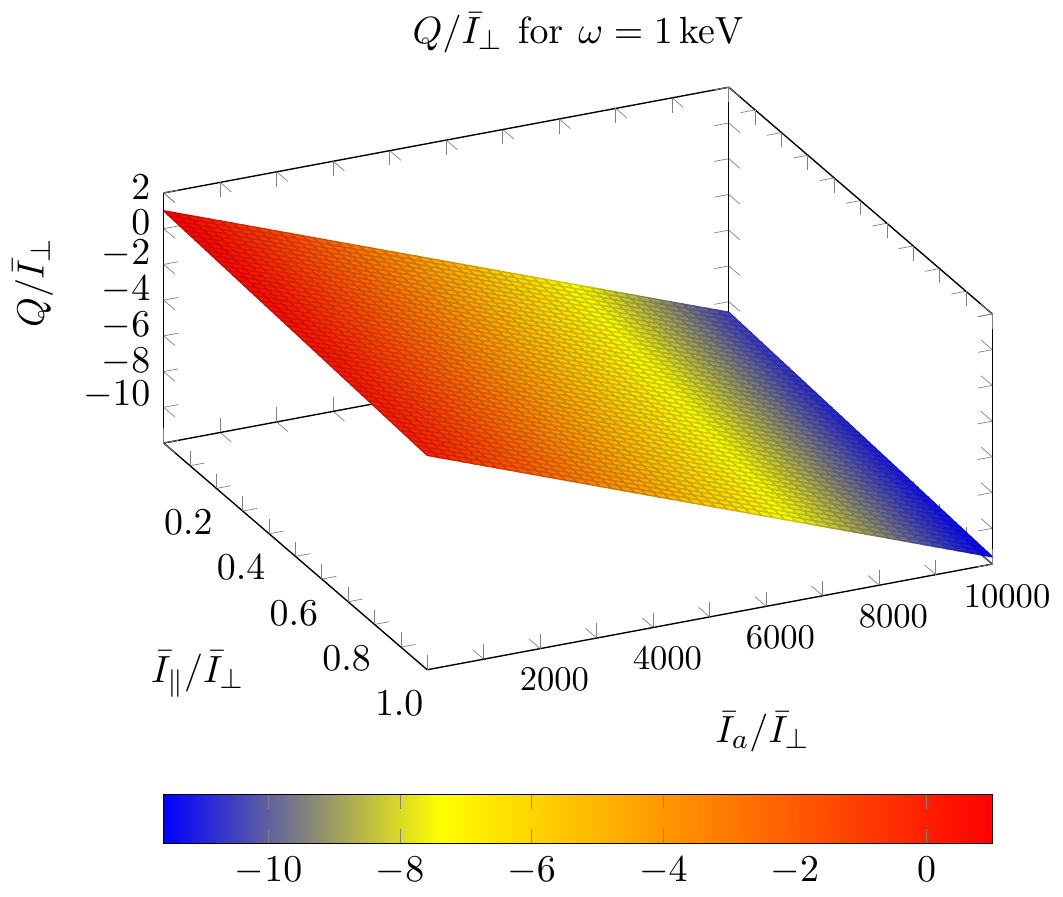}
\includegraphics[width=0.45\textwidth]{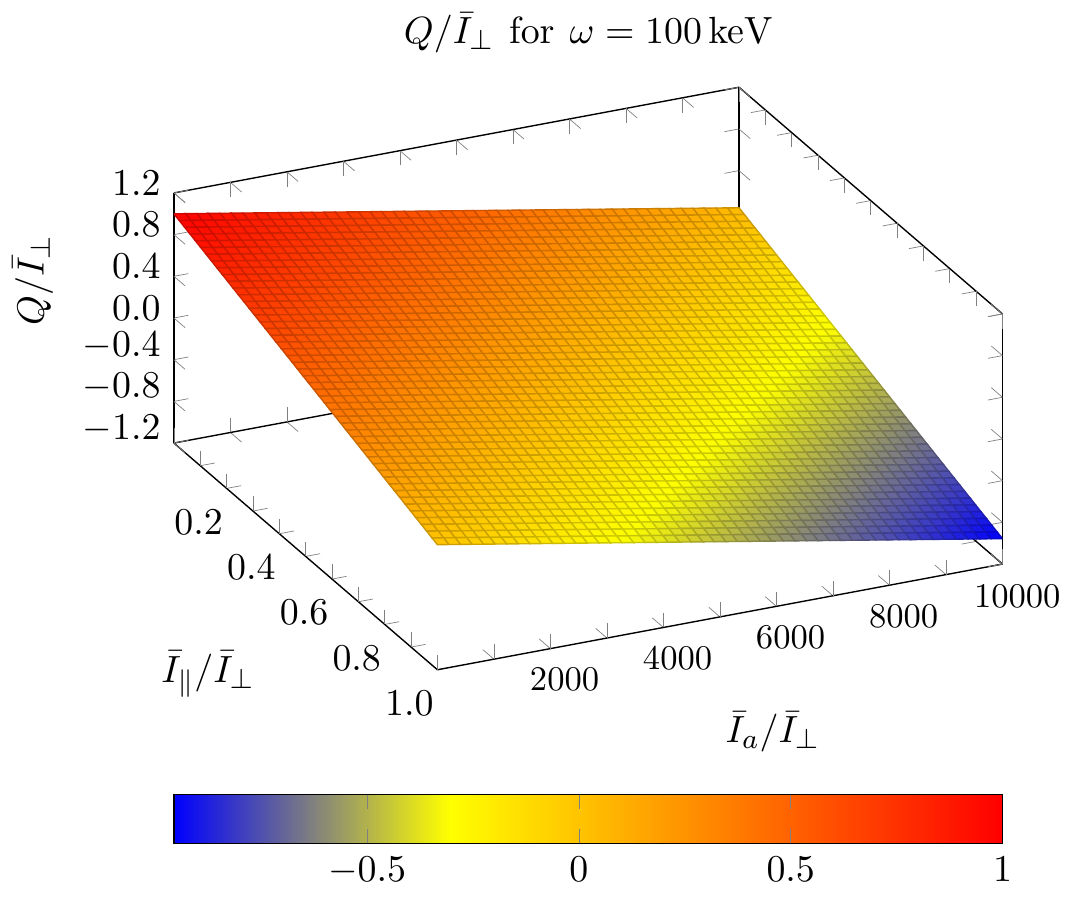}
\caption{O-mode polarization in X-rays from ALP-photon conversion: The normalized Stokes parameter $Q/\bar{I}_\perp$ at infinity is shown for the benchmark point $m_a=10^{-5}\,\text{eV}$, $g/e=5\times10^{-17}\,\text{keV}^{-1}$, $r_0=10\,\text{km}$, $B_0=20\times10^{14}\,\text{G}$ and $\theta=\pi/2$ in the plane $(\bar{I}_\parallel/\bar{I}_\perp,\bar{I}_a/\bar{I}_\perp)$.  The ALP energies are soft X-rays $\omega=1\,\text{keV}$ (left panel) and hard X-rays $\omega=100\,\text{keV}$ (right panel). Astrophysical models predict mostly X-mode polarization and one has $Q/\bar{I}_\perp \sim 1, \bar{I}_\parallel/\bar{I}_\perp \sim 0,$ and $\bar{I}_a/\bar{I}_\perp \sim 0$ in the absence of ALPs. In the presence of ALPs, $Q/\bar{I}_\perp$ may lie anywhere on the plane, depending in the averaged intensity $\bar{I}_a$ of the ALP field. Figures reproduced from \cite{Fortin:2018aom}.
}
\end{figure}


\subsubsection{Polarized X-rays and gamma-rays from magnetars}  

One of the most crucial signatures of ALP-induced emission versus astrophysical background is the difference in polarization.
In the strong magnetic fields near magnetars, photons are  polarized in two normal modes, the ordinary (O) and extraordinary (X) ones, which correspond to the electric field being respectively parallel and perpendicular to the plane containing the external magnetic field and the direction of propagation. For photon energies lower than  the electron cyclotron energy $\sim 11.6 (B/10^{12}\, {\rm G})$ keV, astrophysical models predict that the X-mode opacity is suppressed with respect to the O-mode. Thus, the thermal radiation is almost completely polarized in the X-mode \cite{Heyl:2018kah,Lai:2006af}. Remarkably, resonant inverse Compton upscattering, which is the putative astrophysical model  underlying the hard emission, also predicts a strongly polarized X-mode emission \cite{Wadiasingh:2017rcq}. One therefore expects astrophysical emission to be polarized along the X-mode, in all energy ranges from the soft to the hard spectrum.

Since the magnetic field is dipolar, the net polarization has to be averaged over emissions from regions with different orientations of the field direction. The effect is generally to obscure the overall polarization pattern. 
However,  the QED vacuum birefringence effect can lock the polarization vector to the magnetic field direction up to the polarization radius $r_{PL}$, where the polarization vector stops tracking the magnetic field, which is more uniform \cite{Krawczynski:2019ofl}. The adiabatic tracking of the polarization vector continues up to the polarization radius $r_{PL}$.  This leads to astrophysical predictions of polarization fractions as high as 40\% to 80\%  at the detector \cite{Heyl:2018kah}. Therefore, not only is the astrophysical emission X-mode, the birefringence effect in the magnetosphere near magnetars ``locks in'' this mode as far as detection is concerned. This ``locking in'' is especially efficient in magnetars with their strong magnetic fields. The radius of polarization is given by \cite{Heyl:2018kah}
\begin{equation}
r_{PL}=\left(\frac{\alpha}{45}\frac{\nu}{c}\right)^{1/5}\left(\frac{B_0}{B_c}r_0^3\sin\beta\right)^{2/5}\sim923.4\left(\frac{\omega}{1\,\text{keV}}\right)^{1/5}\left(\frac{B_0}{10^{14}\,\text{G}}\right)^{2/5}\left(\frac{r_0}{10\,\text{km}}\right)^{6/5}\,\text{km},
\end{equation}
where $\beta$ is the angle between the dipole axis and the line of sight.

The unique observational signature of ALP-photon conversion is the change in the predicted polarization pattern.  ALPs add to the astrophysical picture described above by producing O-mode photons, for which the electric field is parallel to the plane containing the external magnetic field and the direction of propagation. The radius of conversion, where the probability of conversion becomes significant, is typically larger than the polarization radius, implying an overall O-mode superposed on the X-mode coming purely from astrophysics \cite{Fortin:2018aom}. The ratio is given by
\be\frac{r_{PL}}{r_{a\to\gamma}}\sim0.57\left(\frac{1\,\text{keV}}{\omega}\right)^{2/15}\left(\frac{B_0}{10^{14}\,\text{G}}\right)^{1/15}\left(\frac{m_a}{10^{-8}\,\text{keV}}\right)^{1/3}\left(\frac{r_0}{10\,\text{km}}\right)^{1/5}\,\,.
\ee
We note that $r_{PL}\ll r_{a\to\gamma}$ for typical ALP benchmark points. The effect of ALPs on the observed polarization pattern will thus be to add an O-mode intensity to the purely astrophysical X-mode intensity.

The relative strength of the O-mode component depends on the intensity of ALPs produced in the core and the probability of conversion.  This can be quantified by considering X-ray emission produced both by astrophysical processes and by ALP-photon conversion, in an uncorrelated fashion, and in different relative proportions. The uncorrelated production of photons and ALPs allows one to average over the initial phase difference $\Delta \phi$ between the fields $E_{\parallel}(x)$ and $a(x)$ (the sum phases decouples). It is convenient to introduce the following quantities, all obtained by averaging over the  phase difference $\Delta \phi$: the sum ($I=\bar{I}_\perp+\bar{I}_\parallel$) and difference ($Q=\bar{I}_\perp-\bar{I}_\parallel$) of the phase-averaged photon intensities in the parallel and perpendicular planes, and the phase averaged ALP intensity $\bar{I}_a$. These intensities can be normalized by the total  amplitude of the $E_{\parallel}(x)$ and $a(x)$ fields.

The resulting Stokes parameter $Q$ is shown in Fig.~\ref{polfig}, for a photon energy of 1 keV and 100 keV. The astrophysical prediction is at the top left corner, where $Q/\bar{I}_\perp = 1$, $(\bar{I}_\parallel/\bar{I}_\perp = 0$, and $\bar{I}_a/\bar{I}_\perp)= 0$. With ALPs, on the other hand, $Q/\bar{I}_\perp$ can take values anywhere on the planes, depending on the ALP intensity emitted. We can thus see a significant departure from the astrophysical prediction.

  There are several missions that may be able to explore these features.  The Imaging X-ray Polarimetry Explorer (IXPE), Small Explorer (SMEX) and the Enhanced X-Ray Timing and Polarimetry Mission (eXTP) missions will launch in the next few years, and look for signals in the $2-10\,\text{keV}$ range \cite{Krawczynski:2019ofl}.  The next generation of X-ray polarimeters and increasingly sophisticated modeling of the astrophysics of magnetars provide an opportunity to investigate ALPs using polarization.

\subsection{Photon-to-ALP conversion in galactic and extra-galactic magnetic fields} \label{egalp}


We finally circle back to the topic of axion conversions in large-scale magnetic fields. Photons produced by distant sources will convert to ALPs in large-scale magnetic fields as they propagate towards the earth. This is a long-standing topic of interest in ALP searches with a vast amount of associated literature that merits its own dedicated review. We only mention some broad ideas and a sliver of the recent literature. 

\vspace{4mm}

\noindent{\textbf{X-ray Sources:} } \quad A number of studies have concentrated on X-ray point sources located within or behind galaxy clusters. The intracluster magnetic field serves as the medium in which ALP-photon interconversion occurs, as X-rays travel from the source to the earth. The result is that there are spectral distortions of the X-ray source, which can be leveraged to place constraints on the ALP-photon coupling.

As we have indicated before, a major challenge is the limited knowledge that we have about the morphology of the magnetic field along our line of sight to the source. Typically, a domain-like configuration is assumed, with the field being assumed to be constant in each such domain. The expression for the conversion probability in the constant magnetic field within a given domain is standard and was given in the classic paper by Raffelt and Stodolsky \cite{Raffelt:1987im}. We follow the notation of, for example, \cite{Chen:2017mjf}:
\be
P_{\gamma \rightarrow a} \, = \, 4 \vartheta^2\sin^2(\Delta_{osc}L/2),
\ee
%
where $L$ is the domain length,  $\vartheta\approx\frac{1}{2}\tan(2\vartheta)=\frac{\Delta_{g_{a\gamma\gamma}}}{\Delta_\gamma-\Delta_a}$ and $\Delta_{osc}=\Delta_\gamma-\Delta_a$.
Here the expressions for the various quantities are  $\Delta_\gamma=-\frac{\omega^2}{2\omega_p}$, $\Delta_a=-\frac{m_a^2}{2\omega_p}$, $\Delta_{g_{a\gamma\gamma}}=\frac{1}{2}B g_{a\gamma\gamma}$, and $\omega_p=\sqrt{\frac{4\pi\alpha n_e}{m_e}}$
is the plasma frequency. The intracluster magnetic field is expected to have a central strength of $B_0 \sim 2\,-\,30$ $\mu$G and typical coherence lengths of  $L \sim 1-10$ kpc.

The simplicity of the conversion probability in a given domain belies the complexity of obtaining the total probability and finally the signatures of ALP-photon interconversion. This involves several steps, which we describe in some detail following, for example, Ref.~\cite{Conlon:2018iwn};  similar methods are employed for other X-ray and gamma-ray sources.

$(i)$ A radial dependence for the magnetic field is assumed, with a benchmark central strength $B_0$ in the range stated above; $B(r) \propto B_0 [n_e(r)]^{\eta}$, with $\eta$ expected to be between 0.5 and 1. For example, for the Perseus AGN NGC1275 $B(r) \propto [n_e(r)]^{0.7}$  \cite{Conlon:2018iwn}. The electron density $n_e(r)$ is given as a function of radial distance and has functional form $n_e(r) \propto n_0 [1+(r/r_c)^2]^{\beta}$, where $n_0$ is the central electron density, $r_c$ is the radius of the core, and $\beta$ is a constant. A double $\beta$ dependence is appropriate for cool-core clusters, for example in the specific case of Perseus $\beta_1 = -1.8, n_{0,1} = 3.9 \times 10^{-2}, r_{c,1} = 80$ kpc and  $\beta_2 = - 0.87, n_{0,2} = 4.05 \times 10^{-3}, r_{c,1} = 280$ kpc. For more details, we refer to \cite{Conlon:2018iwn}.

$(ii)$ The total propagation distance is divided into a large number of domains ($\sim 300$), with domain lengths $L$ drawn as a random parameter in the range 1-10 kpc. Within each domain, the magnetic field is assumed to be constant at a value given by the radial dependence described above, with $r$ being taken as the distance from the center of the cluster to the near end of the domain. The direction of the magnetic field is taken from a flat distribution between $[0,2\pi]$.

$(iii)$ A simulation is then run for a range of ALP couplings $g_{a\gamma\gamma}$ and masses $m_a$; for each choice of  $(g_{a\gamma\gamma},m_a)$ there are $\sim 300$ choices for domain lengths, and $\sim 100$ choices of the magnetic field in each domain. 

The above steps can be used to obtain the final survival probability. The observed spectrum is  the source spectrum modulated by the survival probability. These quasi-sinusoidal modulations can be analyzed using global goodness-of-fits or machine learning techniques.  A variety of X-ray sources have been studied in this framework \cite{Conlon:2018iwn, Conlon:2017qcw, Reynolds:2019uqt, Conlon:2015uwa}.

\vspace{4mm}

\noindent{\textbf{Gamma-ray Sources:} } \quad In parallel with distant X-ray sources, there is a substantial amount of literature on studying gamma-ray sources within the same framework of photon-ALP conversion. The sources studied include flat spectrum radio quasars and BL Lacs, supernova remnants, galactic pulsars, etc.  \cite{Dobrynina:2014qba, Kartavtsev:2016doq, Galanti:2019sya, Hooper:2007bq}. The intervening magnetic field is  that of the host  galaxy or cluster \cite{Hochmuth_2007, Simet_2008, De_Angelis_2008, Chelouche_2009, Wouters_2014}, extragalactic space between galaxy clusters \cite{Kartavtsev:2016doq, S_nchez_Conde_2009, Choi:2018mvk}, the field within the blazar jet \cite{Dobrynina:2014qba}, etc. As in the previous case, a domain-like modeling of the large-scale magnetic field is employed, with each domain length being equal to the field's coherence length, and the field being assumed to be constant in a given domain. The ALP-photon transfer function is computed by performing simulations of various magnetic field configurations and computing the mean and variance of the gamma-ray flux  \cite{Mirizzi:2009aj, Mortsell:2002dd}.

\vspace{4mm}

\noindent{\textbf{ALPs from Supernovae:} } \quad The galactic-scale scenarios described above study photon-to-ALP conversions. One could also study the opposite scenario of ALP-to-photon conversion within the same framework. ALPs produced from supernovae will convert to photons in the magnetic field of the Milky Way on their way to the earth. The result will be, for example, a gamma-ray flash during a supernova \cite{Brockway:1996yr, Grifols:1996id, Payez:2014xsa} or a diffuse supernova flux with energies $\sim \mathcal{O}(50)$ MeV \cite{Calore:2020tjw}. Using the absence of such a gamma-ray flash in observations of SN1987a or using data from the Fermi satellite to constrain a diffuse supernova flux leads to constraints on ALP couplings.

The properties of the interstellar medium of the Milky Way affect the ALP-to-photon conversion probability in such scenarios, and hence the final constraints on ALP parameter space. A recent modeling of the galactic magnetic field which has been used in the ALP literature is the one by  Jansson and Farrar \cite{Jansson_2012}. The field in this model is composed  of a sum of coherent, random, and striated components. On the other hand, the electron density of the interstellar medium, which controls the photon mass, can be modeled using the thin and thick disc components of the NE2001 model. For more details, we refer to  \cite{Day:2015xea}.

\vspace{4mm}

\noindent{\textbf{ALPs from other Astrophysical Sources:} } \quad 
It is known that CMB photons can convert to axions on their way to the earth \cite{Mirizzi:2009nq,Tashiro:2013yea}. This causes the CMB spectrum to deviate from being a perfect black body, ultimately putting constraints on $g_{a\gamma\gamma} B_{IGM}$. Assuming some large value for $B_{IGM}$, such as $B_{IGM} \sim \mathrm{nG}$, the constraint on ALPs lighter than $10^{-12}\; \mathrm{eV}$ is $g_{a\gamma\gamma} \lesssim 10^{-13}\; \mathrm{GeV}^{-1}$.

In a recent study \cite{Buen-Abad:2020zbd}, different cosmic distance measurements were combined to constrain the axion-photon coupling. The constraints included the luminosity distance of type Ia supernovae and angular diameter distance of galaxy clusters. The idea is that due to the photon-to-axion conversion, the apparent magnitude of the luminosity is decreased compared to the would-be value if there had been no axion conversion. Any distance measurement that relies on the apparent magnitude is affected. The constraints rely on either intergalactic magnetic field or the intracluster magnetic field (ICM), depending on where the conversion happens.  Therefore this provides a constraint that scales differently with the intergalactic magnetic field alone as in \cite{Mirizzi:2009nq,Tashiro:2013yea}. It was shown that for the axion masses in the range  $m_a \lesssim 10^{-12}\;\mathrm{eV}$, a constraint of the photon-axion coupling can be put in the range of $g_{a\gamma\gamma} \lesssim 5\times 10^{-13}\;\mathrm{GeV}^{-1}$ if optimistic ICM model is adopted, or $5\times 10^{-12}(\mathrm{nG}/B)\; \mathrm{GeV}^{-1}$ if ICM conversion is neglected altogether.

\section{Neutron Star Mergers \label{sec:merger} }
In this section, we review the inspiral and postmerger stages of a binary neutron star merger, and then discuss what each stage of the merger can tell us about the physics of axions.
\subsection{Dynamics of a neutron star merger}
\label{sec:ns_merger_generalities}
As a binary neutron star system evolves in time, the individual stars emit gravitational radiation as they orbit each other.  The gravitational waves carry energy away from the system (see \cite{Misner:1974qy,Andersson:2019yve} for the relevant calculations), causing the distance between the two stars to decrease.  The orbit is gradually circularized if it had an initial eccentricity.  The frequency of the emitted gravitational waves is twice the orbital frequency, and thus as the two stars move closer together, the gravitational wave frequency increases.  After tens or hundreds of millions of years \cite{Andersson:2019yve,Simonetti:2019evn}, the two stars get sufficiently close to each other that they begin to tidally deform, which takes additional energy away from their orbit, bringing the system closer to the eventual merger.  During most of this ``inspiral'' phase of the neutron star merger, the internal condition of one neutron star is unaffected by the other.  The nuclear matter within the stars is cold due to millions of years of neutrino cooling \cite{Yakovlev:2004iq}, and it remains cold until the stars collide \cite{Arras:2018fxj}.

On August 17, 2017, the LIGO-Virgo collaboration observed the gravitational wave signal from the inspiral of the binary neutron star merger, GW170817 \cite{TheLIGOScientific:2017qsa}.  The signal was detected when the frequency of the emitted gravitational waves from the inspiral rose above 24 Hz, and continued for about 100 seconds until the two stars collided.  No postmerger gravitational wave signal was detected.  To date, only gravitational waves from the inspiral phase of a neutron star merger have been measured\footnote{The LIGO-Virgo collaboration has also measured the gravitational wave signal from many binary black hole mergers, the first of which was GW150914 \cite{Abbott:2016blz}.} \cite{Abbott:2020niy,Abbott:2020uma}.  

From the early stages of the inspiral, little about the neutron stars can be learned, because they are sufficiently separated to be indistinguishable from point masses.  Once the stars are close enough to tidally deform each other, the deformation alters the phase of the gravitational wave signal, from which information about the nuclear equation of state can be inferred \cite{Baiotti:2019sew,Read:2013zra}.  In Sect.~\ref{sec:axions_inspiral} we will review the information that can be learned about axions from the neutron star inspiral.

From the beginning of tidal deformation onward, numerical simulations are necessary to understand the dynamics of the neutron star merger.  Numerical simulations evolve Einstein's equations to determine the spacetime metric throughout the merger.  The nuclear matter in the neutron stars is treated as a fluid and evolved according to the equations of relativistic hydrodynamics.  The fluid has an equation of state derived from nuclear theory \cite{Oertel:2016bki}.  As their mean free path in hot and dense matter is macroscopic \cite{Roberts:2016mwj,Alford:2018lhf}, neutrinos are treated separately from the nuclear matter, either with a kinetic formalism or as a separate fluid with a loss term \cite{Ardevol-Pulpillo:2018btx,Sekiguchi:2012uc}.  Merger simulations are reviewed in \cite{Baiotti:2016qnr,rezzolla2013relativistic,Duez:2018jaf,Andersson:2019yve} and will be featured in Sect.~\ref{sec:axion_cooling_of_mergers}.  

Snapshots of the configuration of the merger at various times in its evolution, obtained from numerical simulations, can be found in \cite{Radice:2016dwd,Baiotti:2008ra,Gieg:2019yzq,2013PhRvD..87b4001H,Price:2006fi,East:2015vix,Rezzolla:2010fd}.  A few orbits after the stars begin to tidally deform, they collide with a relatively large impact parameter, which generates a shear interface that may lead to a Kelvin-Helmholtz instability \cite{rezzolla2013relativistic,Kiuchi:2015sga}.  When the two neutron stars merge, unless there is prompt collapse to a black hole, a differentially rotating mass of hot nuclear matter emerges.  Right at the time of merger, the nuclear matter at the collision interface heats to several tens of MeV.  The differential rotation \cite{Baumgarte:1999cq}, but also perhaps the thermal pressure,\footnote{The influence of the thermal pressure on the lifetime of the remnant is complicated - see the discussion in \cite{Kaplan:2013wra}.} keep the star from collapsing for a period of time, ranging from a few milliseconds to as much as a few seconds \cite{Lucca:2019ohp,Bernuzzi:2020tgt}.\footnote{It is believed that the remnant created in GW170817 lasted for approximately one second before collapsing to a black hole \cite{Gill:2019bvq}.}  If the mass of the spinning remnant is below the maximum mass of a stable neutron star \cite{glendenning2000compact,2010arXiv1012.3208L}, the remnant will survive indefinitely.

\begin{figure*}[t!]
\begin{minipage}[t]{0.5\linewidth}
\includegraphics[width=.95\linewidth]{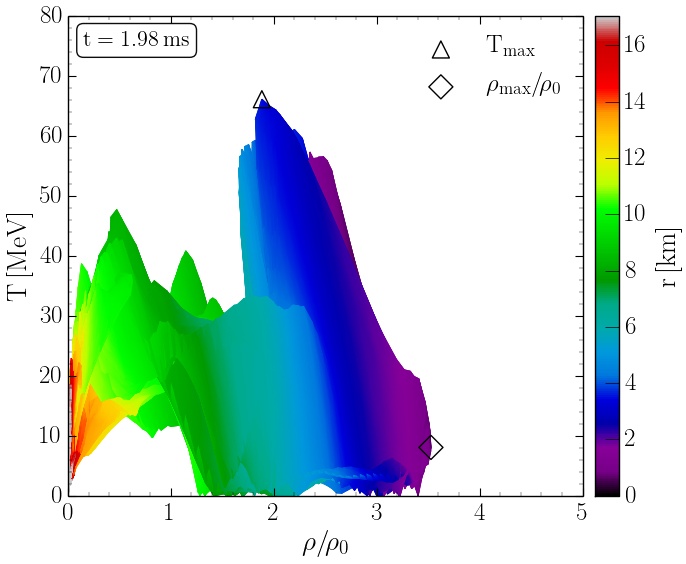}
\end{minipage}\hfill%
\begin{minipage}[t]{0.5\linewidth}
\includegraphics[width=.95\linewidth]{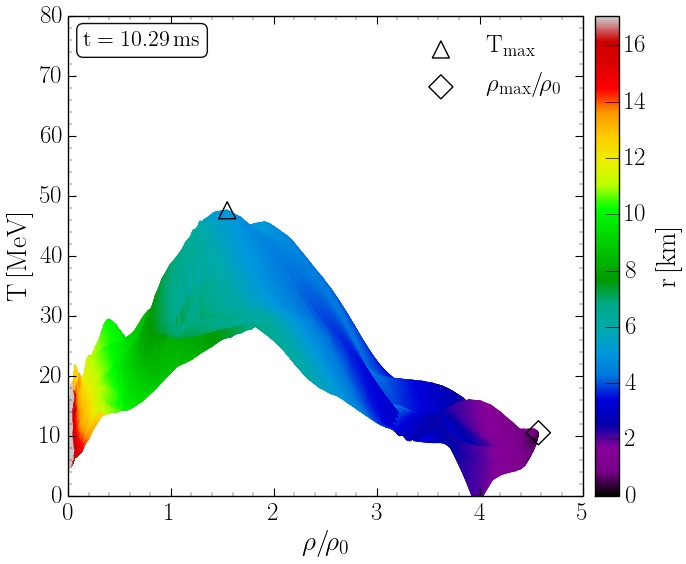}
\end{minipage}%
\caption{Temperatures and densities attained during a neutron star merger, according to a numerical simulation.  The simulation does not include axions.  Each point corresponds to a fluid element, colored by its distance from the center of the merger.  While two milliseconds after the merger (left panel) there is little correlation between the temperature and density of fluid elements in the merger, after ten milliseconds (right panel) the nuclear matter has settled down to a configuration with warm, low density matter on the outskirts of the merger, hot medium-density matter in the outer core, and cooler, high density matter in the core.  Figure courtesy of M.~Hanauske and the Rezzolla group, see also \cite{Hanauske:2019qgs}.}
\label{fig:T_nb_plane}
\end{figure*}

In Fig.~\ref{fig:T_nb_plane} we show results from a merger simulation conducted by the Frankfurt group, where they have plotted the temperature and density of fluid elements in the merger at two snapshots in time.  We see that 2 milliseconds after the stars touch, the nuclear matter is still in a state of flux, and there is no clear correlation between the density and temperature of a fluid element.  However, we can see that significant heating has occurred, with some fluid elements reaching temperatures as high as 60-70 MeV.  After ten milliseconds, the merger remnant has settled into a configuration which has a high density, relatively cool core and warm, low-density matter at the outskirts.  In between, the matter is quite hot, reaching temperatures of 40-50 MeV.  The central density of the two merging neutron stars was about $2.5n_0$, but the right panel shows that the maximum density reached in the remnant is almost twice this.  The trends seen in this simulations are seen in many other simulations as well \cite{Perego:2019adq,Lalit:2018dps,Camelio:2020mdi}, although the specifics depend on the initial masses of the two stars and the nuclear equation of state.  After the collision of the two stars, very little -- if any -- of the nuclear matter is superfluid, since the temperature of the vast majority of the remnant is above 1 MeV.  The thermodynamic conditions of the nuclear matter in the merger have a large impact on the role of axions in the merger remnant, which we discuss in Sect.~\ref{sec:axions_in_merger_remnants}.

\subsection{Ultralight axions in the neutron star inspiral}
\label{sec:axions_inspiral}
It is unlikely that the inspiral of a binary neutron star system can tell us much more about ``particle-like'' axions (axions with a Compton wavelength $\lambda_{\text{Compton}} = 1/m_a$ much smaller than the size of a neutron star) than we already know from isolated neutron stars (Sect.~\ref{sec:magnet}), since for most of the inspiral the internal conditions of one star are not modified by the presence of its binary companion.\footnote{An exception would be the f-mode oscillations that are induced in binaries with eccentric orbits \cite{Baiotti:2019sew,Pratten:2019sed}.}  However, axions with very small masses have long Compton wavelengths, and when the Compton wavelength reaches the size of a neutron star or larger (corresponding to axion masses of less than $10^{-11} \text{ eV}$) the dynamics of the inspiral could potentially be impacted by the existence of ultralight axions.

The authors of \cite{Hook:2017psm} found that ultralight axions can be sourced by high-baryon density objects like neutron stars, endowing those objects with a scalar charge.  They studied a ``tuned'' version of the QCD axion\footnote{As part of a detailed extension of the QCD axion to finite density, the authors of \cite{Balkin:2020dsr} note that it is unlikely that the standard QCD axion will be able to significantly alter the inspiral dynamics of a merger.  Tuning of the axion potential (as done in \cite{Hook:2017psm}) is likely required for the axion to modify the inspiral dynamics.  In \cite{Hook:2017psm}, the authors provide an example scenario that gives rise to the tuned axion potential.  See also the very recent work \cite{DiLuzio:2021pxd}.} and showed that this type of axion can mediate a force \cite{Huang:2018pbu}
\begin{equation}
    \mathbf{F} = -\frac{Q_1Q_2}{4\pi r^2}e^{-m_a r}\hat{r},
\end{equation}
between two neutron stars with scalar charge.  Here, $m_a$ is the axion mass and the scalar charge is $Q=\pm 4\pi^2 f_a R$, where $f_a$ is the axion decay constant and $R$ is the radius of the neutron star.  The force between the two stars can be either attractive or repulsive.  When the separation between the two neutron stars drops to about $1/m_a$, deviations from the standard general-relativistic inspiral dynamics will become evident and could be seen in the gravitational wave signal.  In addition, once the frequency of the inspiral rises above $m_a$, scalar Larmor radiation can occur, dissipating orbital energy from the merging system. 

The authors of \cite{Huang:2018pbu} provided a forecast for how well Advanced LIGO would be able to constrain such an ultralight axion from a measurement of the gravitational wave signal from the neutron star inspiral.  They also speculated that by measuring the scalar charge of the neutron star (from the deviation of the inspiral dynamics from their expected general-relativistic behavior), one could determine the compactness $M/R$ of the star, which constrains the nuclear equation of state \cite{Chen:2015zpa}.  

The work in \cite{Poddar:2019zoe,Seymour:2020yle} sets upper limits on the axion decay constant by comparing the measured decay of the period of several binary neutron star and neutron star - white dwarf systems to the expected result from the energy loss due to gravitational radiation, allowing them to constrain the amount of scalar Larmor radiation that could have been produced by the compact objects.  
\subsection{Axions in merger remnants}
\label{sec:axions_in_merger_remnants}
Following in the path of studying the role of ``particle-like'' axions in supernovae \cite{Raffelt:1996wa,Raffelt:2006cw,Bollig:2020xdr,Carenza:2020cis,Bar:2019ifz,Lucente:2020whw,Fischer:2016cyd} and isolated neutron stars \cite{Beznogov:2018fda,Sedrakian:2018kdm,Sedrakian:2015krq,Fortin:2021sst,Dessert:2019dos}, recent progress has been made in understanding the role of axions in neutron star merger remnants, which will be discussed in this section.  The study of ultralight axions in merger remnants \cite{Sagunski:2017nzb} is less developed, and we will not discuss it here.
\subsubsection{Axion mean free path in hot, dense matter}
The potential role of axions in a neutron star merger after inspiral is determined by their mean free path in the hot, dense conditions present in the merger remnant.  If their mean free path is small compared to the size of the remnant, they would form a thermally equilibrated Bose gas and would be able to contribute to transport processes inside the star, for example, thermal conductivity or shear viscosity (see Sect.~\ref{sec:trapped_axions}).  If the axion mean free path is comparable to or larger than the size of the remnant, then the axions would free-stream through the system, taking some energy with them.  This would serve to cool down the merger remnant \cite{Harris:2020qim} (see Sect.~\ref{sec:axion_cooling_of_mergers}).  The mean free path of axions is determined by how often they are absorbed by the inverse bremsstrahlung process $N+N'+a\rightarrow N+N'$ where $N$ and $N'$ are either neutrons or protons.

\begin{figure}
\centering
\includegraphics[width=.45\textwidth]{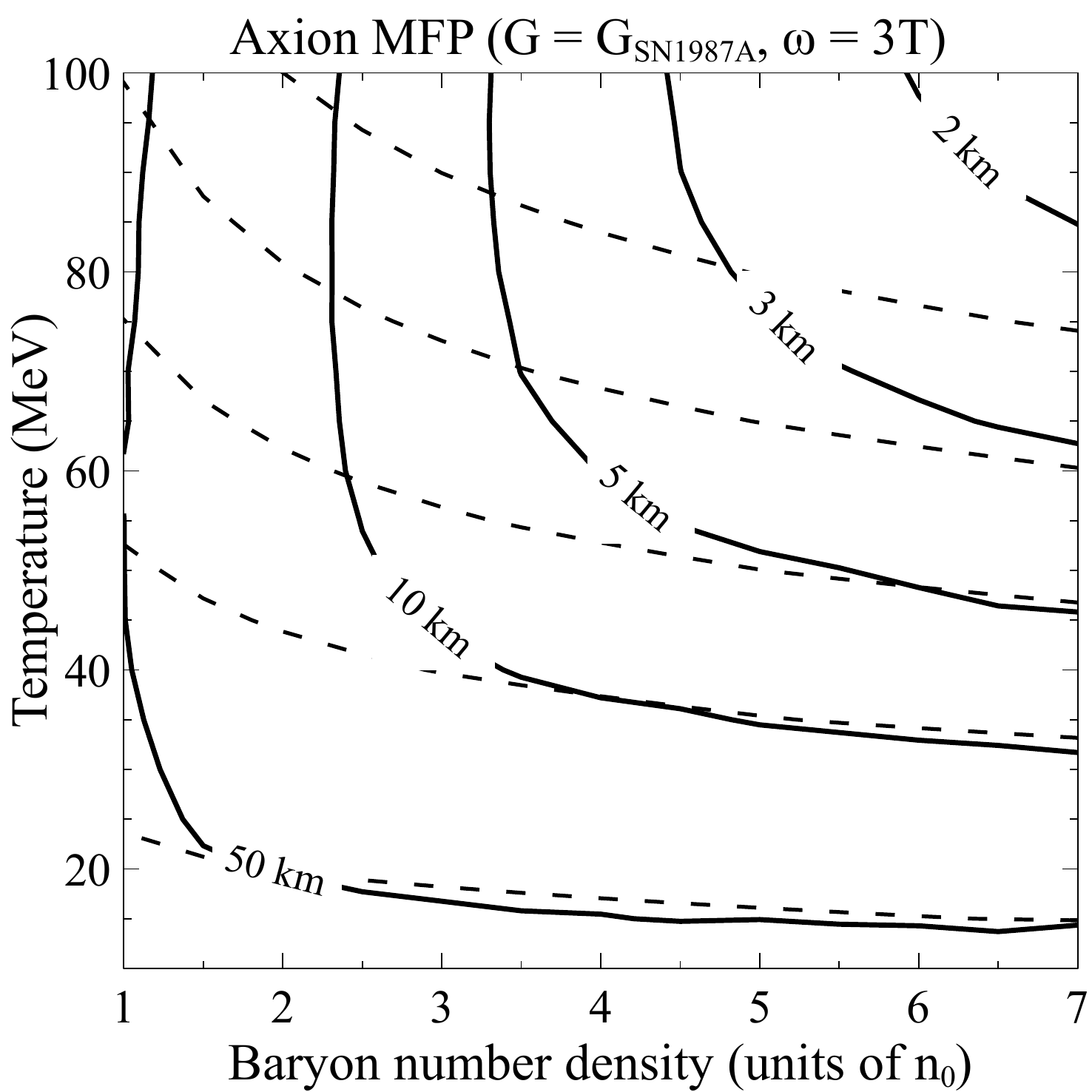}
\caption{Axion mean free path (for an axion with energy $\omega = 3T$) due to absorption via $n+n+a\rightarrow n+n$.  The axion-neutron coupling constant is chosen to be at the upper bound set by SN1987a \cite{Graham:2015ouw}, so this plot indicates the minimum mean free path an axion with $\omega=3T$ could have at each density and temperature.  The dashed contours correspond to the axion MFP calculated in strongly degenerate nuclear matter [Eq.~\eqref{eq:axion_mfp_degenerate}], while the solid contours are the result of the full integration over phase space [Eq.~\eqref{eq:axion_mfp_phase_space}].  Figures reproduced from \cite{Harris:2020qim}.}
\label{fig:axionMFP}
\end{figure}

The mean free path of axions in nuclear matter is determined from the individual mean free paths due to each possible absorption process
\begin{equation}
    \lambda^{-1}_a = \lambda^{-1}_{nn}+\lambda^{-1}_{np} + \lambda^{-1}_{pp},
\end{equation}
and is a function of the baryon density and temperature, but also of the axion energy.  The mean free path of a given absorption process, for example, $n+n+a\rightarrow n+n$, is given by the phase space integral
\begin{equation}
    \lambda_{nn}^{-1} = \int \frac{\mathop{d^3p_1}}{(2\pi)^3}\frac{\mathop{d^3p_2}}{(2\pi)^3}\frac{\mathop{d^3p_3}}{(2\pi)^3}\frac{\mathop{d^3p_4}}{(2\pi)^3}\frac{S\sum_{\text{spins}}\vert\mathcal{M}\vert^2}{2^5E_1^*E_2^*E_3^*E_4^*\omega}(2\pi)^4\delta^4(p_1+p_2-p_3-p_4+p_a)f_1f_2(1-f_3)(1-f_4).
    \label{eq:axion_mfp_phase_space}
\end{equation}
The factors in this equation have the same meaning as in Eq.~\eqref{eq:emissivity_integral}.  Using the Fermi surface approximation, the mean free path of an axion with energy $\omega$ has been calculated in degenerate nuclear matter (for example, in cold neutron stars), and has the simple form
\begin{equation}
    \lambda_{nn}^{-1}=\frac{1}{18\pi^5}f^4G_{an}^2(m_n/m_{\pi})^4p_{Fn}F(c)\frac{\omega^2+4\pi^2T^2}{1-e^{-\omega/T}}.
    \label{eq:axion_mfp_degenerate}
\end{equation}
Here, $c=m_{\pi}/(2p_{Fn})$ and the derivation and definition of $F(c)$ -- which is a function of the baryon density -- is given in \cite{Harris:2020qim} (the original calculation was done in \cite{Ishizuka:1989ts}, however).  The mean free path of the axion in nondegenerate nuclear matter (seen, for example, in a supernova) is calculated in \cite{Burrows:1990pk,Giannotti:2005tn} and in matter of arbitrary degeneracy (although assuming non-relativistic nucleons) in \cite{Carenza:2019pxu}.  These calculations are reviewed and extended to the case of arbitrary degeneracy and fully relativistic nucleons in \cite{Harris:2020qim}, where the full phase space integration is performed.  In Fig.~\ref{fig:axionMFP}, we show the results of a calculation of the axion mean free path (from \cite{Harris:2020qim}) in the absorption process\footnote{This mean free path calculation considers only $n+n+a\rightarrow n+n$.  The inclusion of $p+p+a\rightarrow p+p$ and $n+p+a\rightarrow n+p$ will shrink the mean free path by a factor of a few.  However, the calculation in Fig.~\ref{fig:axionMFP} also does not include the factor of (approximately) four reduction in the square of the matrix element due to improvements to the one-pion exchange interaction (Sect.~\ref{sec:magnetar_emissivity}).  This leads to a lengthening of the mean free path by a factor of 4.  Thus, the results shown in this figure are likely close to the correct answer.} $n+n+a\rightarrow n+n$.

Fig.~\ref{fig:axionMFP} shows that even for the strongest allowed axion-nucleon coupling, in the thermodynamic conditions likely encountered in neutron star mergers, the axion mean free path is quite long and axion trapping in mergers is unlikely.
\subsubsection{Axion cooling of merger remnants}
\label{sec:axion_cooling_of_mergers}
Since axions likely free-stream from the nuclear matter in a neutron star merger, the temperature of a fluid element decreases due to the energy loss from axion emission according to 
\begin{equation}
    \frac{\mathop{dT}}{\mathop{dt}}=-\frac{Q}{c_V},
    \label{eq:dTdt}
\end{equation}
where $Q$ is the axion emissivity and $c_V$ is the specific heat of the nuclear matter at constant volume (or baryon density).  Since the vast majority of the matter in neutron star mergers has a temperature well above an MeV, we can ignore nucleon superfluidity and thus axions are produced only by the three nucleon bremsstrahlung processes $N+N'\rightarrow N+N'+a$.  The total emissivity is the sum of the emissivity from each axion production process 
\begin{equation}
    Q = Q_{nn}+Q_{pp}+Q_{np}.
\end{equation}
The emissivities are calculated in the degenerate limit in the context of magnetars (Sect.~\ref{sec:magnetar_emissivity}) (one can consider strongly degenerate nuclear matter that is not superfluid by just setting the reduction factors $R_{np}$ and $R_{pp}$ to one).  The emissivity of axions due to bremsstrahlung processes has also been calculated for nonrelativistic and nondegenerate nucleons in \cite{Turner:1987by,Stoica:2009zh,Raffelt:1993ix} and extended to nonrelativistic nucleons with arbitrary degeneracy in \cite{Carenza:2019pxu}.  The full phase space integration was performed in \cite{Harris:2020qim} to obtain the axion emissivity with relativistic nucleons of arbitrary degeneracy.  As in the mean free path calculation, the emissivity calculation in \cite{Harris:2020qim} chose $C_{\pi}=1$ and neglected axion production processes involving protons.

The specific heat is dominated by the particle species with the largest number of low energy excitations, which in dense matter is the neutron.  The specific heat of a degenerate Fermi liquid of neutrons is \cite{1994ARep...38..247L}
\begin{equation}
    c_V = \frac{1}{3}m_n^Lp_{Fn}T,
\end{equation}
where $m_n^L$ is the Landau effective mass of the neutron.
The specific heat does not change significantly even when the neutrons become nondegenerate, which occurs in the regions of the merger which have low density but high temperature \cite{Harris:2020qim}. 

\begin{figure*}[t!]
\begin{minipage}[t]{0.5\linewidth}
\includegraphics[width=.95\linewidth]{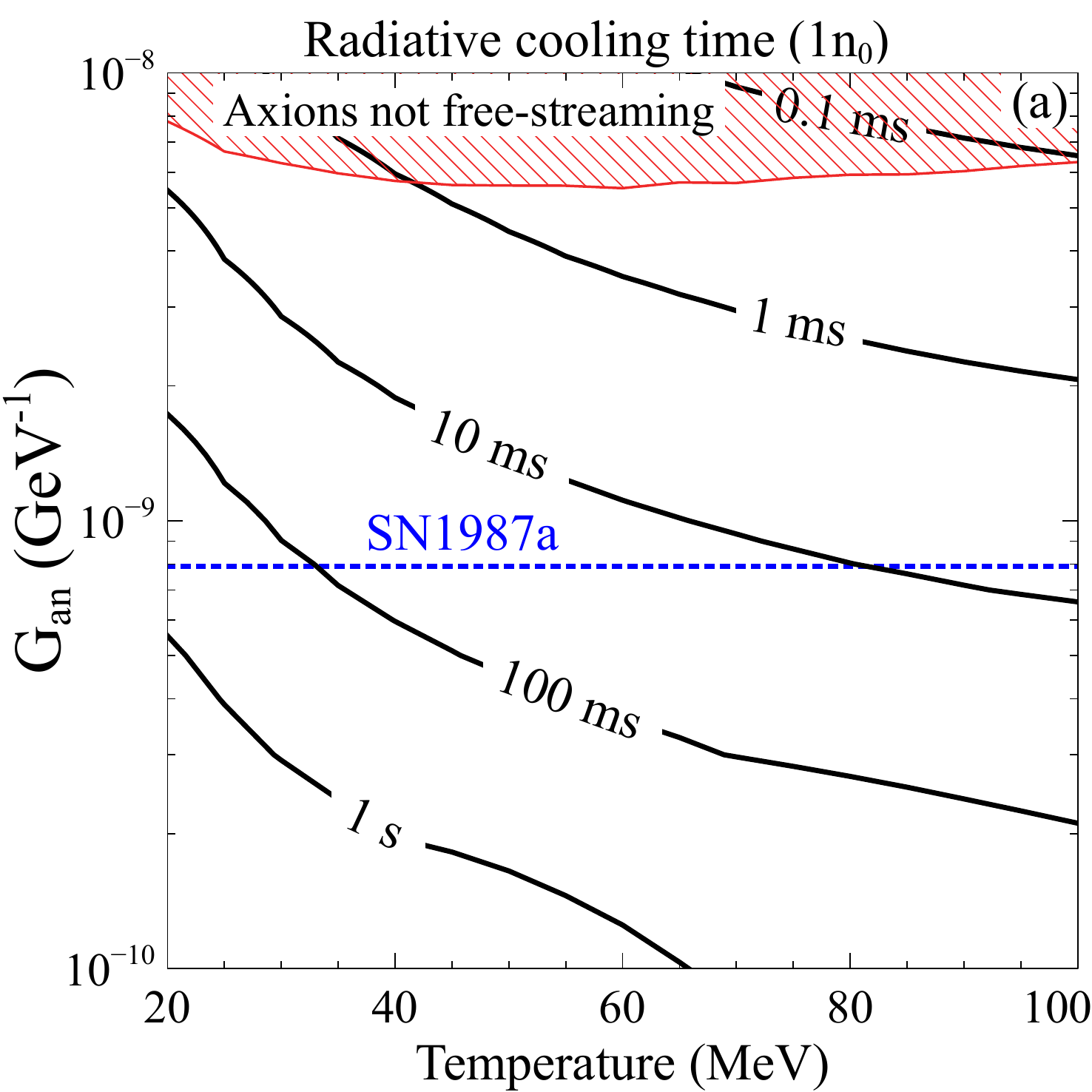}
\end{minipage}\hfill%
\begin{minipage}[t]{0.5\linewidth}
\includegraphics[width=.95\linewidth]{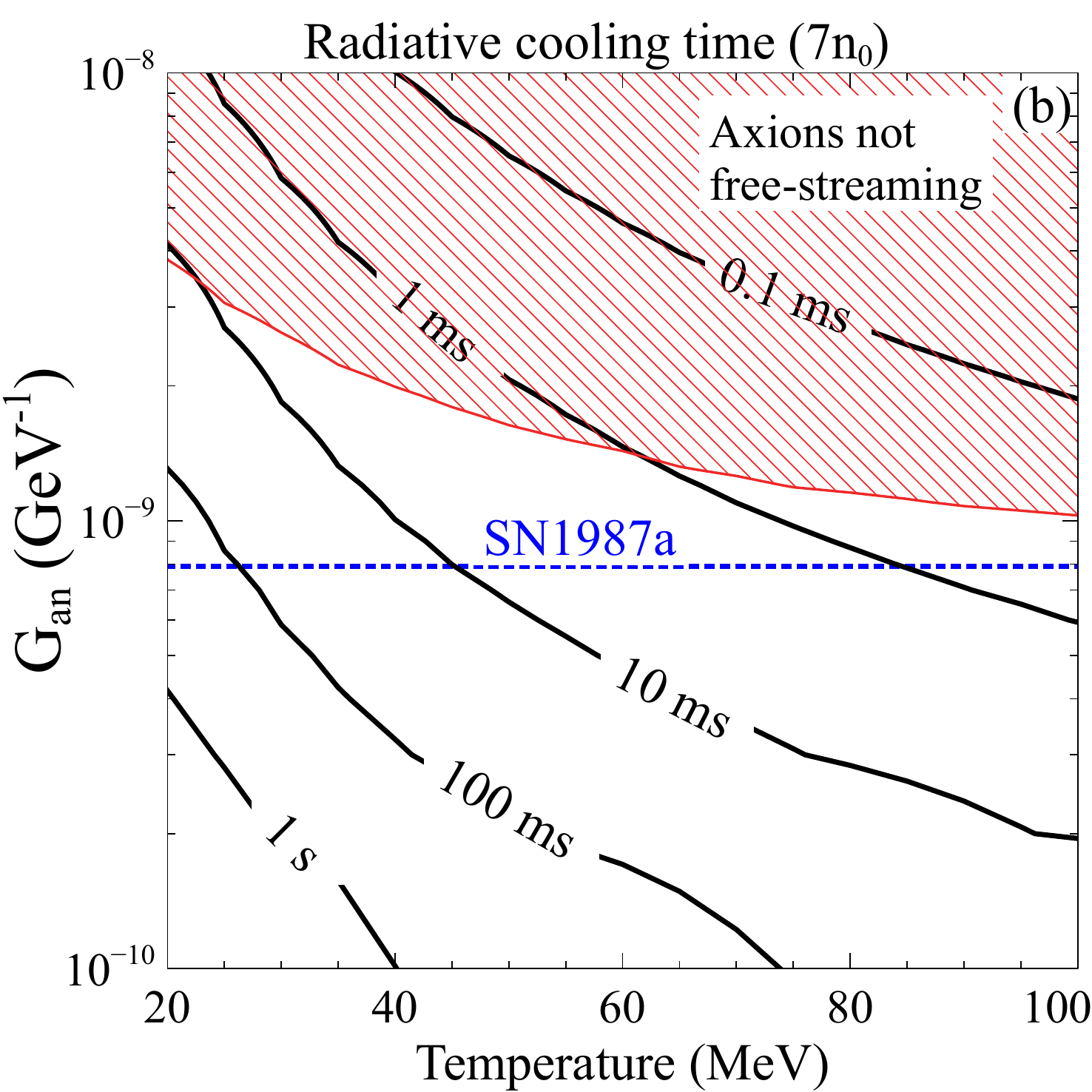}
\end{minipage}%
\caption{Radiative cooling time due to axion emission from the process $n+n\rightarrow n+n+a$ at densities of $1n_0$ (left panel) and $7n_0$ (right panel).  All couplings stronger than the dotted blue line are disallowed by the observation of SN1987a.  As long as the axion-neutron coupling is not too far below the bound set by SN1987a, the lower density regions of the star can cool substantially in tens of milliseconds, while the highest density regions of the merger could possibly cool in under 10 milliseconds (although it is not expected that the densest regions of the merger reach temperatures of more than about 10 MeV - see Fig.~\ref{fig:T_nb_plane}).  The emissivity used in this calculation only took into account axion production from $n+n\rightarrow n+n+a$, neglecting the other two bremsstrahlung processes.  Also, $C_{\pi}=1$ was chosen, not the more realistic value of around $1/4$.  Including the other two bremsstrahlung processes and choosing $C_{\pi}=1/4$ will not significantly change the results shown here.  Figures reproduced from \cite{Harris:2020qim}.}
\label{fig:rad_cool_time_vs_G}
\end{figure*}

The time that it takes for a fluid element to cool to half of its current temperature by radiating axions is calculated by solving the differential equation in \eqref{eq:dTdt}.  The cooling time is calculated in \cite{Harris:2020qim}, where the nuclear matter is described by the NL$\rho$ equation of state \cite{Liu:2001iz} and the emissivity is calculated by performing the full phase space integration [Eq.~\eqref{eq:emissivity_integral}].  The results are plotted in Fig.~\ref{fig:rad_cool_time_vs_G} at saturation density $n_0$ (left panel) and at $7n_0$ (right panel).  Along the y-axis is the strength of the axion-neutron coupling.  Couplings above the blue dotted line are ruled out by SN1987a \cite{Graham:2015ouw}.  For very large (and thus, ruled out) values of the coupling, axions would be trapped inside the merger and cooling would actually occur on the  diffusive timescales of several seconds \cite{1984ApJ...283..848B}, not the radiative timescales shown in the red hatched region of Fig.~\ref{fig:rad_cool_time_vs_G}.  As the coupling becomes weaker, the axion mean free path grows and axions can escape the merger, cooling it.  The calculations shown in Fig.~\ref{fig:rad_cool_time_vs_G} indicate that, as long as the axion-neutron coupling is not too much smaller than the SN1987a bound, at low densities fluid elements can cool significantly in timescales of tens of milliseconds and at high densities, significant cooling can take place on timescales under ten milliseconds.  Simulations (conducted without axion cooling) show that the hottest matter in the merger exists at around $1-2n_0$ and can reach temperatures of at least 50 MeV (see Fig.~\ref{fig:T_nb_plane}).  The calculations shown in Fig.~\ref{fig:rad_cool_time_vs_G} predict axion emission could cool this region of the merger to 25 MeV in approximately 30 milliseconds, well within the lifetime of many merger remnants.  

\begin{figure*}[t!]
\begin{minipage}[t]{\linewidth}
\includegraphics[width=.95\linewidth]{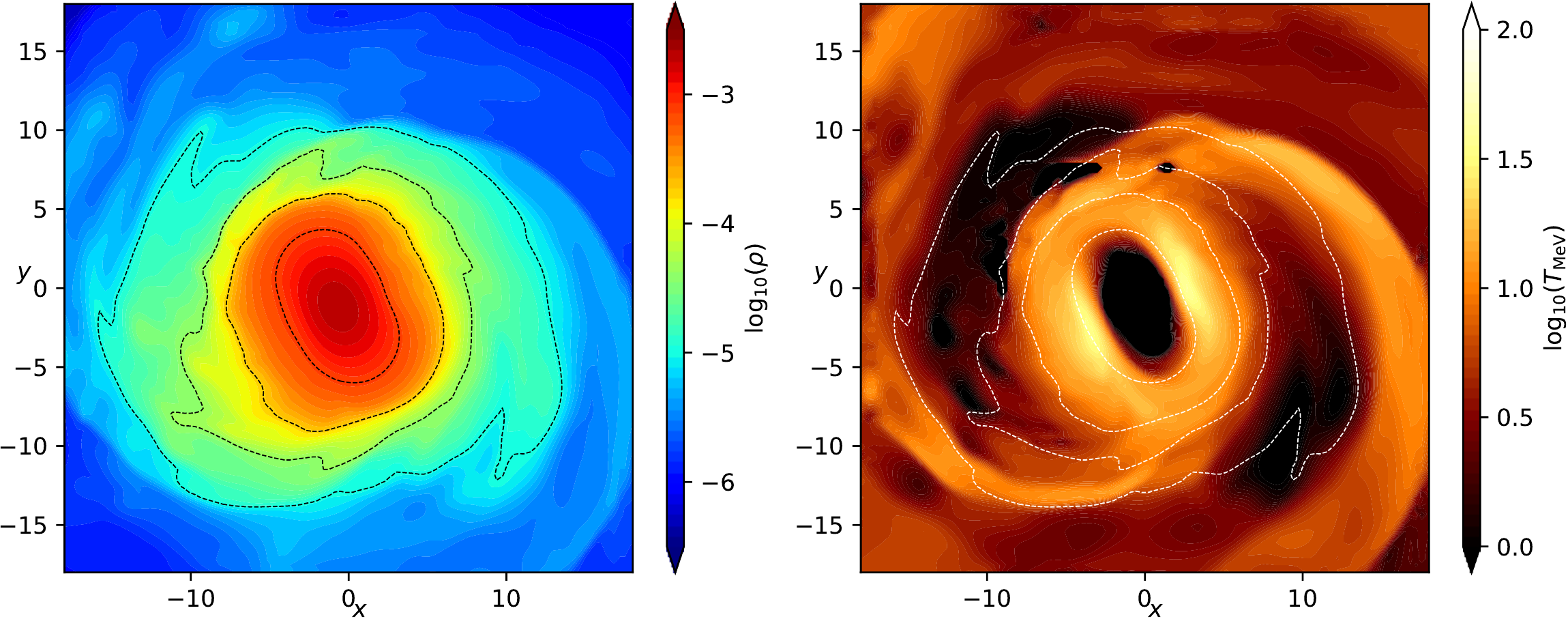}
\end{minipage}\hfill%
\begin{minipage}[t]{\linewidth}
\includegraphics[width=.95\linewidth]{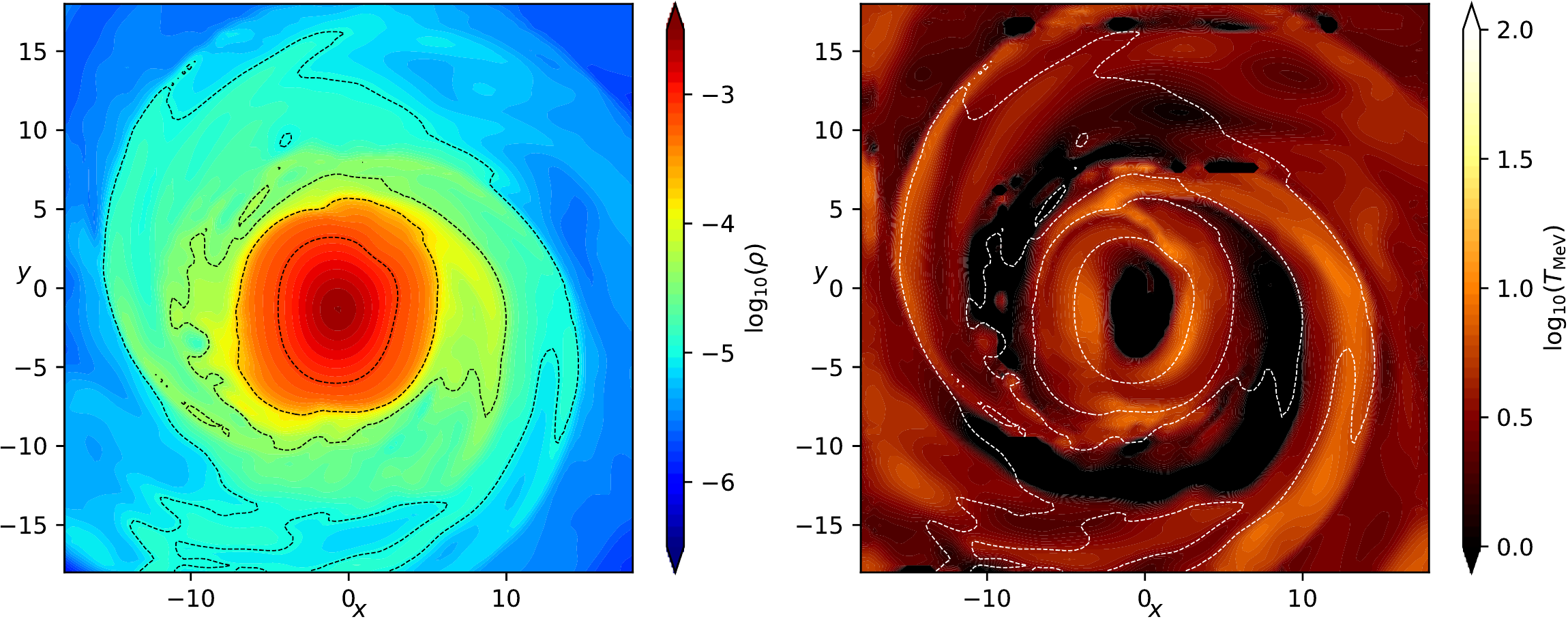}
\end{minipage}%
\caption{Density profile (left panel) and temperature profile (right panel) from the neutron star merger simulation conducted in \cite{Dietrich:2019shr}.  Snapshots are all from 15 ms after merger.  The top row neglects axion cooling, and the bottom row incorporates axion cooling assuming (for illustration) an axion-nucleon coupling approximately 100 times the SN1987a bound \cite{Graham:2015ouw}, significantly overestimating the effect that axions could have on the cooling.  The simulation neglects neutrino cooling.  Both neutron stars in the merger initially had masses of $1.365M_{\odot}$.  Figures adapted from \cite{Dietrich:2019shr}.}
\label{fig:merger_simulation_with_axions}
\end{figure*}

The effects of axion cooling were included in a neutron star merger simulation by the authors of \cite{Dietrich:2019shr}.  They included an energy-loss term in their hydrodynamic equations that was set by the axion emissivities calculated by the authors of \cite{PhysRevD.38.2338}.  This method of incorporating cooling due to particle emission is similar to an early way of treating neutrino cooling in supernovae simulations \cite{1996ApJ...472..308S,1984oup..book.....M} and, later, compact star mergers \cite{Paschalidis:2011ez}.  In Fig.~\ref{fig:merger_simulation_with_axions} we display results of the simulations including cooling from axion emission.  The left column plots are colored by density and the right column plots are colored by temperature.  The top row corresponds to a simulation without axion cooling, while the bottom row corresponds to a simulation where axion cooling is included, with a value of the axion-nucleon coupling constant approximately 100 times the value allowed by SN1987a.  While this coupling is unrealistically large, and thus overestimates the effects of cooling, it was chosen so that the effects of axion cooling can be easily seen by eye.
It is easy to see from the right column of Fig.~\ref{fig:merger_simulation_with_axions} that the axions indeed cool the remnant significantly in 15 milliseconds (the time between the collision and the snapshots in Fig.~\ref{fig:merger_simulation_with_axions}) compared to the case where no cooling mechanism exists -- note that the simulations discussed here neglect cooling of the remnant due to neutrinos so as to focus on the effects of axion cooling.  In addition, the energy loss due to axions sphericalizes the remnant, eliminating the bar shape that was present when axion cooling was neglected.  The density of the central region is observed to be slightly larger when the axion-neutron coupling is stronger.  In general, the authors of \cite{Dietrich:2019shr} observed that with increased axion-neutron coupling, the lifetime of the remnant decreases, as is expected because the remnant is sustained partially by thermal pressure, as discussed in Sect.~\ref{sec:ns_merger_generalities}.  However, the lifetime is only shortened by a couple milliseconds between the top and bottom rows of Fig.~\ref{fig:merger_simulation_with_axions}, indicating that even strong axion emission is unlikely to significantly induce gravitational collapse.  This is consistent with the expectation that differential rotation is the dominant influence on the lifetime of the remnant \cite{Baumgarte:1999cq}.  The difference in the gravitational wave signal predicted by the simulation with and without axions was small.  

The conclusion of the work including cooling from axion emission in neutron star mergers \cite{Dietrich:2019shr,Harris:2020qim} is that when the axion-neutron coupling is not too much smaller than the SN1987a bound, axion emission can lead to substantial cooling of the merger remnant.  However, this cooling only has minor changes on the dynamics of the merger at the level of the simulations conducted in \cite{Dietrich:2019shr}, raising the possibility that, at present, it may be difficult to use the postmerger phase to constrain axion physics any further than has been done through other astrophysical environments.  This is especially true given the current uncertainties in, for example, the nuclear equation of state, which has a large impact on the post-merger dynamics.

However, if additional complexities are considered, axions again have the chance to substantially influence the dynamics of neutron star mergers.  As merger simulations develop, there is interest in including a possible phase transition to quark matter \cite{Annala:2019puf,Hanauske:2019qgs,Weih:2019xvw,Most:2019onn,Chesler:2019osn,Gieg:2019yzq,Dexheimer:2019mhh,Most:2018eaw,Bauswein:2018bma,Oechslin:2004yj} -- or another exotic phase\footnote{For example, pions \cite{Fore:2019wib,Pethick:2015jma}, hyperons \cite{Sekiguchi:2011mc,Alford:2020pld}, or quarkyonic matter \cite{McLerran:2018hbz}.} -- at high density or high temperature.  In this case, cooling on merger timescales that is due to axion emission could trigger an unexpected phase transition in the nuclear matter, which would almost certainly have observable consequences.  In addition, efforts are starting to be made to include transport processes in merger simulations \cite{Harris:2020rus}.  Many transport processes are strongly temperature dependent,\footnote{For example, bulk viscosity \cite{Alford:2017rxf,Alford:2019qtm,Alford:2019kdw,Alford:2020pld,Alford:2020lla}, shear viscosity \cite{Alford:2017rxf,Schmitt:2017efp,1982ApJ...253..816G}, and thermal conductivity \cite{Alford:2017rxf,Schmitt:2017efp,1982ApJ...253..816G}.} and thus cooling from axion emission would impact transport during the merger.
\subsubsection{Transport from trapped dark sector particles}
\label{sec:trapped_axions}
Thermal transport is due to particles with long mean free paths.  If the mean free path is long, but still shorter than the system size, the particle will contribute to evening out thermal gradients in the system.  In a neutron star merger, neutrinos -- if they are trapped -- dominate the thermal conductivity because their mean free path is always longer than that of the neutron, proton, or electron.  The authors in  \cite{Alford:2017rxf} calculated that a temperature gradient over a distance of 100 meters could be eliminated by neutrino-driven conduction in tens of milliseconds.  In a region of the star where neutrinos are not trapped, it would take months for electron-driven conduction to eliminate the same temperature gradient.

Dark sector particles, if they are trapped inside the merger remnant, could provide a more expedient path towards thermal equilibration.  While the axion is not likely to be trapped in a merger, as an illustration Ref.~\cite{Harris:2020qim} calculated the thermal equilibration timescale due to axion emission and absorption via nucleon bremsstrahlung processes in a merger remnant.  The timescale is highly dependent on the axion-neutron coupling.  It would be useful to consider the possibility of thermal transport due to other dark sector particles, including CP-even scalars \cite{Dev:2020eam,Dev:2020jkh}, dark photons \cite{Mahoney:2017jqk,Dent:2012mx}, or other light dark matter candidates.  It may also be possible for dark sector particles to contribute to shear or bulk viscosity inside the merger remnant. 
\section{Laboratory-Produced ALP Searches \label{sec:labsearch}}

If an axion or ALP exists, it could be produced in the laboratory, without relying on any extraterrestrial sources, and leave experimental signatures within a detector. Unlike the cases in Sects.~\ref{sec:magnet} and \ref{sec:merger}, environmental parameters associated with production of axions or ALPs are more under control, and therefore, the relevant searches can set more conservative and model-independent constraints on models of axion or ALP.  
In particular, in regards to the PVLAS anomaly~\cite{Zavattini:2005tm} various models or mechanisms to avoid astrophysical and/or cosmological bounds have been proposed. They typically introduce features in the hidden sector that suppress or turn off Primakoff production in stellar environments but restore them elsewhere: for example, choices of couplings to facilitate trapping~\cite{Jain:2005nh}, phase transitions~\cite{Mohapatra:2006pv,Masso:2006gc}, and chameleon-like screening effects~\cite{Brax:2007ak} (see \cite{Jaeckel:2006xm} for a more detailed discussions on these and other studies). 
More recently, the EDGES~\cite{Bowman:2018yin} and Xenon1T~\cite{Aprile:2020tmw} anomalies have stimulated similar efforts in constructing viable ALP models, e.g., \cite{DeRocco:2020xdt} and \cite{Bloch:2020uzh}, and the importance of laboratory-based searches receives attention again as robust consistency checks of those models. 

In the following subsections, we elaborate the ALP searches at beam-dump and reactor neutrino experiments and review related studies, followed by a survey on the recent developments in the collider ALP probes. 
Light-shining-through-wall (LSW) experiments, e.g., ALPS I/II~\cite{Ehret:2010mh,Bahre:2013ywa}, CROWS~\cite{Betz:2013dza}, and OSQAR~\cite{Ballou:2015cka}, and polarization experiments, e.g., PVLAS~\cite{Zavattini:2005tm}, 
are other important classes of laboratory-produced ALP probes, for which comprehensive reviews can be found in \cite{Irastorza:2018dyq,Sikivie:2020zpn}. 

\subsection{Beam-dump and reactor neutrino experiments}

\begin{table}[t]
    \centering
    \resizebox{\columnwidth}{!}{%
    \begin{tabular}{c| c c c c c c c c}
    \hline \hline
    \multirow{2}{*}{Experiment} & \multirow{2}{*}{Beam} & $E_{\rm beam}$ & POT/EOT & \multirow{2}{*}{Target} & \multirow{2}{*}{Detector} & \multirow{2}{*}{Mass} & Distance & \multirow{2}{*}{Angle} \\
    & & [GeV] & [yr$^{-1}$] &  &  &  & [m] &  \\
    \hline
    CCM~\cite{CCM,dunton_2019,Dutta:2020vop} & $p$ & 0.8 & $1.5\times 10^{22}$ & W & LAr & 7~t & 20 & 90$^\circ$ \\
    \multirow{2}{*}{COHERENT~\cite{Akimov:2017ade,Akimov:2018ghi,Akimov:2019xdj}} & \multirow{2}{*}{$p$} & \multirow{2}{*}{1} & \multirow{2}{*}{$1.5\times 10^{23}$} & \multirow{2}{*}{Hg} & CsI[Na] & 14.6~kg & 19.3 & $90^\circ$ \\
      &  &  &  &  & LAr & 24~kg (0.61~t) & 28.4 & $137^\circ$ \\
    JSNS$^2$~\cite{Ajimura:2015yux,Ajimura:2017fld,Dutta:2020vop} & $p$ & 3 & $3.8\times 10^{22}$ & Hg & Gd-LS & 17~t & 24 & $29^\circ$ \\
    \hline
    MiniBooNE~\cite{AguilarArevalo:2008qa} & $p$ & 8 & ($\sim3\times10^{21}$) & Be & Mineral oil& 450~t & 541 & On-axis \\
    MicroBooNE~\cite{Antonello:2015lea,Acciarri:2016smi} & $p$ & 8 & $6.6\times 10^{20}$ & Be & LArTPC & \underline{89}~t & 470 & On-axis \\
    SBND~\cite{Antonello:2015lea} & $p$ & 8 & $6.6\times 10^{20}$ & Be & LArTPC & \underline{112}~t & 110 & On-axis \\
    ICARUS~\cite{Antonello:2015lea} & $p$ & 8 & $6.6\times 10^{20}$ & Be & LArTPC & \underline{476}~t & 600    & On-axis \\
    \multirow{3}{*}{T2K~\cite{Abe:2011ks}} & \multirow{3}{*}{$p$} & \multirow{3}{*}{30} & \multirow{3}{*}{$4.8\times10^{21}$} & \multirow{3}{*}{Graphite} & Water & $\sim1.9$~t & \multirow{3}{*}{280} & \multirow{3}{*}{$2.5^\circ$} \\
      & & &  & & Gas TPC & $9$~kL & & \\
       & & &  & & Water$\,+\,$PS & 2.2~t & & \\
    NO$\nu$A~\cite{Ayres:2007tu} & $p$ & 120 & $6.0\times 10^{20}$ & Graphite & PVC-LS & \underline{125}~t & 1,000 & 0.84$^\circ$ \\
    \multirow{2}{*}{DUNE~\cite{Abi:2020wmh,Abi:2020evt}} & \multirow{2}{*}{$p$} & \multirow{2}{*}{120} & \multirow{2}{*}{$1.1\times 10^{21}$} & \multirow{2}{*}{Graphite} & LArTPC & 67.2~t & \multirow{2}{*}{574} & \multirow{2}{*}{Movable} \\
     &  &  &  &  & GArTPC & \underline{1.8}~t &  &  \\
     \multirow{2}{*}{SHiP~\cite{Anelli:2015pba}} & \multirow{2}{*}{$p$} & \multirow{2}{*}{400} & \multirow{2}{*}{$0.4\times 10^{20}$} & \multirow{2}{*}{TZM} & Pb-ECC & \underline{9.6}~t & $\sim50$ & \multirow{2}{*}{On-axis} \\
      &  &  &  &  & ECAL/HCAL & -- & $\sim110$ &  \\
    \hline
    LDMX~\cite{Akesson:2018vlm} & $e^-$ & 4 -- 16 & $10^{16}$ & W & ECAL/HCAL & -- & $\mathcal{O}(1)$ & On-axis  \\
    BDX~\cite{Battaglieri:2016ggd,Battaglieri:2019nok} & $e^-$ & 10.6 & $\sim10^{22}$ & Al & ECAL & -- & 20  &  On-axis \\
    NA64~\cite{Banerjee:2020fue} & $e^-$ & 100 & ($2.84\times 10^{11}$) & PRS/ECAL & HCAL & -- & $\mathcal{O}(1)$ &  On-axis \\
    \hline \hline
    \end{tabular}}
    \caption{Key specifications of existing and future beam-dump type (neutrino) experiments. In the first four experiments, charged pions and their decay product muons created in the target are stopped inside the target, whereas in the other proton-beam experiments, a large fraction of them decay outside the target, resulting in a neutrino flux in the forward region. The POT of MiniBooNE corresponds to the data collection in-between 2002 and 2019, the EOT of LDMX is for Phase 2, and the EOT of NA64 corresponds to the data collection in-between 2016 and 2018. The mass value of the COHERENRT-LAr detector in parentheses is for a future upgrade. The (underlined) values in the Mass column are fiducal (active) masses. [POT/EOT: protons/electrons on target, TZM: titanium-zirconium doped molybdenium, LAr/GAr: liquid/gaseous argon, Gd-LS: gadolinium-loaded liquid scintillator, PVC-LS: PVC cells filled with liquid scintillators, TPC: time projection chamber, Pb-ECC: emulsion cloud chamber with lead plates, PRS: pre-shower detector, ECAL/HCAL: electromagnetic/hadronic calorimeter, PS: plastic scintillators] \label{tab:listdump}}
    \end{table}
    
    \begin{table}[t]
    \resizebox{\columnwidth}{!}{%
    \begin{tabular}{c|c c c c}
    \hline \hline
    Experiment~ & Thermal power [GW] & ~Detector~ & ~Mass~ & ~Distance [m] \\
    \hline
    CONNIE~\cite{Moroni:2014wia,Aguilar-Arevalo:2016khx}  & 3.95 & Skipper CCD & 52~g & 30 \\
    CONUS~\cite{Hakenmuller:2019ecb}  & 3.9 & Ge & 3.76~kg & 17.1 \\
    MINER~\cite{Agnolet:2016zir,Dent:2019ueq} & 0.001 & Ge$\,+\,$Si & 4~kg & 1 -- 2.25 \\
    NEON~\cite{neon} & 2.82 & NaI[Tl] & $\sim10/50/100$~kg (Ph1/2/3) & 24 \\
    $\nu$-cleus~\cite{Strauss:2017cuu}  & 4 & CaWO$_4\,+\,$Al$_2$O$_3$ & ~6.84~g$\,+\,$4.41~g~ & ~15/40/100 (N/M/F) \\
    $\nu$GeN~\cite{Belov:2015ufh}  & $\sim 1$ & Ge & 1.6 -- 10~kg & 10 -- 12.5\\
    RED-100~\cite{Akimov:2017xpp,Akimov:2017hee} & $\sim1$ & DP-Xe & $\sim100$~kg & 19 \\
    Ricochet~\cite{Billard:2016giu}  & 8.54 & Ge$\,+\,$Zn & 10~kg & 355/469  \\
    SBC-CE$\nu$NS~\cite{sbc,AristizabalSierra:2020rom}& 0.68 & LAr[Xe] & 10~kg &  3/10 \\
    SoLid~\cite{Abreu:2020bzt}  & 40 -- 100 & PVT$\,+\,^6$LiF:ZnS(Ag)  & 1.6~t &  5.5 -- 12\\
    TEXONO~\cite{Wong:2016lmb}  & 2.9 & Ge & 1.06~kg & 28\\
    vIOLETA~\cite{Fernandez-Moroni:2020yyl} & 2 & Skipper CCD & 1 -- 10~kg & 8 -- 12 \\
    \hline \hline
    \end{tabular}
    }
    \caption{Key specifications of existing and future reactor (neutrino) experiments. [CCD: charge couple device, DP-Xe: dual-phase xenon, Ph1/2/3: phase1/phase2/phase3, N/M/F: near/medium/far]\label{tab:listreactor}}
\end{table}

The laboratory-produced ALP searches are extensively performed in beam-dump type experiments (i.e., fixed target experiments) including E137~\cite{Bjorken:1988as}, E141~\cite{Riordan:1987aw}, reanalyses~\cite{Dobrich:2015jyk} of CHARM~\cite{Bergsma:1985qz} and NuCal~\cite{Blumlein:1990ay} ALP searches, NOMAD~\cite{Astier:2000gx}, and NA64~\cite{Banerjee:2020fue}, and reactor experiments~\cite{Vuilleumier:1981dq,Zehnder:1982bg,Datar:1982ef,Alekseev:1982iq,Cavaignac:1982ek,Ananev:1983ki,Ketov:1986az,Koch:1986aq,Chang:2006ug} for the past decades, and an increasing number of studies have pointed out that the next-generation beam neutrino experiments~\cite{Dobrich:2015jyk,Alekhin:2015byh,Berlin:2018pwi,Berlin:2018bsc,Akesson:2018vlm,Dobrich:2019dxc,Bonivento:2019sri,Brdar:2020dpr,Kelly:2020dda,Dev:2021ofc} and high-power reactor neutrino experiments~\cite{Dent:2019ueq,AristizabalSierra:2020rom} can explore a wider range of ALP parameter space.\footnote{See also \cite{deNiverville:2020qoo} proposing new physics searches at reactor neutrino experiments, using the ALP-dark photon-photon coupling.} 
We provide summaries tabulating key specifications of existing and future beam-dump type (neutrino) experiments in Table~\ref{tab:listdump}\footnote{See also \cite{Lanfranchi:2020crw} for an extensive list of accelerator-based experiments including the decommissioned ones.} and reactor (neutrino) experiments in Table~\ref{tab:listreactor}. 
The first four experiments in Table~\ref{tab:listdump} utilize charged pions and their decay product muons that are stopped inside the target in order to produce an isotropic neutrino flux. By contrast, in the other proton-beam experiments, a large fraction of the charged pions and muons escape from the target and decay outside, resulting in a neutrino flux orienting in the forward direction. 
While proton or electron beam-dump experiments and their physics opportunities have been extensively investigated, photon-beam experiments are receiving attention as a complementary avenue of exploring ALP parameter space, e.g., PrimEx~\cite{Larin:2010kq} and GlueX~\cite{Shepherd:2009zz}.
In addition, proposed muon beam experiments, e.g., NA64$_\mu$~\cite{Gninenko:2640930} and M$^3$~\cite{Kahn:2018cqs}, would provide opportunities for the ALP search.  

The calculation of the expected signal rate in these experiments essentially comprises of three parts: i) production of ALPs at a source point \eqref{sec:prod}, ii) transportation of ALPs from the source point to the detector of interest \eqref{sec:tran}, and iii) detection of ALPs at the detector \eqref{sec:det}. Here the production takes into account only the portion of the ALP flux directed toward the detector. 
Given a source particle (say, $i$th particle)  which would be converted to an ALP, one can understand that its associated signal rate $n_i$ is a product of the probabilities corresponding to the three parts~\cite{Dev:2021ofc}:
\begin{equation}
    n_i = P_{\rm prod} \times P_{\rm tran} \times P_{\rm det}\,, \label{eq:probprod}
\end{equation}
where $P_{\rm prod}$, $P_{\rm tran}$, and $P_{\rm det}$ are the probabilities associated with i), ii), and iii) above, respectively. Here we omitted the number of source particles in the right-hand side of Eq.~\eqref{eq:probprod}, since it is unity. So, $n_i$ can be interchangeably used with $P_i$, the signal detection probability with respect to the given source particle. 
If $N_{\rm src}$ such source particles are available for a given exposure, the total number of ALP signal events detected at the detector of interest (denoted by $N_{\rm sig}$) is given by
\begin{equation}
    N_{\rm sig}=\sum_i^{N_{\rm src}} n_i = N_{\rm src} \langle n_i \rangle\,,
\end{equation}
where the second equality implies that equivalently, one may evaluate the average of $n_i$, $\langle n_i\rangle$, for a sufficiently large sub-set out of $N_{\rm src}$ and multiply it by $N_{\rm src}$.
This approach essentially allows us to factorize the rate calculation into ALP physics and source physics. 

%
\subsubsection{Production of ALP \label{sec:prod}} 
Depending on the underlying ALP model details, ALPs can be produced in various ways inside the target of beam-dump type experiments or the reactor core of reactor neutrino experiments. Among possible source particles, the photon is particularly interesting as it is one of the most copiously produced particle species in the target or the reactor core.  
Photons are typically produced by decays of mesons including $\pi^0$, $\eta$, etc. and bremsstrahlung of the incoming beam particle and secondary charged particles inside the target in the beam-dump experiments, and produced by nuclear transitions and neutron captures inside the reactor core. While standard event generators, e.g., \texttt{PYTHIA}~\cite{Sjostrand:2006za,Sjostrand:2014zea}, \texttt{HERWIG}~\cite{Bahr:2008pv}, and \texttt{SHERPA}~\cite{Gleisberg:2008ta}, can describe the meson decays and beam-induced bremsstrahlung, photon production by secondary particles such as ionized electrons, nuclear transitions, and neutron captures require a dedicated detector simulation, e.g., \texttt{GEANT4}~\cite{Agostinelli:2002hh} and \texttt{FLUKA}~\cite{Ferrari:2005zk}.
One may estimate the photon fluxes induced by the first two mechanisms (semi-)analytically. For example, production of pions through energetic proton beams on target is empirically parametrized by the Sanford-Wang description~\cite{Sanford:1967zzb} and the BMPT model~\cite{Bonesini:2001iz}, and their decay photons can be used as an injection photon flux~\cite{Bonivento:2019sri}.
For another example, the equivalent photon approximation, also known as Fermi-Weizs\"{a}ker-Williams method~\cite{Fermi:1924tc,vonWeizsacker:1934nji,Williams:1934ad}, provides a convenient framework to estimate bremsstrahlung photons from an energetic charged particle. See e.g., \cite{Dobrich:2015jyk} for a more systematic discussion in the context of ALP searches in the beam-dump experiments. 

Once a photon emerges, it can be converted to an ALP via the Primakoff process, i.e., $\gamma+ A \to a+ A$, with $A$ denoting the atomic system of interest in the target or the core material, if the ALP-photon coupling $g_{a\gamma\gamma}$ in \eqref{eq:intlag} is non-zero. The differential cross-section in the angle of ALP with respect to the incoming photon direction $\theta_a$~\cite{Tsai:1986tx} is 
\begin{equation}
    \frac{d\sigma_{\rm prod}^{\rm Prim}}{d\cos\theta_a} = \frac{1}{4}g_{a\gamma\gamma}^2\alpha Z^2 F^2(t)\frac{|\vec{p}_a|^4\sin^2\theta_a}{t^2}\,, \label{eq:xsprimakoff}
\end{equation}
where $\alpha$, $Z$, and $\vec{p}_a$ are the electromagnetic fine structure constant, the atomic number of the target or reactor core material, and the ALP three-momentum, respectively. Here $t$ denotes the square of the four-momentum transfer:
\begin{equation}
    t= (p_\gamma -p_a)^2= m_a^2 -2 E_\gamma(E_a-|\vec{p}_a|\cos\theta_a)\,.
\end{equation}
Note that the typical momentum transfer to the target nucleus is much smaller than $E_\gamma$, so that $E_\gamma \approx E_a$ valid under the collinear limit can provide a good approximation and simplify calculational procedures.
Finally, $F(t)$ describes a form factor as a function of $t$. Depending on the coherency length scale, an atomic or nuclear form factor has to be taken into account. In the beam-dump experiments where the coherency length is usually at the nuclear scale, $F(t)$ will be a nuclear form factor such as the Helm parametrization~\cite{Helm:1956zz}. 

If the ALP-electron coupling $g_{aee}$ is non-vanishing, ALPs can be produced via $s$-channel and $t$-channel Compton-like scattering processes on electrons inside an target atom, i.e., $\gamma +e^- \to a+ e^-$. Its differential cross-section has the form of
\begin{equation}
    \frac{d\sigma_{\rm prod}^{\rm Comp}}{dx}=\frac{g_{aee}^2 \alpha Z x}{4(s-m_e^2)(1-x)} \left[ x- \frac{2m_a^2 s}{(s-m_e^2)^2}+\frac{2m_a^2}{(s-m_e^2)^2}\left(\frac{m_e^2}{1-x}+\frac{m_a^2}{x} \right) \right]\,, \label{eq:xscompton}
\end{equation}
where $s$ is a Mandelstam variable given by
\begin{equation}
    s = (p_\gamma+p_e)^2 = m_e^2+2m_e E_\gamma\,. 
\end{equation}
Here $x$ is the fractional light-cone momentum whose value lies in-between 0 and 1. In the laboratory frame, one may perform a change of variables, using the relation 
\begin{equation}
    x = 1-\frac{E_a}{E_\gamma }+\frac{m_a^2}{2m_e E_\gamma}\,.
\end{equation}

The fiducial (total) production cross-section $\sigma_{\rm prod}^{\rm fid}$ ($\sigma_{\rm prod}^{\rm tot}$) can be obtained by integrating \eqref{eq:xsprimakoff} and/or \eqref{eq:xscompton} over the phase space in which the produced ALPs are directed toward the detector of interest (over the entire phase space).  
However, the Primakoff and/or Compton-like processes do not always arise, but they basically compete with standard interactions such as pair production, photoelectric absorption, etc that the photons usually get through.  
Denoting the total cross-section of the standard interactions by $\sigma_{\rm SM}$, we therefore write the probability of ALP production $P_{\rm prod}$ as follows:
\begin{equation}
    P_{\rm prod}=\frac{\sigma_{\rm prod}^{\rm fid}}{\sigma_{\rm SM}+\sigma_{\rm prod}^{\rm tot}} \approx \frac{\sigma_{\rm prod}^{\rm fid}}{\sigma_{\rm SM}} \,.
\end{equation}
where the approximation is usually valid due to $\sigma_{\rm SM}\gg \sigma_{\rm prod}^{\rm tot}$ in most of the well-motivated ALP parameter space. 
Note that $\sigma_{\rm SM}$ is generally a function of photon energy $E_\gamma$ for which the measurement data is available in e.g., \cite{xcom}.

While we have focused on $P_{\rm prod}$ of photon-initiated Primakoff or Compton-like ALP production, this approach is straightforwardly applicable to other production mechanisms. For example, if a photon originates from neutron captures or nuclear deexcitations and an ALP is produced via its coupling to nucleons, $P_{\rm prod}$ is given by the branching ratio of ALP emissions in the transitions~\cite{Chang:2006ug}.
For another example, if ALP is mixed with pseudoscalar mesons such as $\pi^0$, $\eta^{(')}$, etc in the presence of the ALP-gluon coupling~\cite{Kelly:2020dda}, the source particle and the probability can be replaced by the mesons and the associated mixing angle squared. 

%
\subsubsection{Transportation of ALP \label{sec:tran}}
Once an ALP is produced in the target or the reactor core, it should neither decay before reaching the detector of interest nor interact with target material. 
The usual decay law defines the former probability $P_{\rm tran}^{\rm decay}$:
\begin{equation}
    P_{\rm tran}^{\rm decay} = \exp\left(- \frac{d}{\bar{\ell}_a^{\rm lab}} \right)\,,
\end{equation}
where $d$ is the distance between the source point and the detector. 
Here $\bar{\ell}_a^{\rm lab}$ stands for the laboratory-frame mean decay length of ALP which is a function of the total decay width of ALP, $\Gamma_a^{\rm tot}$, and the boost factor of ALP, $\gamma_a$,
\begin{equation}
    \bar{\ell}_a^{\rm lab}=\frac{\sqrt{\gamma_a^2-1}}{\Gamma_a^{\rm tot}}\,.
\end{equation}
The partial particle widths of the diphoton and the $e^+e^-$-pair decay modes are, respectively,  
\begin{eqnarray}
    \Gamma_{a\to2\gamma} &=& \frac{g_{a\gamma\gamma}^2 m_a^3}{64\pi}, \\
    \Gamma_{a\to e^+e^-} &=& \frac{g_{aee}^2m_a}{8\pi}\sqrt{1-\frac{4m_e^2}{m_a^2}}\,.
\end{eqnarray}

The latter probability that no ALP interactions arise in the target (say, $P_{\rm tran}^{\rm int}$) can be evaluated in a similar fashion. Assuming that the ALP interactions are dominated by the ALP scattering and the distance between the source point and the target end is denoted by $D$, we can express $P_{\rm tran}^{\rm int}$ as
\begin{equation}
    P_{\rm tran}^{\rm int}=\exp(-\rho_T \sigma_{\rm scat}^{\rm tot} D)\,,
\end{equation}
where $\rho_T$ is the number density of target particles in the beam target with respect to $\sigma_{\rm scat}^{\rm tot}$. 
Here $\sigma_{\rm scat}^{\rm tot}$ describes the total scattering cross-section of ALP for which the related formulation is discussed in the next subsection.
Therefore, the transportation probability is 
\begin{equation}
    P_{\rm tran} = P_{\rm tran}^{\rm decay}\times P_{\rm tran}^{\rm int}.
\end{equation}
However, since $D$ is much smaller than the mean free path of ALPs defined by $1/(\rho_T \sigma_{\rm scat}^{\rm tot})$ in most of the experiments, $P_{\rm tran}^{\rm int}$ becomes approximately unity, and in turn, $P_{\rm tran}\approx P_{\rm tran}^{\rm decay}$. 

%
\subsubsection{Detection of ALP \label{sec:det}}
There are broadly three channels to detect ALPs at a detector: through their decay, through their scattering with detector material, and through their conversion to photons. 
Most of the beam-dump type experiments have performed  ALP searches in the first channel~\cite{Bjorken:1988as,Riordan:1987aw,Bergsma:1985qz,Blumlein:1990ay,Banerjee:2020fue,Dobrich:2015jyk,Alekhin:2015byh,Berlin:2018pwi,Berlin:2018bsc,Akesson:2018vlm,Dobrich:2019dxc,Brdar:2020dpr,Kelly:2020dda}. 
Since the ALP should decay to visible particles before escaping from the detector fiducial volume, in order to leave an experimental signature, the probability of ALP detection in this decay channel, $P_{\rm det}^{\rm decay}$, is again described by the decay law and has the form of
\begin{equation}
    P_{\rm det}^{\rm decay} = 1-\exp\left(-\frac{L_{\rm det}}{\bar{\ell}_a^{\rm lab}} \right)\,,
\end{equation}
where $L_{\rm det}$ is the cord length along which the ALP would sweep in the detector fiducial volume. Since the decay within the detector is crucial, this channel is suited for ALPs with a relatively sizable decay width. 
If $m_a$ is too small, the associated decay length of ALP is too large to have decay signals. 

\begin{figure}[t]
    \centering
    \includegraphics[width=7.9cm]{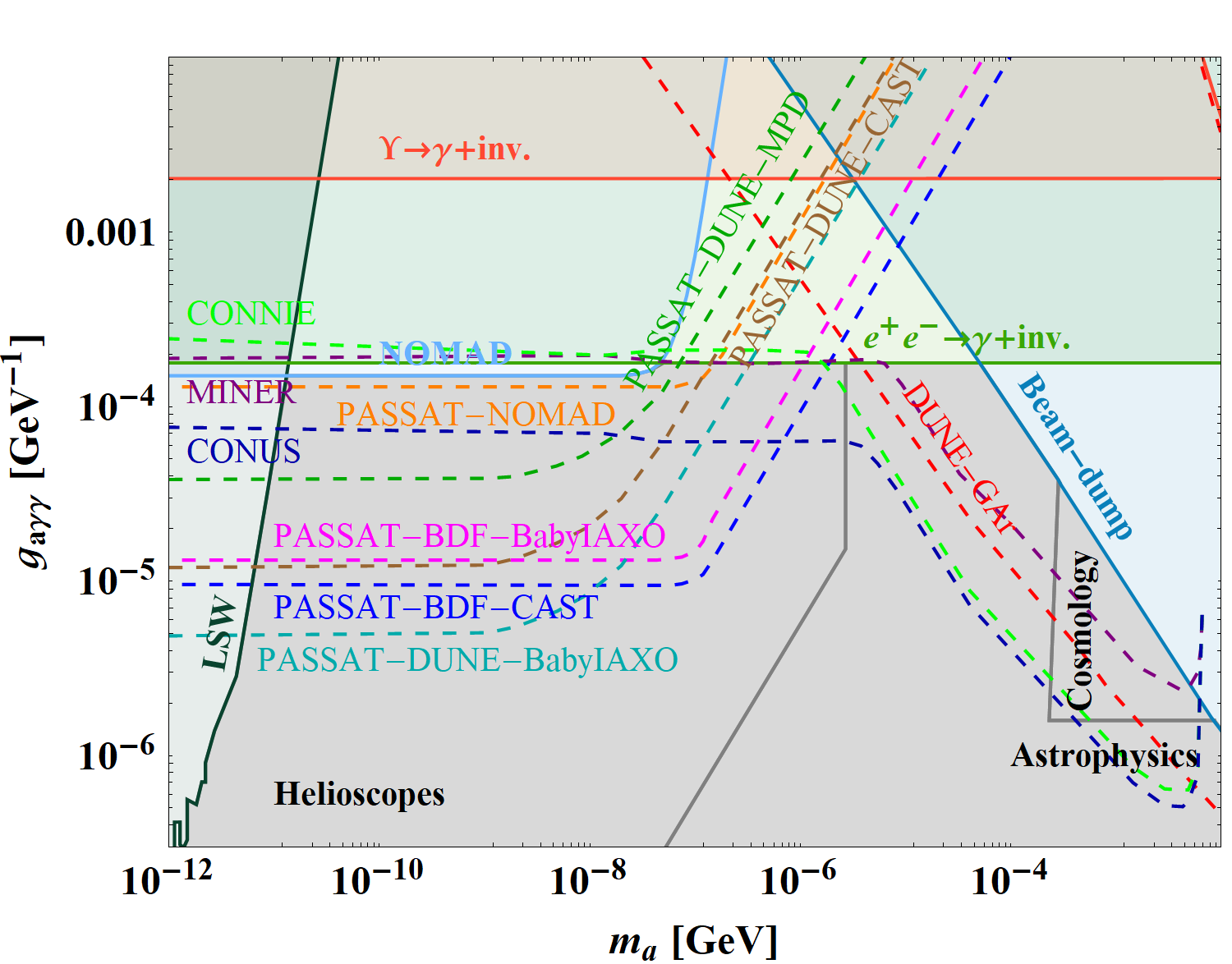}
    \includegraphics[width=7.9cm]{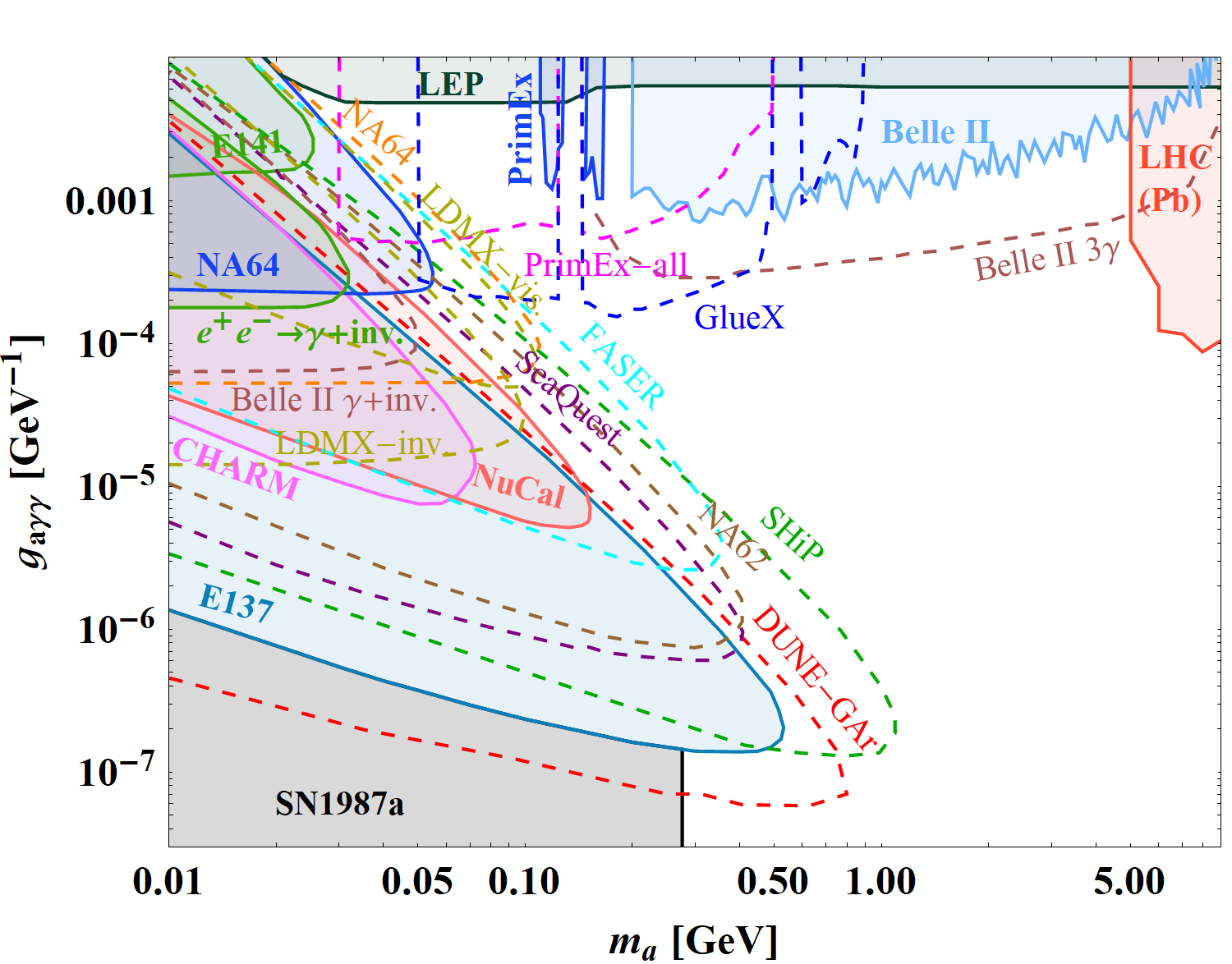}
    \caption{Existing (color-shaded regions) and expected (dashed lines) limits of laboratory-produced ALPs in the $(m_a, g_{a\gamma\gamma})$ parameter plane. 
    For reference purposes, we show astrophysical and cosmological limits (gray-shaded regions) compiled in \cite{Bauer:2018uxu}.  
    Left panel: Limits for $m_a <10$~MeV to which the scattering and the conversion channels in beam-dump and reactor neutrino experiments are relevant. The current constraints include $e^+ +e^-\to\gamma+{\rm inv.}$~\cite{Abbiendi:2000hh,Heister:2002ut,Achard:2003tx,Abdallah:2003np,Hewett:2012ns}, $\Upsilon \to \gamma+{\rm inv.}$~\cite{Balest:1994ch,delAmoSanchez:2010ac}, LSW-type experiments~\cite{Redondo:2010dp}, NOMAD~\cite{Astier:2000gx}, and beam-dump experiments~\cite{Bjorken:1988as}. Future expected limits include a PASSAT interpretation~\cite{Bonivento:2019sri} of NOMAD~\cite{Astier:2000gx}, PASSAT implementations at the BDF facility with the CAST or BabyIAXO magnets with $2\times 10^{20}$ POTs~\cite{Bonivento:2019sri}, a PASSAT implementation at the DUNE MPD with a 7-year exposure~\cite{Dev:2021ofc}, PASSAT implementations at DUNE with the CAST or BabyIAXO magnets with a 7-year exposure~\cite{Dev:2021ofc}, and reactor searches at CONNIE, CONUS, and MINER~\cite{Dent:2019ueq}.
    Right panel: Limits for $m_a>10$~MeV to which the decay channels in beam-dump and reactor neutrino experiments are relevant. The current constraints include $e^++e^-\to\gamma+{\rm inv.}$~\cite{Abbiendi:2000hh,Heister:2002ut,Achard:2003tx,Abdallah:2003np,Hewett:2012ns}, Belle-II~\cite{BelleII:2020fag}, CHARM~\cite{Bergsma:1985qz}, E137~\cite{Bjorken:1988as}, E141~\cite{Riordan:1987aw}, LEP~\cite{Abbiendi:2002je}, LHC (Pb)~\cite{Sirunyan:2018fhl,Aad:2020cje}, NA64~\cite{Banerjee:2020fue}, NuCal~\cite{Blumlein:1990ay}, and PrimEx~\cite{Aloni:2019ruo}. Future prospects shown here include Belle II in the $3\gamma$ and the $\gamma+{\rm inv.}$ modes with 20~fb$^{-1}$ and $g_{a\gamma Z}=0$~\cite{Dolan:2017osp}, DUNE-GAr with one year exposure~\cite{Brdar:2020dpr}, FASER~\cite{Feng:2018pew}, GlueX with 1~pb$^{-1}$~\cite{Aloni:2019ruo}, LDMX-vis. and LDMX-inv. with $10^{16}$ EOTs of an 8 GeV beam~\cite{Berlin:2018bsc}, LHC (Pb) with 10~nb$^{-1}$~\cite{Knapen:2016moh}, NA62 with $10^{18}$ POTs~\cite{Dobrich:2019dxc}, NA64 with $5\times 10^{12}$ EOTs~\cite{Dusaev:2020gxi}, PrimEx with all runs~\cite{Aloni:2019ruo}, SeaQuest in Phase I~\cite{Berlin:2018pwi}, and SHiP with $2\times 10^{20}$ POTs~\cite{Alekhin:2015byh}.}
    \label{fig:dumplimits}
\end{figure}

In Fig.~\ref{fig:dumplimits}, we display existing (color-shaded regions) and future expected (dashed lines) limits of $g_{a\gamma\gamma}$ in $m_a$ for laboratory-produced ALPs. The laboratory-based experiments contributing to the existing limits are listed in the figure caption, and we include astrophysical and cosmological limits (gray-shaded regions) compiled in \cite{Bauer:2018uxu} for reference purposes. 
Future prospects of beam-dump experiments in the decay channel are shown in the right panel of Fig.~\ref{fig:dumplimits}: DUNE-GAr (one year exposure)~\cite{Brdar:2020dpr}, LDMX-vis. and LDMX-inv. ($10^{16}$ EOTs of an 8 GeV beam)~\cite{Berlin:2018bsc}, NA62 ($10^{18}$ POTs)~\cite{Dobrich:2019dxc}, NA64 ($5\times 10^{12}$ EOTs)~\cite{Dusaev:2020gxi}, SeaQuest (Phase I)~\cite{Berlin:2018pwi}, and SHiP ($2\times 10^{20}$ POTs)~\cite{Alekhin:2015byh}.
We further include expectations of photon-beam experiments, e.g., GlueX (Pb target with 1~pb$^{-1}$)~\cite{Aloni:2019ruo} and PrimEx (all runs)~\cite{Aloni:2019ruo}, and a forward-physics experiment, FASER~\cite{Feng:2018pew}. 
The expected limits of the reactor neutrino experiments in the decay channel are shown in the left panel of Fig.~\ref{fig:dumplimits}, including CONNIE, CONUS, and MINER~\cite{Dent:2019ueq}. Since the energy of photons in the reactor core is limited, the reactor experiments are not sensitive to the region of $m_a \gtrsim 10$~MeV. However, the close proximity to their detector enables them to explore the regions of smaller $m_a$ and $g_{a\gamma\gamma}$ than what typical beam-dump experiment would reach. 

\begin{figure}[t]
    \centering
    \includegraphics[width=8cm]{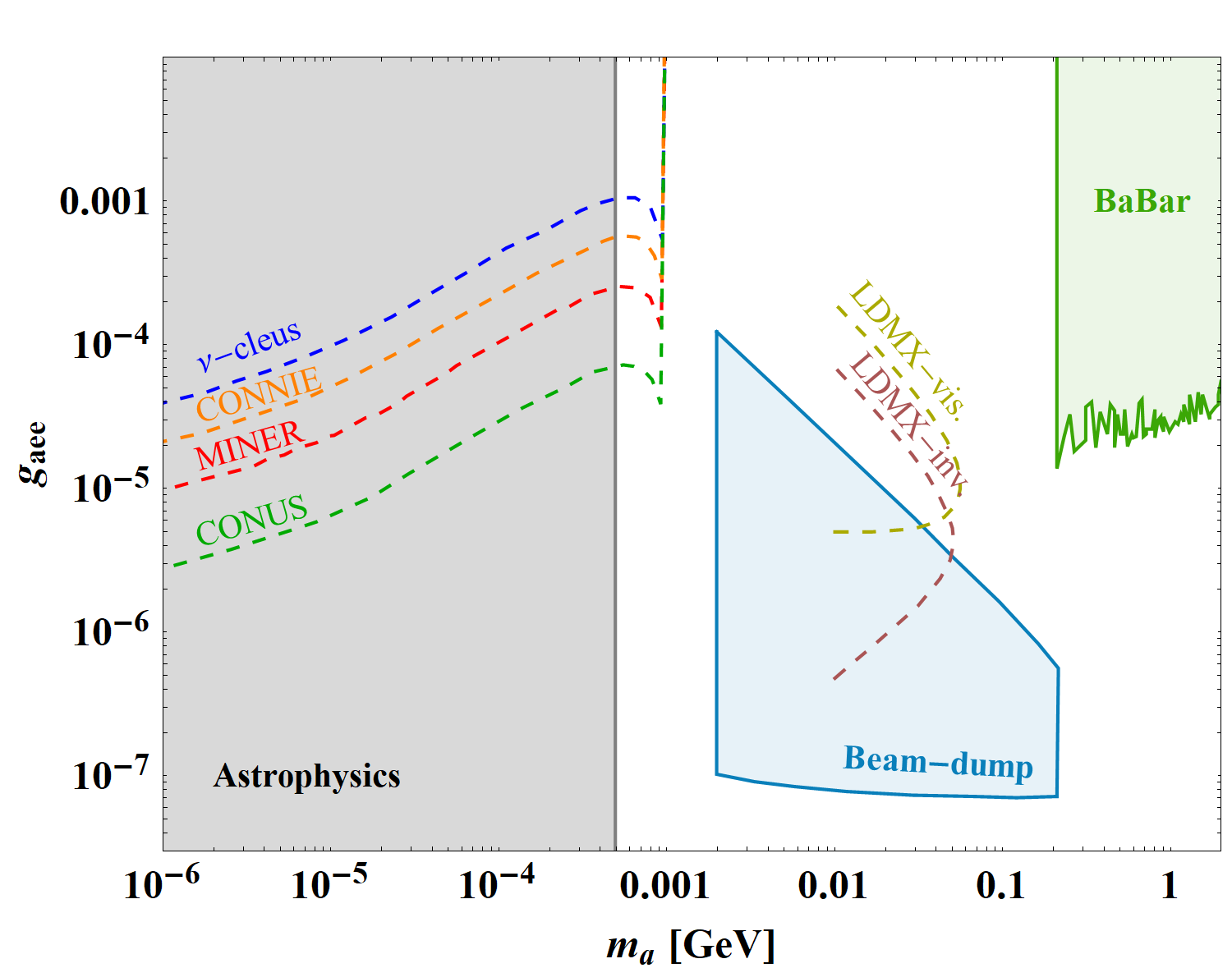}
    \caption{Existing (color-shaded regions) and expected (dashed lines) limits of laboratory-produced ALPs in the $(m_a, g_{aee})$ parameter plane. 
    The current constraints from the laboratory-produced ALP searches include beam-dump experiments performed at SLAC~\cite{Essig:2010gu} and limits derived from the dark photon search of BaBar~\cite{TheBABAR:2016rlg,Bauer:2017ris} under the assumption of the approximate universality of the ALP-lepton couplings. 
    For reference purposes, we show the astrophysical limits (gray-shaded regions) set by EDELWEISS-III.~\cite{Armengaud:2018cuy}.
    Future prospects shown here include LDMX-vis. and LDMX-inv. with $10^{16}$ EOTs of an 8 GeV beam~\cite{Berlin:2018bsc} and reactor searches at CONNIE, CONUS, MINE, and $\nu$-cleus~\cite{Dent:2019ueq}. }
    \label{fig:gaeelimits}
\end{figure}

Similarly, we exhibit existing (color-shaded regions) and future expected (dashed lines) limits on $(m_a, g_{aee})$ in Fig.~\ref{fig:gaeelimits}, for ALPs purely coupled to electrons. Future prospects include LDMX-vis. and LDMX-inv. ($10^{16}$ EOTs of an 8 GeV beam)~\cite{Berlin:2018bsc}, but the reactor searches are sensitive to the ALP signals of the decay channel only within a narrow mass range since most of the energy values of the photons created in their reactor core are not big enough to overcome the production threshold $s\geq (m_a+m_e)^2$~\cite{Dent:2019ueq}. 

Light ALPs which do not decay inside a detector may leave an experimental signature via its scattering off detector material. This possibility has been recently pointed out in \cite{Dent:2019ueq}, and ALP detection prospects in the scattering channel have been investigated in reactor neutrino experiments~\cite{Dent:2019ueq,AristizabalSierra:2020rom} and neutrino beam experiments~\cite{Brdar:2020dpr}. For example, in the presence of the ALP-photon interaction, an ALP can induce a photon via an inverse Primakoff process, i.e., $a +A \to \gamma+ A$. Depending on the detector capability, one may observe a photon and potentially a nucleus recoil. The differential scattering cross-section in the outgoing photon angle $\theta_\gamma$ is almost the same as in \eqref{eq:xsprimakoff}: 
\begin{equation}
    \frac{d\sigma_{\rm det}^{\rm Inv~Prim}}{d\cos\theta_\gamma} =\frac{1}{2}g_{a\gamma\gamma}^2\alpha Z^2 F^2(t) \frac{|\vec{p}_a|^4\sin^2\theta_\gamma}{t^2}\,, \label{eq:xsdetInvPrim}
\end{equation}
where $Z$ and $F$ are associated with the detector material and where $\vec{p}_a$ is the three-momentum of the incoming ALP. The alteration that the prefactor 1/4 becomes 1/2 is because the initial spin states include a spin-0 ALP rather than a spin-1 photon. 
For another example, the ALP could scatter off electrons through the (inverse) Compton-like process, i.e., $a +e^-\to \gamma +e^-$, in the presence of a non-zero $g_{aee}$, and the resulting final state involves a recoiling electron and a photon.
The differential cross-section for this scattering process in the solid angle of the outgoing photon~\cite{Avignone:1988bv,Bellini:2008zza} is given by
\begin{equation}
    \frac{d\sigma_{\rm det}^{\rm Inv~Comp}}{d\Omega_\gamma}=\frac{g_{aee}^2 \alpha Z E_\gamma}{8\pi m_e^2 |\vec{p}_a|}\left[1+\frac{4m_e^2E_\gamma^2}{(2m_eE_a+m_a^2)^2}-\frac{4m_eE_\gamma}{2m_eE_a+m_a^2}-\frac{4m_a^2|\vec{p}_a|^2 m_e E_\gamma}{(2m_eE_a+m_a^2)^3}\sin^2\theta_\gamma \right], \label{eq:xsdetInvComp}
\end{equation}
where again $Z$ is the atomic number of the detector material. 

As for the production cross-section, the fiducial scattering cross-section $\sigma_{\det}^{\rm fid}$ can be obtained by integrating \eqref{eq:xsdetInvPrim} and/or \eqref{eq:xsdetInvComp} over the phase space consistent with detector thresholds, cuts, etc. 
In well-motivated regions of ALP parameter space, the total scattering cross-section is sufficiently small that the scattering probability for a given ALP $P_{\rm det}^{\rm scat}$ is
\begin{equation}
    P_{\rm det}^{\rm scat} = n_T \ \sigma_{\rm det}^{\rm fid}\ L_{\rm det}\,,
\end{equation}
where $n_T$ is the number density of target particles in the detector under consideration and where $L_{\rm det}$ is the length through which the ALP would travel in the detector fiducial volume. 

The expected sensitivity reaches of reactor neutrino experiments in this channel are shown in the left panel of Fig.~\ref{fig:dumplimits}: the horizontal pieces of the sensitivity curves of CONNIE, CONUS, and MINER experiments~\cite{Dent:2019ueq}.
The scattering channel allows for the exploration of the ALP parameter space to which the decay channel would be insensitive, providing complementary information in the ALP search. 
Similar analyses also appear in \cite{AristizabalSierra:2020rom} by taking sets of benchmark ALP production and detection parameters, not specifying concrete experiments. When it comes to the beam-dump type neutrino experiments, the authors of \cite{Brdar:2020dpr} provide a preliminary discussion on the experimental reach of ALP parameter space in the scattering channel, taking the DUNE LArTPC near detector and assuming the ALP-photon coupling. 

Similarly, the expected sensitivity reaches of reactor neutrino experiments in the $(m_a,g_{aee})$ plane are shown in Fig.~\ref{fig:gaeelimits}: the horizontal through rising pieces of the sensitivity curves of CONNIE. CONUS, MINER, and $\nu$-cleus~\cite{Dent:2019ueq}. As in the $g_{a\gamma\gamma}$ case, the scattering channel provides a complementary probe. 
    
Finally, if a detector carries a magnetic field region, it can be sensitive to models of ALP interacting with photons through the ALP-to-photon conversion. 
For an ALP traveling a distance $L_B$ in a magnetic field $B$, the conversion probability $P_{\rm det}^{\rm conv}$~\cite{Irastorza:2018dyq} is
\begin{equation}
    P_{\rm det}^{\rm conv} = \left(\frac{g_{a\gamma\gamma}BL_B}{2}\right)^2\left(\frac{2}{qL_B}\right)^2\sin^2\left(\frac{qL_B}{2}\right)^2\,,
\end{equation}
where the product of the second and third factors parametrizes the form factor describing the decoherency of the conversion. In vacuum and in the relativistic limit, $q$ is given by
\begin{equation}
    q=2\sqrt{\left(\frac{m_a^2}{4E_a}\right)^2+\left(\frac{g_{a\gamma\gamma}B}{2} \right)^2}.
\end{equation}
In the limit of $m_a,g_{a\gamma\gamma}\to 0$, this $q$ becomes small and the conversion probability is simplified to 
\begin{equation}
    P_{\rm det}^{\rm conv} \approx \left(\frac{g_{a\gamma\gamma}BL_B}{2}\right)^2. \label{eq:convsimp}
\end{equation}

This hybrid type experiment of beam-dump and helioscope approaches was first proposed in \cite{Bonivento:2019sri}, dubbed Particle Accelerator helioScopes for Slim Axion-like-particle deTection or PASSAT as shorthand; the ALP decay is replaced by the ALP conversion from the perspective of conventional beam-dump ALP searches, while the sun is substituted by a target from the perspective of a traditional helioscope.  
As is implied by the decoherency form factor, the ALP-to-photon conversion can be maximally effective for light ALPs. Therefore, this channel enables the exploration of ALP parameter space to which the ALP searches in the decay channel are not sensitive, and complements the scattering channel~\cite{Bonivento:2019sri}.

Expected sensitivity reaches under the PASSAT framework are shown in the left panel of Fig.~\ref{fig:dumplimits}. The authors in \cite{Bonivento:2019sri} interpreted the ALP search result in NOMAD~\cite{Astier:2000gx} by a LSW-like regeneration method (ALP production in the magnetic field of the beam focusing horn and ALP detection in the magnetic field of the detector) in terms of PASSAT (see the line for PASSAT-NOMAD). They further proposed a realization of PASSAT by recycling the magnets of the CAST or the proposed BabyIAXO experiments (after decommissioned) and placing them at the proposed beam-dump facility~\cite{Kershaw:2018pyb} at CERN (see the lines for PASSAT-BDF-CAST and PASSAT-BDF-BabyIAXO). Similar proposals were made for DUNE or DUNE-like neutrino facilities~\cite{Dev:2021ofc}. Since the multi-purpose detector (MPD) of DUNE~\cite{Abi:2020wmh} carries a magnetic field region, the idea of PASSAT is readily applicable (see the line for PASSAT-DUNE-MPD). Alternatively, one could reuse the magnets of CAST or BabyIAXO and place them near the dump area, reducing the distance between the target (i.e., the source point) and the magnetic field region to improve the signal sensitivity (see the lines for PASSAT-DUNE-CAST and PASSAT-DUNE-BabyIAXO). 
As mentioned above, this conversion channel is better suited for the regimes of small $m_a$ where the decoherency form factor gets negligible hence the conversion probability becomes maximized, and 
Eq.~\eqref{eq:convsimp} suggests that the conversion probability is essentially governed by $B$ and $L_B$. Therefore, large-scale stronger magnets enable the investigation of ALP parameter space that the decay and scattering channels would never access. 

\subsection{Collider searches}

Collider probes have formed an important branch of the ALP search efforts as they are sensitive to ALP signals directly and indirectly~\cite{Kleban:2005rj}, and have mostly constrained models of ALP interacting with the SM photon. Due to their relatively large center-of-mass energy, colliders have provided particular opportunities in the search for MeV-to-TeV mass-range ALPs.

For example, the authors of \cite{Jaeckel:2015jla} interpreted  LEP measurement data available in \cite{Acciarri:1994gb,Akrawy:1990zz,Abreu:1991rm,Abreu:1994du,Acciarri:1995gy} using a process of ALP interacting with photons, $e^+ +e^- \to \gamma^* \to \gamma +a$, $a\to \gamma+\gamma$, and filled the missing $m_a$ gap between $\sim{\rm MeV}$ and $\sim10~{\rm GeV}$. In this mass regime, the produced ALP would be so significantly boosted that its decay products, two photons would be merged hence appear single-photon-like. Therefore, the null signal observation in the diphoton channel can set the limits in the $(m_a, g_{a\gamma\gamma})$ parameter space.

 LEP data in the mono-photon channel (i.e., $e^+ +e^- \to \gamma+{\rm inv.}$)~\cite{Abbiendi:2000hh,Heister:2002ut,Achard:2003tx,Abdallah:2003np,Hewett:2012ns} and radiative decays of Upsilon mesons (i.e., $\Upsilon(1S) \to \gamma+{\rm inv.}$)~\cite{Balest:1994ch,delAmoSanchez:2010ac} have set the limits up to $m_a\sim 50$~MeV, as no excesses were observed. Proton colliders have also provided the bounds: limits were derived from CDF data in the triphoton channel~\cite{cdf,Mimasu:2014nea} and $pp$ collision data of ATLAS and CMS in channels involving $\geq2$ photons in the final state~\cite{Chatrchyan:2012tv,Aad:2014ioa,Aad:2015bua,Aaboud:2016tru}, covering up to the $\sim {\rm TeV}$ scale. Beyond $pp$ collisions, it was pointed out that light-by-light (LBL) scattering in heavy ion collisions enables the investigation of unexplored regions of parameter space below $m_a\sim 100$~GeV~\cite{Knapen:2016moh}, and ATLAS and CMS Collaborations have reported their search results~\cite{Sirunyan:2018fhl,Aad:2020cje}. Recently, Belle-II searched for an ALP process, $e^+ +e^-\to \gamma + a$, $a\to \gamma+\gamma$ using data corresponding to 445~pb$^{-1}$, and set new limits as no evidence was found~\cite{BelleII:2020fag}.

Future sensitivity reaches of existing collider experiments in the $(m_a,g_{a\gamma\gamma})$ plane have been investigated in various parts of the literature. Example studies include Belle II in the $3\gamma$ and the $\gamma+{\rm inv.}$ modes with 20~fb$^{-1}$ and 50~ab$^{-1}$~\cite{Dolan:2017osp} and LHC Pb-Pb collisions with 10~nb$^{-1}$~\cite{Knapen:2016moh}, as also shown in the right panel of Fig.~\ref{fig:dumplimits}. 
For the cases where the produced ALPs are long-lived, the two photons from the ALP decay would be appreciably displaced so that existing searches may not be sensitive enough to the associated signature and a more dedicated search would be needed~\cite{Mimasu:2014nea}.\footnote{Recent studies on the displaced vertex signature of ALPs interacting with gluons appear in \cite{Hook:2019qoh,Gershtein:2020mwi}, using the dedicated displaced vertex trigger at the HL-LHC.}  If ALPs decay even outside the detector because they are very light and/or very weakly coupled, their decay signature could be observed by a future surface-based detector specifically designed for the purpose of searching for very long-lived particles, e.g., MATHUSLA~\cite{Chou:2016lxi}. 
Future lepton colliders, e.g., FCC-$ee$~\cite{Abada:2019zxq}, ILC~\cite{Fujii:2017vwa}, CEPC~\cite{Tang:2015qga}, and CLIC~\cite{CLIC:2016zwp,Charles:2018vfv}, hadron colliders, e.g., HL-LHC~\cite{Apollinari:2284929}, HE-LHC~\cite{Zimmermann:2651305}, FCC-$hh$~\cite{Zimmermann:2651305}, and SPPC~\cite{Tang:2015qga}, and electron-hadron colliders, e.g., LHeC~\cite{Agostini:2020fmq} and FCC-$he$~\cite{Agostini:2020fmq}, can provide unprecedented sensitivities to ALP signals especially in the high mass regime, and we provide a summary of key parameters of future (proposed) energy-frontier colliders in Table~\ref{tab:colliders} for reference purposes. 
Related investigations of ALP phenomenology have been performed, including \cite{Mimasu:2014nea,Jaeckel:2015jla,Brustein:2018mpn,Bauer:2018uxu,Buttazzo:2018qqp,Biswas:2019lcp,Coelho:2020saz,Inan:2020aal,Inan:2020kif,Steinberg:2021iay}.

\begin{table}[t]
    \centering
    \begin{tabular}{c|c c c}
    \hline \hline
      Collider   & ~Particles collided~ & Center-of-mass energy [TeV]& Integrated $\mathcal{L}$ [ab$^{-1}$]  \\
      \hline
      HL-LHC~\cite{Apollinari:2284929} & $pp$ & 14 & 3  \\
      HE-LHC~\cite{Zimmermann:2651305} & $pp$ & 27 & 10   \\
      SPPC~\cite{Tang:2015qga,Zou:2016mqo} & $pp$ & 75, 100 & 3\\
      FCC-$hh$~\cite{Zimmermann:2651305} & $pp$ & 100 & 30 \\
      \hline 
      FCC-$ee$~\cite{Abada:2019zxq} & $e^+e^-$ & 0.091, 0.161, 0.24, 0.34 -- 0.365 & 145, 120, 50, 15  \\
      ILC~\cite{Fujii:2017vwa} & $e^+e^-$ & 0.25 & 2 \\
      CEPC~\cite{Tang:2015qga} & $e^+e^-$ & 0.25 & 5 \\
      CLIC~\cite{CLIC:2016zwp,Charles:2018vfv} & $e^+e^-$ & 0.35 (and 0.35), 1.5, 3.0 & 1, 2.5, 3  \\
      \hline
      LHeC~\cite{Agostini:2020fmq} & $e^-p$ &  1.3 ($E_e=0.05$, $E_p=6.5$) &  1\\
      FCC-$he$~\cite{AbelleiraFernandez:2012cc,Agostini:2020fmq} & $e^-p$ &  4.8 ($E_e=0.06$, $E_p=20$), 12 ($E_e=0.06$, $E_p=50$) & 3, 3  \\
      \hline \hline
    \end{tabular}
    \caption{Key parameters of future (proposed) energy-frontier colliders. }
    \label{tab:colliders}
\end{table}


For models of ALPs interacting with photons, a limited number of search channels are available at colliders.
By contrast, for ALPs with couplings to other SM particles like gluons, massive gauge bosons, Higgs, and leptons, a number of search channels become available and richer phenomenology is expected.  
For example, for ALPs coupling to the hypercharge boson, the ALP-$Z$-photon coupling is non-zero so that ALPs can be produced by the decay of on-shell $Z$ gauge bosons.  Studies of ALPs using on-shell decays $h \rightarrow a+a$,  $h \rightarrow Z+a$ and $Z \rightarrow \gamma+ a$ have been conducted by \cite{Jaeckel:2015jla, Brivio:2017ije, Bauer:2018uxu, Alves:2016koo}. As an example of these studies, we briefly discuss the work of  \cite{ Bauer:2018uxu}.  Figure 14 of \cite{Bauer:2018uxu} contains the projected discovery contours on the plane of $(m_a, g_{a\gamma\gamma})$ in the channel $Z \rightarrow a+\gamma$ with Br($a \rightarrow \gamma+\gamma$) $=1$, and assuming 3,000 fb$^{-1}$ of $pp$ collision data at the LHC. The authors calculate the relevant cross-sections and determine the signal significance by requiring a minimum signal yield of 100 events, and find that  $g_{a\gamma\gamma} \sim 5 \times 10^{-6}$ GeV$^{-1}$ can be probed for $m_a \sim 1-100$ GeV.  Far more conservative results are obtained once one incorporates $3\gamma$ and $4\gamma$ backgrounds at the LHC, which are certainly not negligible, with important contributions arising due to genuine  $3\gamma$ production, ``fake $\gamma$'' backgrounds due to neutral pion decays and large bremsstrahlung from electrons \cite{Florez:2021zoo}.  The situation is even worse if one considers realistic experimental uncertainty. Conservatively, the best systematic uncertainties on genuine tri-gamma backgrounds at CMS and ATLAS are probably about 15\%.
Realistic background modeling and uncertainty estimation were accounted for by the authors of \cite{Florez:2021zoo}, who presented a feasibility study for the detection of ALPs produced through VBF processes and decaying via $a \rightarrow \gamma + \gamma$. For 3,000 fb$^{-1}$ of data, the discovery reach was found to be  $g_{a\gamma\gamma} \sim 5 \times 10^{-4}$ GeV$^{-1}$, for $m_a$ from 10 MeV to 100 GeV.

There are several other studies along these lines: mono-gauge-boson (including mono-photon) signatures were proposed in the search for ALP signals by~\cite{Mimasu:2014nea,Brivio:2017ije};  ALPs interacting with electroweak gauge bosons have been studied by for example~\cite{Alonso-Alvarez:2018irt}; it has been suggested that non-resonant ALP-mediated diboson production could be a promising channel due to the derivative nature of ALP couplings~\cite{Gavela:2019cmq}; and it has been  claimed that a triboson search would be useful in the search for ALPs giving rise to boosted four photons which appear diphoton-like~\cite{Sokolenko:2018btg}.

Since the LHC features a large luminosity of gluons, it can provide particular opportunities for  ALPs coupled to gluons~\cite{Jaeckel:2012yz}. For example, stronger limits between $m_a \sim 10$~GeV and $ \sim 65$~GeV can be derived in the diphoton channel together with the ALP-photon coupling~\cite{Mariotti:2017vtv}; new search channels are available such as ALP production in association with a gluon-induced jet~\cite{Haghighat:2020nuh} and ALP production in association with a jet and a photon~\cite{Ebadi:2019gij}; and gluon-fusion-induced ALP production in the $pp$, Pb-Pb collisions in the next LHC runs~\cite{Goncalves:2020bqi}.
ALP production via light-by-light (LBL) scattering has also been  investigated: LBL in $pp$ collisions with proton tagging~\cite{Baldenegro:2018hng}; LBL in heavy-ion collisions~\cite{Bruce:2018yzs,Baldenegro:2019whq}; and \texttt{SuperChic}, a simulation package for LBL in $pp$, $p$-heavy ion, heavy ion-heavy ion collisions~\cite{Harland-Lang:2018iur}.
Exotic decays of $Z$ or Higgs also can be good channels to look for ALP signals~\cite{Aad:2015bua,Bauer:2017nlg,Bauer:2017ris}.   
Finally, the scenario that ALP couples to sterile neutrinos and its collider signature were investigated~\cite{Alves:2019xpc}.  
\section{Bose-Einstein Condensates }
\label{sec:BEC-sec}
It is known that ALP could form Bose-Einstein Condensate (BEC) if they constitute a fraction of dark matter (DM). In this section, we discuss the BEC properties and their implications on both the galactic scale and stellar scales. In what follows, we assume the scalar to be {as light as sub-eV}, 
while the reasons for it will be elaborated shortly.

\subsection{Theoretical estimate}
\label{sec:theoretical-estimate}

Due to its bosonic nature, ultra-light bosonic dark matter can exhibit collective behaviors at the macroscopic level that are not obvious at the Lagrangian level. 
It has been observed and well understood in condensed matter physics that for bosons there exists a phase, BEC phase, once the ensemble is cooled to below the critical temperature. In the case of ultra-light dark matter, one can estimate the mass range one requires for it to be in the BEC phase in a fashion similar to the estimate of the critical temperature of BEC \cite{Guth:2014hsa,Berezhiani:2015bqa,Fan:2016rda,Ferreira:2018wup}.
We start with a potential of the following type, partly motivated by the axion cosine potential
\begin{align}
  \mathcal{L}
  & =
    \frac{1}{2} \partial_\mu \phi  \partial^\mu \phi -
    \left ( \frac{1}{2} m^2 \phi^2 + \frac{\lambda}{4!} \frac{m^2}{f^2} \phi^4 + ... \right  ).
\end{align}
By requiring the de Broglie wavelength to be longer than the inter-spacing between dark matter particles, one has
$   \frac{2 \pi}{m v}
   \gtrsim
    \left (\frac{m}{\rho
    } \right )^{1/3}, $
which gives 
\begin{align}
  m &\lesssim
      1\; \mathrm{eV} \left (\frac{10^{-3}}{v} \right )^{3/4}
      \left ( \frac{\rho}{\rho_{DM}} \right )^{1/4},
\end{align}
where the average of the dark matter density is $\rho_{DM} = \rho_c \Omega_{DM} \approx 1.3 \times 10^{-6} \; \mathrm{GeV/cm^3}$, with $\rho_c$ being the critical density of the universe. Assuming negligible self-interaction,  numerically solving the equation of motion in the non-relativistic regime leads to a BEC mass \cite{Bar:2018acw}
\begin{align}
  M_{BEC}
  & \approx
    2.79 \times 10^{12}  \; \mathrm{M_{\odot}}
    \; \left ( \frac{m}{10^{-22} \; \mathrm{eV}} \right )^{-1}
    \; \left ( \frac{\phi(0)}{M_{Pl,r}} \right ),    
\end{align}
where $M_{Pl,r}$ is the reduced Planck mass, $\phi(0)$ the value of the wave function at $r=0$. One can see that $m\sim 10^{-22} \; \mathrm{eV} \; (10^{-10} \; \mathrm{eV})$ corresponds to galactic (stellar) scale BEC structures. The mass profile is parametrized by two variables, the scalar mass $m$ and the central density related to $\phi(0)$. While there is no simple analytical expression for the mass profile, a few approximations exist that are in good agreement with the numerical solutions, such as \cite{Schive:2014dra,Eby:2015hyx,Eby:2017teq,Kling:2017hjm,Kling:2017mif}. 

\subsubsection{Cosmological evolution}
\label{sec:cosm-evol}

It is pointed out that ultra-light dark matter relieves the core-cusp problem \cite{Hu:2000ke,Hui:2016ltb}. This can be easily seen through the equations of motion of the inhomogeneous perturbation of the field.
In Newtonian gauge, we denote the metric perturbation following the notation of \cite{Ma:1995ey}
\begin{align}
  ds^2 = -a^2(\tau) ( 1+ 2\Psi) d\tau^2 + a^2(\tau) ( 1 - 2 \Phi) d x^2,
\end{align}
where $a$ is the scale factor. 
 Writing $\phi(t, x) = \phi(t) + \delta \phi(t, x)$, minimizing the action order by order gives the equations of motion for both the homogeneous background and the fluctuation. In the momentum space they are \cite{Zhang:2017flu,Arvanitaki:2019rax}
\begin{align}
  \phi'' + 2 \mathcal{H} \phi' + V_{,\phi} a^2
  &= 0,
    \cr
    \delta \phi'' + 2 \mathcal{H} \delta \phi' + (k^2  + V_{,\phi\phi} a^2 )\delta \phi
  & = -2 V_{,\phi} a^2 \Psi + (\Psi' + 3 \Phi') \phi',
\end{align}
where $()'$ is the derivative with respect to the conformal time $\tau$, $\mathcal{H} = a'/a$, and $V_{,\phi} = dV/d\phi, V_{,\phi\phi} = d^2V/d\phi^2$, and $k$ is the comoving momentum.
In the non-relativistic limit, the equation of motion for the perturbation reduces to 
\begin{align}
  \delta \phi'' + 2\mathcal{H} \delta \phi'
  +  \frac{c_s^2 k^2}{a^2}\delta \phi
  & =
    4 \pi G \bar \rho \; \delta \phi,
\end{align}
where $c_s$ is the sound speed of the $\phi$ fluid. In the limit $k^2 \ll m^2$, it is given by \cite{Fan:2016rda,Salehian:2020bon}
\begin{align}
  c_s^2
  & \approx \frac{k^2 }{4m^2a^2} + \frac{\lambda \bar \rho_\phi}{4 m^2 f^2},
\end{align}
where $\bar \rho_\phi \approx m^2 \phi^2$. Comparing the pressure term with the gravity term gives an estimate of the Jeans scale. There are a few competing forces in question. When $\lambda$ is negligible, i.e., $\lambda \phi^4 \ll m^2 \phi^2$, gravity competes with the quantum pressure, which gives
\begin{align}
  \label{eq:Jeans-scale-fuzzy}
  k_J
  & = (16 \pi G_N m^2 \bar \rho_\phi a^4  )^{1/4}
    \cr
  & \approx 9 \left (\frac{m}{10^{-22}\; \mathrm{eV}}  \right )^{1/2}
    \left ( \frac{\bar \rho_{\phi,0}}{1.3 \times 10^{-6} \; \mathrm{GeV/cm^3}}\right )^{1/4}
    \left ( \frac{a}{a_{eq}} \right )^{1/4} \; \mathrm{Mpc^{-1}},
\end{align}
where $a_{eq}$ is the matter-radiation equality, $\rho_{\phi,0}$ the $\phi$ energy density today.
When $\lambda > 0$ and the self-interaction term is non-negligible, gravity needs to compete directly with $\lambda \phi^4$ term, which gives a Jeans scale
\begin{align}
  k_J
  & =
    (16 \pi G_N/\lambda)^{1/2} mfa\
    \cr
  & =
    2.7 \; \lambda^{-1/2} \left ( \frac{m}{10^{-20} \; \mathrm{eV}} \right )
    \left ( \frac{f}{10^{13} \; \mathrm{GeV}} \right )
    \left ( \frac{a}{a_{eq}} \right ) \; \mathrm{Mpc}^{-1}.
\end{align}
When $\lambda < 0$ and non-negligible, the quantum pressure competes with both gravity and the attractive self-interaction. This leads to
\begin{align}
  k_J^2
  & =
    \frac{a^2}{2} \left (
    \left [
    64 \pi G_N m^2 \bar \rho_\phi
    + \left (\frac{\lambda \bar \rho_\phi}{f^2} \right )^2
    \right ]^{1/2}
    - \frac{\lambda \bar \rho_\phi}{f^2} 
    \right ), 
\end{align}
which leads to possible gravitational collapse assisted by the attractive self-interaction at even smaller scales.
As a special case, in \cite{Zhang:2017flu,Arvanitaki:2019rax}, it is pointed out that if $\phi$ starts  to roll with a specific initial condition, such as from the hilltop part of a cosine potential, $(V_{,\phi\phi} + k^2/a^2)$ term can be negative all together for certain $k$ modes, without the need of gravity. This allows a quick growth of these $k$ modes before matter domination.

Alternatively, one can understand the deviation of ultralight dark matter from cold dark matter (CDM) as whether the density contrast of a given $k$ mode (in comoving frame), $\delta \rho_\phi(k)/\rho_\phi$,  grows the same as $\delta \rho_{CDM}(k)/\rho_{CDM}$. Roughly speaking, modes that enter the horizon after $m \sim H(z)$ grows the same as CDM. Those that enter the horizon before this do not immediately grow as $\delta \propto a$ during matter domination, because the background $\rho_\phi$ is still frozen, hence the deviation from CDM. This leads to the same estimate as comparing the pressure and gravity.

\subsubsection{Stable self-gravitating structures}
\label{sec:stable-self-grav}
Spherically symmetric self-gravitating BEC structure is verified to be stable against radial perturbation~\cite{Schiappacasse:2017ham,Guo:2020tla}. There are many studies of the stability of stellar scale BEC's  \cite{Colpi:1986ye,PhysRevD.38.2376,Chavanis:2011zi,Eby:2015hyx,Eby:2016cnq,Eby:2017teq,Schiappacasse:2017ham,Croon:2018ybs} and galactic scale BEC's \cite{Deng:2018jjz} to name a few. Instead of using approximations of high precision \cite{Eby:2017teq,Eby:2018dat}, here we briefly outline the estimate using a simple ansatz, which is shown to agree with the numerical solution reasonably well:
\begin{align}
  \label{eq:exponential-ansatz}
  \phi(r) & \approx
         \sqrt{\frac{N}{m\pi R^3}} \mathrm{e}^{-r/R}, 
\end{align}
where $N$ is the total number of particles, and $R$ is the radius where $\phi$ starts to drop exponentially. In other words, $R$ parametrizes the characteristic size of the BEC object.
The energy of such a system can be broken into kinetic, self-interaction, and gravitational energies, 
\begin{equation}
  H = H_{kin} + H_{int} + H_{grav}, 
    \label{eq:energy-of-NR-system}
\end{equation}
where
\begin{eqnarray}
  H_{kin}
 & = &
   -   \frac{1}{2}\int d^3 x \;  \phi(r) \nabla^2 \phi(r)
   = \frac{N}{2m R^2}   , \cr
   H_{int}
  & =& 
    \frac{    \lambda m^2  }{4 f^2} \int d^3 x \; \phi(r)^4
    =      \frac{    \lambda     N^2}{32 \pi f^2 R^3}
, \cr
    H_{grav}
  & = &
    -G_Nm^4 \int_0^\infty \left (\int_0^{r}
    \phi(r) ^2
    \; 4\pi r^{\prime 2} \;dr' \right )
    \frac{1}{r}
    \phi(r) ^2
    \; 4\pi r^2 \;dr
    =
    - \frac{5 G_Nm^2N^2}{16 R},
    \label{eq:Hflat}
\end{eqnarray}
and the full Hamiltonian is 
\begin{align}
  H & =
      \frac{N}{2m R^2}       \mp \frac{|\lambda| N^2}{32 \pi f^2 R^3}          - \frac{5 G_Nm^2N^2}{16 R},
     \label{eq:HNR}
\end{align}
where the upper (lower) sign in the second term corresponds to the attractive (repulsive) case.
\begin{figure}[t]
  \centering
  \includegraphics[width=.5\textwidth]{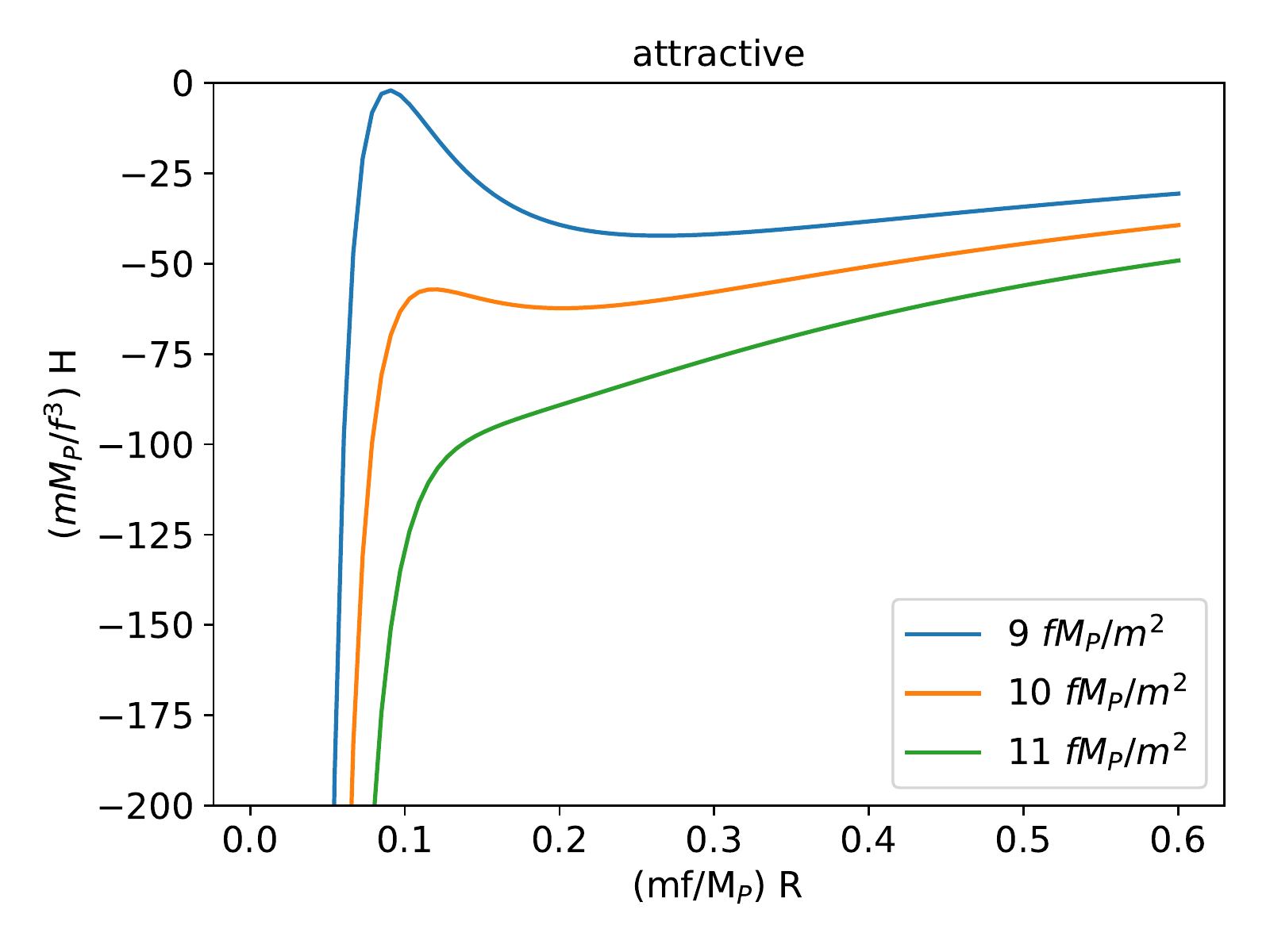}
  \includegraphics[width=.49\textwidth]{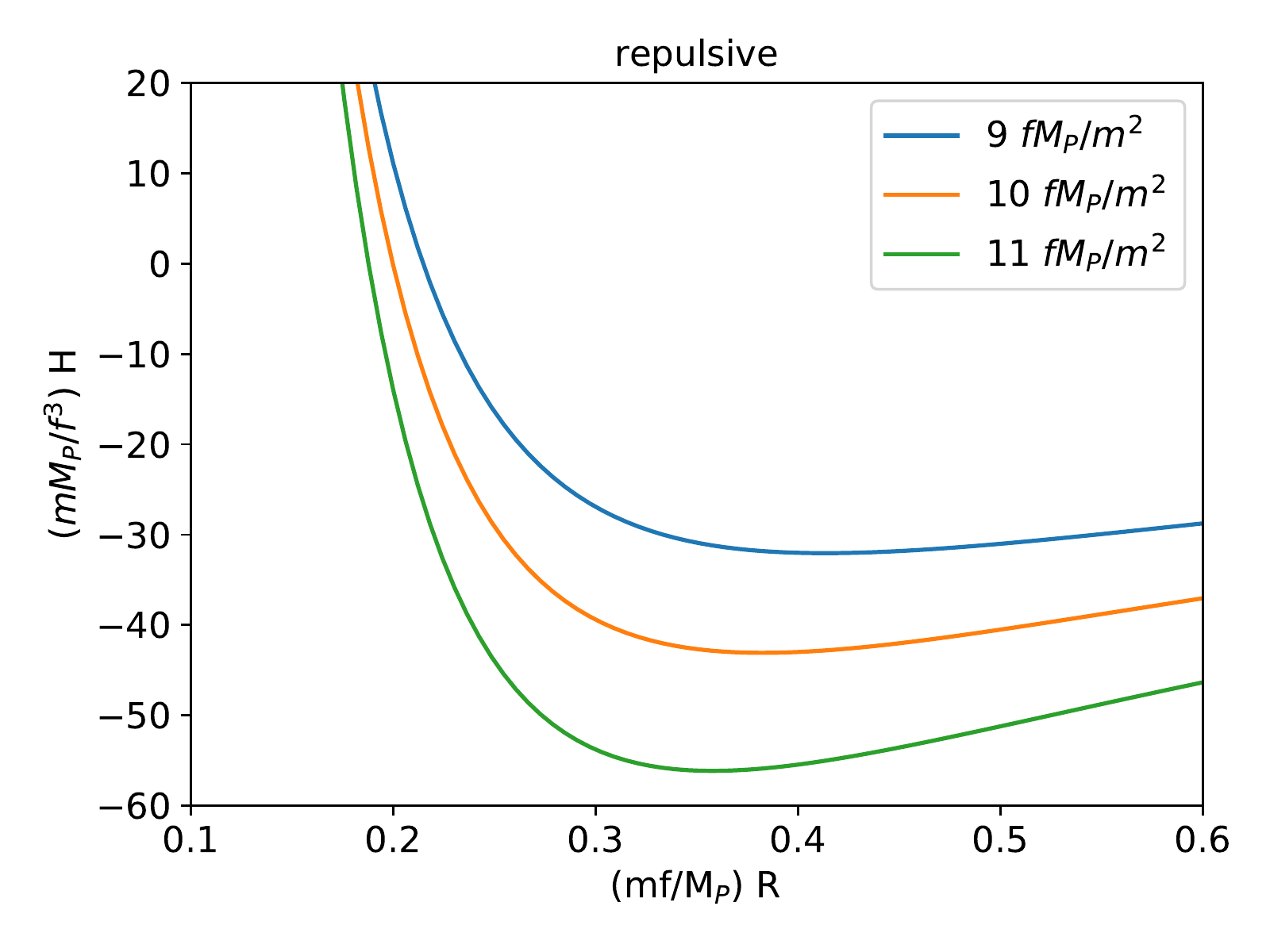}
  \caption{We show the Hamiltonian in the case of attractive self-interaction (left) and repulsive self-interaction (right). In the left panel, from $N=9 fM_P/m^2$ to $N=10 fM_P/m^2$, the local maximum moves to the right and the local minimum
    moves to the left. At $N=11 fM_P/m^2$, the local minimum is lost,
    so there is no stable dark star beyond this number of particles.
  In the right panel, there is always a minimum (corresponding to a stable solution) at any $N$.
  }
  \label{fig:HR}
\end{figure}
A few benchmarks of $H(N,R)$ are plotted out with arbitrarily rescaled units in Fig.~\ref{fig:HR}. From this one can easily see that for $\lambda>0$ the system is stablized while for $\lambda < 0$ the system can be de-stabilized when $\lambda \phi^4$ term becomes large compared to the quantum pressure from the $k^2 \phi^2$ term. When $\lambda$ is negligible, a given $m$ leads to a class of solutions that are related to each other by a $M \sim 1/R$ scaling. This scaling breaks down when general relativity effect kicks in at large density as shown in \cite{Croon:2018ybs}.

It is noted that there are a few variations beyond the simplest spherical setup.
There are studies on self-gravitating BEC structures without the assumption of a spherical symmetry. The spherical symmetry can either be broken by a non-spherical source, such as baryons disk \cite{Bar:2019bqz}, dark disk \cite{Alexander:2019qsh}, or due to rotational excitations \cite{Visinelli:2017ooc,Eby:2017teq,Hertzberg:2018zte,Kling:2020xjj}, to give a few examples. 

Beyond the single scalar assumption, there are studies of the BEC with multiple species. To name a few, assuming non-interacting ultralight scalars, \cite{Broadhurst:2018fei} proposes a solitonic origin of the Nuclear Star Cluster in the Milky Way. Assuming non-zero self-interactions, \cite{Eby:2020eas} analytically studies the mass-radius scaling relation in the context of multiple scalars; \cite{Guo:2020tla} studies the stability and scaling behavior numerically and provides some analytical interpretation. Besides the self-interaction, \cite{Guo:2020tla} also studies the case where non-gravitational interaction is present between the two species. It is observed that a repulsive interaction (+$\phi_1^2 \phi_2^2$) between the two scalars can stabilize the BEC structure even if each has attractive ($-\phi^4$) self-interactions. 

\subsection{Simulations of the ultra-light dark matter}
\label{sec:simul-ultra-light}

There have been a few simulations showing evidence of BEC structures forming on the galactic scales from  gravitational relaxation \cite{Schive:2014dra,Schive:2014hza,Schwabe:2016rze,Veltmaat:2016rxo,Mocz:2017wlg}. The simulations are performed with the scalar mass to be $m \sim 10^{-22}\;\mathrm{eV}$. It is observed that a BEC core, i.e., soliton, forming at the center of simulated galaxies, which matches to {the Navarro-Frenk-White (NFW) dark matter halo profile~\cite{Navarro:1995iw,Navarro:1996gj}} at large radius. While the soliton core profiles are consistent with each other, the NFW tail and the transition between the BEC core and NFW tail are slightly different. For a direct comparison of a few simulated profiles, see \cite{Bar:2018acw}.

While the relation between the BEC core and the halo remains an open question, in particular, the authors of \cite{Schive:2014hza} observe a relation between the two, which can be phrased as
\begin{align}
  M_{BEC}
  \approx \left (\frac{|E_h|}{M_h} \right )^{1/2} \frac{M_{Pl}^2}{m},
\end{align}
where $E_h$ is the virial energy of the halo, $M_h$ the virial mass. Utilizing the scaling of the BEC core, without loss of any information, one can express this empirical relation as \cite{Bar:2018acw}
\begin{align}
  \frac{E_{BEC}}{M_{BEC}}
  \approx
  \frac{E_{h}}{M_{h}}.
\end{align}
In \cite{Mocz:2019pyf}, the interplay between BEC formation and baryonic physics is studied. It is observed that the BEC formation is largely unaffected by the baryonic feedback, while the BEC imprints the distribution of gas and stars with cored structure. 
In \cite{Amin:2019ums}, the role of attractive self-interaction in BEC formation is verified in simulations. 

\subsection{Experimental probes}
\label{sec:exper-constr}

From bullet cluster, potential self-interaction is constrained to be $\sigma/m \lesssim 1 \; \mathrm{cm^2/g}$ if $\phi$ makes up all dark matter. This translates to a constraint on the self-interaction \cite{Fan:2016rda}
\begin{align}
  \lambda \left ( \frac{m}{f} \right )^2 \lesssim 10^{-11} \; \left ( \frac{m}{\mathrm{eV}} \right )^{3/2}.
\end{align}
If ultralight dark matter takes upon a significant fraction of the total dark matter density, from Eq.~(\ref{eq:Jeans-scale-fuzzy}) one can see that structures will be suppressed at a scale that is below the Jeans scale. In the range of $m < 10^{-25}\;\mathrm{eV}$, \cite{Hlozek:2014lca} shows that there is suppression in the linear regime of the matter power spectrum, starting as small as $k \sim 0.03\; \mathrm{Mpc}^{-1}$, which leads to significant change of CMB anisotropy compared to CDM. This constrains the fraction of ultralight dark matter of the total dark matter to be below $\sim 5\%$ in the range of $10^{-33}\;\mathrm{eV}$ to $10^{-26}\; \mathrm{eV}$. In \cite{Hlozek:2016lzm}, a study of weak gravitational lensing for the CMB at high $\ell$ shows future experiment CMB-S4 can probe ultralight dark matter with mass up to $10^{-24}\;\mathrm{eV}$.
At higher mass, nonlinear perturbations or simulations are needed to distinguish between CDM scenario and ultralight dark matter scenario. The work in \cite{Marsh:2013ywa,Marsh:2016vgj,Du:2016zcv} shows that for $m= 10^{-22} \;\mathrm{eV}$, the halo mass function is affected at scale as large as $10^{10} M_\odot$, and sharply cut off at $10^{7} - 10^{8} \; M_\odot$.

Probes of structure suppression at this scale include Lyman-$\alpha$ (excluding $10^{-22}\;\mathrm{eV} \lesssim m \lesssim 2.3 \times 10^{-21} \; \mathrm{eV}$) \cite{Armengaud:2017nkf,Irsic:2017yje,Kobayashi:2017jcf,Nori:2018pka,Hui:2016ltb,Garzilli:2019qki}, halo mass function from stellar stream \cite{Banik:2019smi, Schutz:2020jox} and strong lensing \cite{Dalal:2001fq,Vegetti:2008eg,Li:2015xpc,Penarrubia:2017nzw,Asadi:2017ddk,Mao:2017auo,Minor:2016jou,Despali:2016meh,Daylan:2017kfh,Gilman:2019nap,Schutz:2020jox} ($m \gtrsim 2.1\times 10^{-21} \; \mathrm{eV}$), from Milky Way satellite counting \cite{Maccio:2009isa,Polisensky:2010rw,Lovell:2013ola,Jethwa:2016gra,Kim:2017iwr,Nadler:2019zrb}, and galaxy UV luminosity function \cite{Bouwens:2014fua,Livermore:2016mbs,Schive:2015kza,Corasaniti:2016epp,Menci:2017nsr}. Alternatively, if one takes the empirical soliton-halo relation from simulation \cite{Schive:2014dra,Schive:2014hza}, the soliton-halo profile can be constrained by rotation curve data \cite{Bar:2018acw,Bar:2019bqz} and stellar orbits around Sgr A* (to exclude $2\times 10^{-20} \; \mathrm{eV} \lesssim m \lesssim 8 \times 10^{-19}\; \mathrm{eV}$) and M87* (to exclude $m\lesssim 4 \times 10^{-22} \; \mathrm{eV}$) \cite{Bar:2019pnz}.

On the stellar scale, $m \sim 10^{-10} \; \mathrm{eV}$, possible BEC structures can exist in the form of exotic compact objects $\sim M_{\odot}$ named boson stars \cite{Colpi:1986ye,PhysRevD.38.2376,Chavanis:2011zi,Eby:2015hyx,Eby:2016cnq,Eby:2017teq,Schiappacasse:2017ham,Croon:2018ybs,Colpi:1986ye,PhysRevD.38.2376,Bernal:2009zy,Liebling:2012fv,Kling:2017mif,Kling:2017hjm,Eby:2018dat,Eby:2017teq,Eby:2018dat,Eby:2017teq,Croon:2018ybs,Kling:2020xjj,Chavanis:2011zi,Chavanis:2011zm,Chen:2020cef,Visinelli:2017ooc}. 
That includes gravitational wave from boson star mergers \cite{Cardoso:2016oxy,Giudice:2016zpa,Palenzuela:2017kcg,Bezares:2018qwa,Croon:2018ybs,Croon:2018ftb,Hertzberg:2020dbk}, boson stars in an extreme mass ratio inspiral system \cite{Guo:2019sns}, and boson star decay products \cite{Hertzberg:2020dbk,Hertzberg:2018zte}.

\section{Conclusions}

We bring our review to a conclusion by discussing, in turn, the current state and future prospects of each of the topics we have covered.

High density astrophysical environments have been used quite successfully to constrain axions since the 1980s.  While powerful, at present many of these constraints are subject to uncertainties coming from astrophysics and nuclear theory.  The production rate of axions in the core of a neutron star is very sensitive to the types of pairing in nuclear matter and, although not the focus of our review, the presence of exotic phases deep in the neutron star.  The effects of axions on neutron star mergers - either in the inspiral or postmerger (see Sec.~\ref{sec:ns_merger_generalities}) - appear to be subtle and will likely require (at the very least) a much better understanding of the nuclear equation of state, the behavior of the neutrinos, and the role of transport in the remnant in order to measure.  Fortunately, there are a myriad of recent observations that have been able to constrain some of these uncertainties, and more observations and experiments are on the horizon.  Much progress has been made in constraining the size of the superfluid gaps, for example, by studying the cooling of neutron stars (as discussed in Sect.~\ref{sec:magnetar_emissivity}).  Some constraints on the particle content of dense matter (including the presence of some exotic phases) have been found through studying the thermal relaxation of neutron stars \cite{Brown:2017gxd,Cumming:2016weq,Alford:2019oge}.  We have learned about the stiffness of the nuclear equation of state at different densities as a result of our observation of several $2M_{\odot}$ neutron stars \cite{Demorest:2010bx,Antoniadis:2013pzd,Cromartie:2019kug} and through a recent measurement of the neutron star radius by the NICER experiment \cite{Riley:2019yda,Raaijmakers:2019qny,Miller:2019cac}.  The NICER experiment also found evidence that the magnetic field near the surface of PSR J0030+0451 deviates from the expected dipole behavior \cite{Bilous:2019knh}, which may have implications for axion-photon conversion near the surface of the star.  On the merger side, numerical simulations and gravitational and electromagnetic observations of neutron star mergers are used in concert to develop an understanding of these complex events.  Future runs of Advanced LIGO and Advanced Virgo should provide us with many more merger events to study, and future gravitational wave detectors will enable us to see the postmerger gravitational wave signal which has so far been hidden.  

Magnetars, covered in Sec.~\ref{sec:magnet}, are a promising tool to constrain axions because of their relatively high core temperature combined with a high magnetic field strength, creating the possibility of a unique photon signal coming from the existence of axions.  However, studying magnetars comes with its own uncertainties.  The surface temperature of magnetars is anomalously high, seemingly violating the core-crust temperature relationship that exists in normal neutron stars, leading to the belief that the magnetic field is somehow involved in heating the neutron star crust.  A consequence of our ignorance about the crustal heating mechanism is that it is difficult to be precisely know the temperature of the magnetar core, upon which the axion luminosity strongly depends.  Beyond the core temperature, the extremely high magnetic fields present in magnetars can lead to other axion production processes, for example, the transition of electrons or protons between Landau levels \cite{Maruyama:2017xzl,Kachelriess:1997kn,Borisov:1994wg}, the scattering of electrons from magnetic flux tubes, among other processes discussed in \cite{Potekhin:2015qsa}.  These processes have yet to be included in efforts to constrain axions using magnetars (or neutron star mergers, where it is possible that hydrodynamic and magnetohydrodynamic instabilities produce extremely strong magnetic fields \cite{Ciolfi:2020cpf,Price:2006fi}).  As theoretical models of magnetars develop (see \cite{Gourgouliatos:2018efn}) and as more magnetars are discovered, magnetars will play an increasingly important role in constraining axions.  

There have been recent developments in expectations of the fundamental properties of axions in dense matter.  In particular, it was recently found that a finite baryon density environment causes the axion mass to decrease and the axion coupling to neutrons to be enhanced by up to one order of magnitude \cite{Balkin:2020dsr}.  Even more significant changes can occur for axions in a kaon-condensate or a color-flavor-locked quark matter phase.  How this finite-density modification affects the phenomenology of axions in neutron stars is yet to be determined. 

Beyond the usual couplings of ALPs with nucleons, electrons, and photons, the possibility of couplings between ALPs and other standard model particles is beginning to be investigated.  For example, the axion-muon coupling can be constrained quite dramatically by SN1987a \cite{Bollig:2020xdr,Croon:2020lrf}.  Recent calculations indicate that a significant thermal population of pions might exist in supernovae \cite{Fore:2019wib}, reviving the possibility of axion production from the process $\pi^- + p \rightarrow n + a$.  This process was found to significantly enhance axion production in supernovae and push the peak of the spectrum of emitted axions to higher energies \cite{Carenza:2020cis}.  

We now turn to the second set of topics discussed in our review: laboratory-produced axion or ALP searches. These searches are less reliant on astrophysical assumptions on which the previously discussed searches are often based; hence they hold out the opportunity of setting the most conservative limits in the axion parameter space. We have discussed this aspect of axion searches in Sect.~\ref{sec:labsearch}, focusing on recent developments in  accelerator-based and reactor-based facilities.  
The first half was devoted to neutrino facilities including beam-dump type (Table~\ref{tab:listdump}) and reactor neutrino (Table~\ref{tab:listreactor}) experiments, where the axions can be copiously produced in addtion to neutrions. 
There are three axion detection channels: through their decay, through their scattering on detector material, and through their conversion to photons in the presence of a magnetic field. 
These three channels are suited for different regions of axion parameter space, so that they can provide information complementary to one another and allow for more exhaustive exploration of axion parameter space in the ongoing and upcoming neutrino experiments. 
By contrast, ALP searches at energy-frontier colliders (Table~\ref{tab:colliders}) were discussed in the other half of Sect.~\ref{sec:labsearch}. Due to their large center-of-mass energy, they will allow for unprecedented opportunities to investigate axion parameter space toward the higher end of the axion mass through various search channels.  

Finally, we turn to the last set of topics we have covered in this review: the topic of BECs. Towards the lighter end of the axion mass spectrum, collective behavior due to BEC formation could imply new probes on stellar and galactic scales. At axion masses around $10^{-10}~\mathrm{eV}$ and $10^{-22}\; \mathrm{eV}$, they can be potentially probed by stellar dynamics and galactic dynamics, respectively. The former indicates great opportunities from gravitational wave astronomy including binary systems consisting of boson stars and observation of its decay products. The latter motivates cosmological measurements of the power spectrum toward even smaller scales. In addition, understanding the behavior of the BEC calls for dedicated simulations. For example, the origin of the relation between the BEC mass and the halo mass, observed empirically in a few simulations, is currently an open question. New simulations with different scalar masses are required to address this and distinguish competing explanations.

As a final remark, we emphasize that axion or ALP physics has served as a well-motivated framework in terms of beyond-Standard Model model building and related new signal searches. We hope that this review will provide guidance to beginning researchers, and be of use to experts as well.

We end this review with a list of detection strategies that we did not discuss. Many of these topics are standard and the latest results are covered in the excellent reviews \cite{Graham:2015ouw, Irastorza:2018dyq, Sikivie:2020zpn}.
\begin{itemize} \itemsep1pt \parskip0pt \parsep0pt
\item[$(i)$] Helioscopes such as   CAST \cite{Zioutas:1998cc,Anastassopoulos:2017ftl} and IAXO \cite{Irastorza:2013dav} constrain the ALP-photon coupling $g_{a\gamma\gamma}$ by searching for photons obtained from conversion of ALPs emitted by the sun. The sensitivity of this search method to the coupling goes as $ \sim g^4_{a\gamma\gamma}$, with a contribution of $g^2_{a\gamma\gamma}$ coming from Primakoff production of ALPs in the sun, and a contribution of $g^2_{a\gamma\gamma}$ coming from the subsequent conversion of the ALPs to photons in the magnetic field of the apparatus. 

\item[$(ii)$] Cavity haloscope experiments that probe cold ALP dark matter: ADMX \cite{Asztalos:2001tf}, HAYSTAC \cite{Brubaker:2016ktl}, etc. These searches can constrain $g_{a\gamma\gamma}\times \sqrt{\rho}$, where $\rho$ is the local ALP dark matter density, for some ALP masses between $10^{-7}$ eV and $\sim \mathcal{O}(\rm few \,\, times \,\, 10^{-5})$ eV. There are also other  ALP searches that exploit the fact that ALP cold dark matter behaves as a classical oscillating field in the current universe: experiments include those using wire arrays (ORPHEUS \cite{Rybka:2014cya}) , dielectric plates (MADMAX \cite{TheMADMAXWorkingGroup:2016hpc}), NMR (CASPEr \cite{JacksonKimball:2017elr}), LC circuits (ABRACADABRA  \cite{Kahn:2016aff,Salemi:2019xgl}), birefringent cavities (ADBC \cite{Liu:2018icu}, DANCE \cite{Obata:2018vvr}) and interferometry \cite{Melissinos:2008vn,DeRocco:2018jwe}.

\item[$(iii)$] Light-shining-through-walls (LSW) experiments including ALPS I/II~\cite{Ehret:2010mh,Bahre:2013ywa}, CROWS~\cite{Betz:2013dza}, and OSQAR~\cite{Ballou:2015cka} which constrain $m_a < 3 \times 10^{-4}$ eV and $g_{a\gamma\gamma} < 3.5 \times 10^{-8}$ GeV$^{-1}$.
\end{itemize}

\section{Acknowledgements}
The work of JFF is supported by NSERC. HG and KS are supported by the U.~S. Department of Energy grant DE-SC0009956.  The work of SPH is supported by the U.~S. Department of Energy grant DE-FG02-00ER41132 as well as the National Science Foundation grant No.~PHY-1430152 (JINA Center for the Evolution of the Elements).  The work of DK is supported by DOE under Grant No. DE-FG02-13ER41976/ DE-SC0009913/DE-SC0010813.
The work of CS is supported by the Foreign Postdoctoral Fellowship Program of the Israel Academy of Sciences and Humanities, partly by the European Research Council (ERC) under the EU Horizon 2020 Programme (ERC-CoG-2015 - Proposal n. 682676 LDMThExp), and partly by Israel Science Foundation (Grant No. 1302/19). 

\bibliography{bibliography}



\end{document}